\documentclass{aa}  

\usepackage[colorlinks=true, linkcolor = blue,
            urlcolor  = blue,
            citecolor = blue,
            anchorcolor = blue]{hyperref}
\usepackage{graphicx}
\usepackage{txfonts}
\usepackage{natbib}
\usepackage{enumitem} 
\usepackage{color}
\usepackage{threeparttable}
\usepackage{cancel}
\bibpunct{(}{)}{;}{f}{}{,} 
\usepackage{siunitx}
\newcommand{\Gaia}{\textit{Gaia}}

\newcommand{\logg}{log \textit{g}}
\newcommand{\En}{$\mathrm{E}_\mathrm{n}$}

\newcommand{\Lz}{$\mathrm{L}_\mathrm{z}$}

\newcommand{\zmax}{$\mathrm{Z}_\mathrm{max}$}

\newcommand{\Jr}{$\mathrm{J}_\mathrm{r}$}
\newcommand{\Jphi}{$\mathrm{J}_\mathrm{\phi}$}
\newcommand{\Jz}{$\mathrm{J}_\mathrm{z}$}
\newcommand{\Jx}{$\mathrm{J}_\mathrm{\bar{x}}$}
\newcommand{\Jy}{$\mathrm{J}_\mathrm{\bar{y}}$}
\newcommand{\Jtot}{$\mathrm{J}_\mathrm{tot}$}

\newcommand{\Vr}{$\mathrm{V}_\mathrm{r}$}
\newcommand{\Vphi}{$\mathrm{V}_\mathrm{\phi}$}

\newcommand{\Msun}{$\mathrm{M}_\odot$}
\newcommand{\Mstar}{$\mathrm{M}_\star$}
\newcommand{\FeH}{$\left[\mathrm{Fe}/\mathrm{H}\right]$}
\newcommand{\BaFe}{$\left[\mathrm{Ba}/\mathrm{Fe}\right]$}
\newcommand{\EuFe}{$\left[\mathrm{Eu}/\mathrm{Fe}\right]$}
\newcommand{\EuBa}{$\left[\mathrm{Eu}/\mathrm{Ba}\right]$}
\newcommand{\EuMg}{$\left[\mathrm{Eu}/\mathrm{Mg}\right]$}
\newcommand{\MgFe}{$\left[\mathrm{Mg}/\mathrm{Fe}\right]$}

\newcommand{\BaEu}{$\left[\mathrm{Ba}/\mathrm{Eu}\right]$}

\newcommand{\MgMn}{$\left[\mathrm{Mg}/\mathrm{Mn}\right]$}
\newcommand{\AlFe}{$\left[\mathrm{Al}/\mathrm{Fe}\right]$}

\newcommand{\MnFe}{$\left[\mathrm{Mn}/\mathrm{Fe}\right]$}
\DeclareSIUnit\annum{a}

\begin{document} 

\title{Exploring the chemodynamics of metal-poor stellar populations}

\titlerunning{Chemodynamics of metal-poor stellar populations}
\authorrunning{da Silva \& Smiljanic}
    
\author{A. R. da Silva
          \inst{1}
          \and
          R. Smiljanic \inst{1}
          }

   \institute{
            Nicolaus Copernicus Astronomical Center, Polish Academy of Sciences, ul. Bartycka 18, 00-716, Warsaw, Poland \\
            \email{arodrigo@camk.edu.pl}
             }

    \date{Received June 19, 2023 / Accepted July 7, 2023}

 
  \abstract
   {Metal-poor stars are key for studying the formation and evolution of the Galaxy. Evidence of the early mergers that built up the Galaxy remains in the distributions of abundances, kinematics, and orbital parameters of its stars. Several substructures resulting from these mergers have been tentatively identified in the literature.}
   {We conduct a global analysis of the chemodynamic properties of metal-poor stars. Our aim is to identify signs of accreted and in situ stars in different regions of the parameter space and to investigate their differences and similarities.} 
   {We selected a sample of about 6600 metal-poor stars with [Fe/H] $\leq$ $-$0.8 from DR3 of the GALAH survey. We used unsupervised machine learning to separate stars in a parameter space made of two normalised orbital actions, plus [Fe/H] and [Mg/Fe], without additional a priori cuts on stellar properties.}
   {We divided the halo stars in four main groups. All groups exhibit a significant fraction of in situ contamination. Accreted stars of these groups have very similar chemical properties, except for those of the group of stars with very retrograde orbits. This points to at most two main sources of accreted stars in the current sample, the major one related to \emph{Gaia}-Enceladus and the other possibly related to Thamnos and/or Sequoia. Stars of \emph{Gaia}-Enceladus are r-process enriched at low metallicities, but a contribution of the s-process appears with increasing metallicity. A flat trend of [Eu/Mg] as a function of [Fe/H] suggests that only core collapse supernovae contributed to r-process elements in \emph{Gaia}-Enceladus.}
   {To better characterise accreted stars in the low metallicity regime, high precision abundances and guidance from chemical evolution models are needed. It is possible that in situ contamination in samples of accreted stars has been underestimated. This can have important consequences for attempts to estimate the properties of the original systems.}

   \maketitle
%

\section{Introduction}

The quest to understand how galaxies form is one of the main endeavours of modern-day astrophysics. The Milky Way, our home galaxy, plays a central role in this quest, as stellar populations can be resolved in finer details than is possible for other galaxies. In the modern $\Lambda$CDM paradigm \citep[see][]{Springel2006, Spergel2007}, large galaxies grow hierarchically through the accretion and merging of smaller systems \citep[e.g.,][]{Searle1978, WhiteFrenk1991, BullockJohnston2005, BlandHawthorn2014}. The sequence of these events leaves its imprint on the distributions of stellar properties of the main host galaxy \citep{Helmi2008, Helmi2020}. 
 
Kinematic and orbital stellar parameters carry signs of this accretion history because the exchange of energy and momenta takes longer than the time that has passed since the formation of the Galaxy \citep{Eggen1962}. This process imbues the velocity space with very cold structures, detectable even after many billions of years \citep[see, e.g.,][and references therein]{Helmi2008, Martin2022, Balbinot2023}. In addition, long-lived low-mass stars retain in their stellar atmospheres the chemical abundances from the time and place of their formation. Because of this, the chemical properties of old stars also offer fundamental information needed to reconstruct the history of the Milky Way \citep{Ken2002}.

The discovery of the ongoing merger with the Sagittarius dwarf galaxy \citep{Ibata1994} demonstrated that the hierarchical assembly of the Milky Way continues to this date. Not long after, \cite{Helmi1999b} found evidence of halo star streams in the solar neighbourhood, now referred to as Helmi streams, which are relics of the Milky Way formation process \citep{Limberg2021c,RuizLara2022}. Further substructures (streams and overdensities) in the halo, including debris left by the Sagittarius dwarf, were later uncovered by large photometric surveys \citep{Newberg2002, Majewski2003, Belokurov2006, Juric2008, Perottoni2019} and more recently with \Gaia \space data \citep[e.g.][]{Necib2020,Ibata2021,Viswanathan2023}.
 
Indeed, thanks to the \Gaia \space mission \space \citep{Gaia}, the substructure of the halo and the merger history of the Milky Way are now being revealed in greater detail. \Gaia \space is bringing about a revolution in the field of Galactic archaeology by providing parallaxes and proper motions for more than $10^9$\space stars. Using data from \Gaia, \citet{Helmi2018} and \citet{Belokurov2018} identified an elliptical-like structure in the Toomre and Lindblad diagrams, which they associated with a previous major merger suffered by the Milky Way (with a satellite galaxy now referred to as \Gaia-Sausage or \Gaia-Enceladus; hereafter GE). This merger happened about \SI{9.5}{\giga\annum}\footnote{The lower-case letter ``a" is the symbol adopted by the International Astronomical Union (IAU) and the bureau of weights and measures of the International System of Units (BIPM/SI) for ``year", and is thus the symbol used here.} ago \citep{Gallart2019, Bonaca2020, Montalban2021, Borre22, GiribaldiSmiljanic23}. The GE progenitor galaxy is estimated to have had a stellar mass in the range of \Mstar $\sim$ 10$^8$ -- \SI{5e9}{} \Msun \space \citep{Vincenzo2019,Feuillet2020,MackerethBovy2020, Limberg2022, Lane2023}, although see \citet{Rey2023} for evidence that it is not straightforward to infer the mass ratio of past mergers.

Observational evidence of additional past mergers has since been discovered. The Sequoia merger is a highly retrograde substructure with weakly bound stars discovered by \citet{Barba2019} and \citet{Myeong2019}. A non-exhaustive list of proposed substructures can also include Thamnos \citep{Koppelman2019}, which could be the low-energy tail of Sequoia \citep[see e.g.][]{Kordopatis2020}, the Koala \citep{Forbes2020} or Kraken \citep{Kruijssen2019a, Kruijssen2019b,Kruijssen2020}, among others \citep[see e.g.][and references therein]{Donlon2020, Naidu2020, Necib2020, Yuan2020, Horta2021, Malhan2022}. However, careful work is still necessary to establish the reality and properties of each of these events \citep[e.g.][]{Buder2022,Donlon2022,Horta2023} and to demonstrate whether or not the identified structures actually correspond to distinct events \citep{JeanBaptiste2017, Koppelman2020, Amarante2022, Pagnini22, Rey2023}.

To understand the chemodynamic properties of the progenitors of accreted systems, disentangling the stars that originated in the Milky Way from those that were accreted is crucial. There have been several approaches to this in the literature \citep[e.g.][]{Carollo2021, Naidu2020, Feuillet2021, BelokurovKravtsov2022}, but doing so is, of course, not trivial. Even more so given that we work with noisy data. As a way to select groups of stars likely dominated by one population or another, several authors have resorted to defining boxes or straight-line cuts on the kinematic, dynamic, and/or chemical parameter space(s). Another approach is to use supervised or unsupervised machine learning methods to separate different groups, overdensities, or clusterings of stars identified as peaks over distributions of properties in multidimensional spaces made up of different chemodynamic quantities \citep[e.g.][]{Buder2022, Lovdal2022, Myeong2022, Shank2022a, Shank2022b, Shank2023, Dodd2023, GiribaldiSmiljanic23, Ou2023, Zepeda2023}. However, in many cases, the subsequent discussions usually focus on the properties of the few highlighted stellar groups. The general population of stars that remains as ``background'' tends to be ignored. Thus, such investigations provide only an incomplete assessment of the characteristics of the Milky Way metal-poor stellar populations. Focussing only on the peaks of the distributions can have a particularly impact on the identification of the trends of chemical evolution of each population part of the observed mixture \citep{GiribaldiSmiljanic23}.

Here, we present an investigation that aims for a wider discussion of the properties of old metal-poor stars to try and improve our understanding of accreted and in situ stellar populations. Our analysis tries to limit to a minimum any a priori cut (in kinematics or chemistry) used to define groups of stars of similar properties. We first used unsupervised machine learning methods to separate groups with different properties in a restricted chemodynamic space. We then discuss the more global orbital and chemical properties of these groups to understand if they can be explained by in situ or accreted stars. We are particularly interested in identifying variations of these global properties that can be a sign of different fractions between accreted and in situ stars in each of the groups that are identified. Possible changes in the chemical patterns can also reveal whether one or several merger events are needed to explain the data. The article is divided as follows. In Section \ref{sec:data}, the observational data are presented. Section \ref{sec:Methods} describes the machine learning methods applied to the data. In Section \ref{sec:results}, we present the main results and discuss the characteristics of the stellar groups identified in our sample. Finally, Section \ref{sec:conclusions} presents a summary of our findings. 

\section{Observational data}\label{sec:data}

\subsection{Chemical abundances}

\begin{figure}
    \centering
    \includegraphics[width=\linewidth]{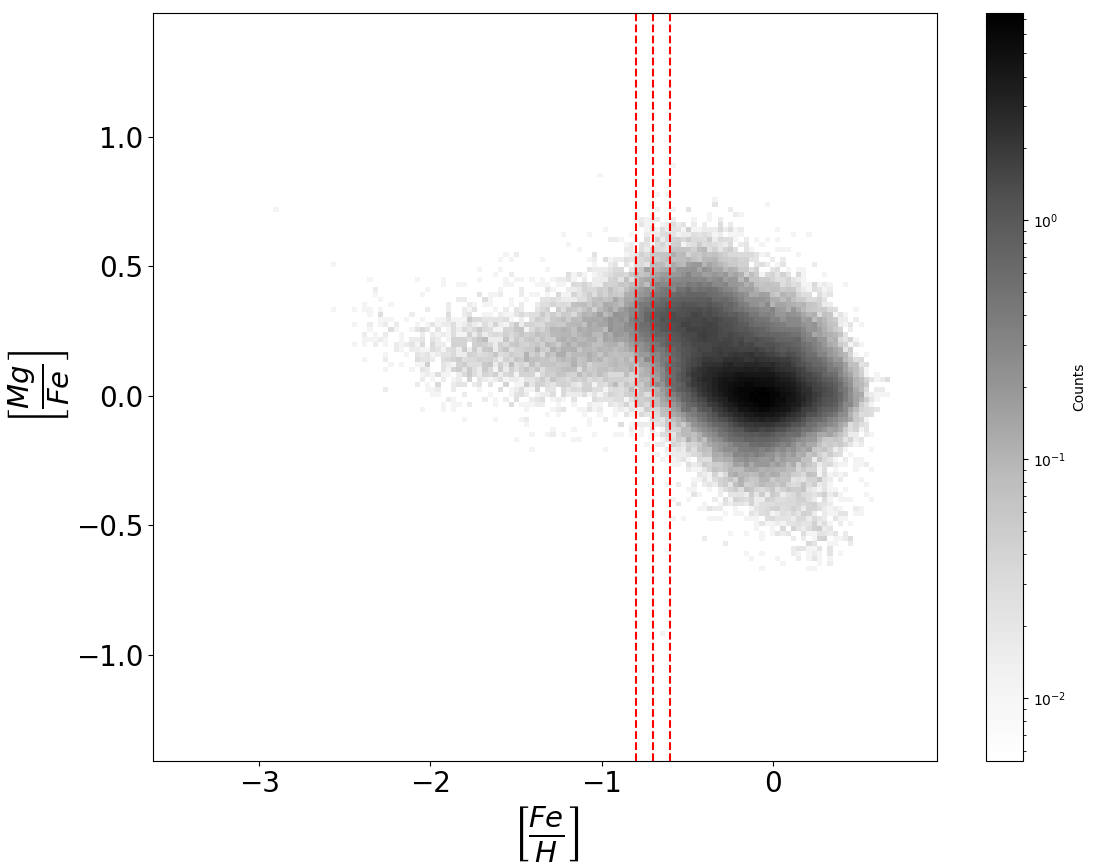}
    \caption{Diagram of [Mg/Fe] as a function of [Fe/H] for the whole GALAH sample. Dashed lines indicate values of [Fe/H] = $-$0.8 dex, $-$0.7 dex and $-$0.6 dex.}
    \label{fig:cuts}
\end{figure}

The stellar sample we analysed here was taken from Data Release (DR) 3 of the GALactic Archaeology with HERMES survey \citep[GALAH,][]{deSilva2015,Buder2021}. GALAH is a spectroscopic survey that provides chemical abundances for up to 30 elements. In particular, GALAH is one of the two current surveys that provides abundances for the heavy neutron-capture elements Ba and Eu \citep[the other one being the \emph{Gaia}-ESO Survey,][]{Gilmore2022,Randich2022}. These elements can be used to trace the contributions of nucleosynthesis through the s- and r-processes. In its DR3, GALAH provides astrophysical parameters (e.g. chemical abundances and radial velocities) for 588\,571 stars.\par
For our analysis, we chose to select only metal-poor stars with [Fe/H] $\leq$ $-$0.8. Our goal with this selection is to eliminate most of the population that is dominated by disc stars and to concentrate the analysis on the old metal-poor stellar populations dominated by the halo, where most accreted stellar populations will be found (see Fig. \ref{fig:cuts}). We expect that this simple metallicity cut will preselect a sample of heterogeneous origin, made up both of accreted and in-situ formed stars. The accreted stars can likely be traced to several mergers, but the in situ population is probably not less complex. The in situ part will include halo stars, such as what was called Erebus by \citet{GiribaldiSmiljanic23}, a fraction of old disc stars \citep[the thick disc and its metal weak component; see][]{Norris1985,Morrison1990,Beers2014}, but also those stars that have been heated to halo orbits \citep{Haywood2018,DiMatteo2019, Gallart2019}, such as what has been called Splash by \citet{Belokurov2020} and Aurora by \citet{BelokurovKravtsov2022}. Although the GE merger was proposed as the agent that heats the orbits of Splash stars \citep{Belokurov2020}, there is the possibility that local interactions between disk stars and gas might also generate such a population \citep{Amarante2020}. Most of what has been identified as the Splash by \citet{Belokurov2020} is more metal rich than our selection criteria, nevertheless, we note that other authors have suggested that this population can extend to lower metallicities, at least down to \FeH \space $\sim -$1.00 \citep{Horta2021,Donlon2022}. Stars formed in a possible starburst related to the GE merger might also be present \citep{Grand2020,Rey2023}. Furthermore, we stress that we are aware that our metallicity selection also removes part of the metal-rich low-$\alpha$ tail of the accreted populations belonging to the halo. However, we consider these low-$\alpha$ stars to have already been extensively discussed in the literature \citep[e.g.][]{Mackereth2019,Buder2022,Myeong2022}. The aim of our work is to explore the chemodynamic structure mainly toward the metal-poor region of the parameter space. Applying the metallicity selection mentioned above to the GALAH DR3 catalogue selected a sample of 24\,817 stars.\par
We decided to use the GALAH Value-Added Catalogue (VAC) which provides a cross-match of GALAH DR3 and \Gaia \space early DR3 (EDR3). This VAC contains the positions and proper motions provided by \Gaia\space \citep[\Gaia\space EDR3;][]{GaiaEDR3} and also the Bayesian distances provided by \citet{Bailer-Jones2021}. To further clean the sample, we followed the recommendations of the GALAH consortium and kept only the stars that have the flags: flag\_sp = 0, snr\_c3\_iraf > 30, flag\_fe\_h = 0 and flag\_Mg\_fe = 0 \citep[see][]{Buder2022}. As we discuss below, the values of [Fe/H] and [Mg/Fe] are used in the unsupervised machine learning analysis. This selection of high-quality results strongly reduces the final sample to 6\,618 stars. We already anticipate here that when abundances of other elements are discussed, the relevant flag for that element is used to restrict the sample to those stars with reliable abundance determination. We do not apply further cuts on abundance errors, but let us mention here that for [Fe/H] and [Mg/Fe] the selected sample has a mean error of $\pm$0.09 dex, in both cases, with a dispersion of 0.03 dex for [Fe/H] and a dispersion of 0.04 dex for [Mg/Fe].

\subsection{Orbital parameters}

We computed stellar orbits using {\sf galpy}\footnote{Available in \url{https://docs.galpy.org/en/}}, a galaxy dynamics tool written in {\sf Python} that provides a range of gravitational potentials for use in calculations of orbital and kinematic parameters \citep[see][]{Bovy2015}. One of the accepted input formats consists of an {\sf astropy} \citep{astropy:2018} object made of the right ascension ($\alpha$), declination ($\delta$), proper motions ($\mu^{*}_\alpha$ and $\mu_\delta$), radial velocity ($\varrho$) and distance ($d$), of each star in any epoch. The vast majority of the sample selected as described above has a fractional error in the parallax value of 20\% or less. We decided not to remove these few stars with larger errors, as they are not present in numbers that can potentially bias the results.

The orbits were integrated for \SI{13}{\giga\annum} assuming the Galactic potential determined by \citet{McMillan2017}. To estimate the uncertainties, a Monte Carlo simulation of 100 random samples was performed on all parameters, assuming that the errors have a Gaussian distribution. For our analysis, we extracted the following parameters: \En, the total binding energy of the orbit; \Jr, the radial action, which is associated with the orbital eccentricity; \Jphi, the azimuthal action, which is related to the rotation around the Galactic Centre; \Jz, the vertical action, which is related to how far the star moves from the galactic plane; and $e$, the eccentricity of the orbit. We adopted a frame of reference in which \Lz = \Jphi. 

\begin{figure}
    \centering
    \includegraphics[width=\linewidth]{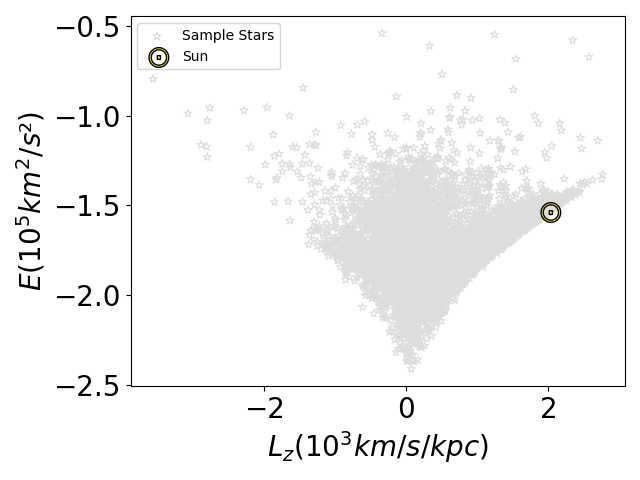}
    \includegraphics[width=\linewidth]{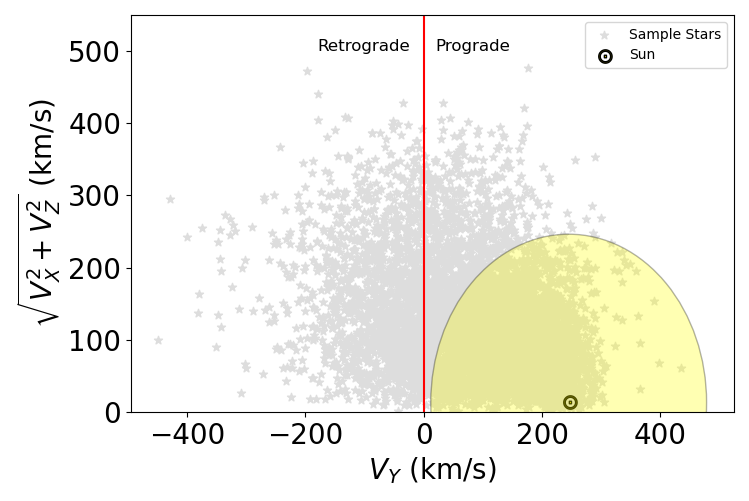}
    \caption{\emph{Top} - Lindblad diagram with our sample stars in gray. The Sun is included to serve as reference. \emph{Bottom} - Toomre diagram with our sample stars in gray. The yellow circle delineates the region of stars with total velocity below 233 km s$^{-1}$ \citep{McMillan2017}. The Sun is also shown for reference.}
    \label{fig:lbam}
\end{figure}

Figure~\ref{fig:lbam} shows diagrams that use these parameters and are commonly used to classify stars into the different Galactic stellar populations. The top panel displays the Lindblad diagram (\Lz $\times$ \En). The bottom panel displays the Toomre diagram with the stellar space velocities in Cartesian coordinates ($V_y$ is the component in the direction of Galactic rotation, $V_x$ in the radial direction, and $V_z$ in the direction of the Galactic north pole). From these plots, we can see that the sample still contains a significant number of stars with disc-like parameters in prograde orbits. The plume-like structure around zero net rotation, which has been associated with GE \citep{Helmi2018}, is also visible.

\section{Analysis methods}\label{sec:Methods}

For our analysis, we sought a way to divide and explore the different stellar populations and/or substructures without resorting to boxes or straight-line cuts in the parameter space. Our aim is to obtain a global overview of the observed distribution functions of the relevant stellar properties. We note that it is not possible to obtain information on the absolute distributions of these properties, since we do not attempt to correct the selection functions of the GALAH and \emph{Gaia} surveys, which are the sources of data we use. We also note that the constraints we applied (like the flags in chemical abundances) introduce additional biases that would need to be taken into account if one wants to recover the absolute distributions. In this first effort, we simply aim to investigate whether there are broad characteristics that can help to differentiate the stellar groups. To some extent, we also want to avoid the possibility that our prior knowledge introduces biases in our investigation of the sample. Thus, we have looked for unsupervised machine learning techniques that would help the sample itself tell how it should be best divided. We decided to build a chemodynamic space with the following quantities: two chemical parameters, \FeH\ and \MgFe, and two dynamic dimensions \Jx \space and \Jy, where \Jx and \Jy\space are the axes of the action map similar to that presented by \citet{Vasiliev2019} \footnote{Where \Jx=\Jphi/\Jtot, \Jy=(\Jr-\Jz)/\Jtot, and \Jtot = |\Jr|+|\Jphi|+|\Jz|.}. 

The diagram \MgFe \space versus \FeH \space has been extensively used to separate accreted from in situ stars \citep[e.g.][and references therein]{Buder2022,Horta2023}. It has been shown that dwarf galaxies have \MgFe \space versus \FeH \space sequences where the so-called ``knee'', i.e. the place in terms of \FeH \space values where the \MgFe \space ratio starts to decrease, occurs at lower metallicity than what is observed in Milky Way stars \citep{Venn2004,Hasselquist2021}. This is caused by a lower star formation efficiency, as the interstellar medium (ISM) cannot achieve higher values of \FeH \space before the extra Fe contribution from Type Ia supernovae becomes important \citep{Kirby2011,Matteucci2021}. Although our metallicity selection does remove a good chunk of low-$\alpha$ stars, the knee itself for accreted populations in the halo has been found between \FeH \space = $-$1.5 to $-$1.00, depending on the author \citep[e.g.][]{Mackereth2019,Feuillet2021,Buder2022}. Therefore, low-$\alpha$ abundance remains useful to define accreted populations. Chemical parameters were normalised as needed, so they have means and variances of the same order of magnitude.

Normalised actions have been used mainly to study globular clusters and their association with halo structures \citep{Vasiliev2019,Myeong2019}. A map built with these quantities is useful to separate objects by the type of orbit (prograde versus retrograde; polar versus radial; in-plane versus out-of-plane). Accreted populations in radial orbits should be easily recognised in such quantities. Nevertheless, as discussed by \citet{Lane2022}, because the actions are normalised, there can be a certain degree of confusion between halo stars and hotter thick-disc stars when their vertical and radial actions become of magnitude comparable to the angular momentum. Nevertheless, thick-disc stars should still have high-$\alpha$ abundances up to higher values of metallicities \citep[e.g.][]{Bensby2005,Fuhrmann2008,Recio-Blanco2014}. From this follows our idea of combining normalised actions and the chemical parameters \MgFe \space and \FeH \space in a multidimensional space. Our initial expectation is that this combination of quantities can break down, at least in part, the degeneracy between populations.

In the first step of the analysis, we used a dimensionality reduction technique called t-distributed stochastic neighbour embedding (t-SNE). In general terms, when t-SNE is applied to data distributed in a multidimensional space, it returns a projected two- (2D) or three-dimensional (3D) map where the neighbourhood of the points is preserved. In our case, we resorted to a projection in 2D. In the second step, we used agglomerative hierarchical clustering with Ward's method \citep{Ward1963} on top of the t-SNE projection, to identify the different groups of neighbouring points. The methodology is described in detail below.

\subsection{t-SNE}\label{sec:tsne}

The t-distributed stochastic neighbour embedding is a manifold learning method that uses the affinity between the data points as a probability. As many other dimensionality reduction tools, such as the better-known principal component analysis (PCA), t-SNE is especially useful for exploring structures from an N-dimensional problem in a 2D map.

In astronomy, t-SNE has been used for a number of applications: spectral reduction and classification \cite[see][]{Traven2017,Traven2020}, selection or derivation of the parameters of metal-poor stars \cite[see][]{Matijevic2017,Hawkins2021,Hughes2022}, identification of star clusters \cite[see][]{Kos2018,Chun2020}, and exploration of properties of stellar populations \citep{Anders2018,Queiroz2023}.

The algorithm works as follows. First, t-SNE computes the similarity between two N-dimensional vectors that represent the properties of two points in the parameter space, $x_i$ and $x_j$:
\begin{equation*}
   p_{j|i} = \frac{\exp{(-||x_i-x_j||^2/2\sigma_i^2)}}{\sum_{k\neq i}\exp{(-||x_i-x_k||^2/2\sigma_i^2)}},
\end{equation*}

where $\sigma_i$ is a generalised uncertainty that can be manually set or estimated by the algorithm. We chose the latter option. The algorithm searches for a value that makes the entropy of the distribution over neighbours equal to $\log~k$, where $k$ is the perplexity parameter. The perplexity is related to the local number of neighbours at that point.

Afterward, the similarity is made symmetric, mainly to avoid outliers:
\begin{equation*}
   p_{ij} = \frac{p_{j|i}+p_{i|j}}{2N}
\end{equation*}

The algorithm then attempts to create a map of smaller dimension, where the two original points are now characterised by the new vectors $y_i$ and $y_j$:
\begin{equation*}
   q_{ij} = \frac{(1+||y_i - y_j||^2)^{-1}}{\sum_{k\neq m}(1+||y_k - y_m||^2)^{-1}}
\end{equation*}

The aim is to make these two neighbourhood distributions, $p_{j|i}$ and $q_{ij}$, match as well as possible. The location of the points on the t-SNE map is finally given by minimising the Kullback-Leibler (KL) divergence \citep{kullback1951}:

\begin{equation*}
    KL(P||Q) = \sum_{i\neq j}{p_{ij}\log{\frac{p_{ij}}{q_{ij}}}}
\end{equation*}

The minimisation of the KL divergence is performed by means of a gradient descent. Note that the method emphasises local distances. It attempts to keep objects that were nearby in the multidimensional space to also end up close by in the final map, at the same time as it tries to separate objects that were far apart. Because of this, it is important to keep in mind that the distances in the final t-SNE map are not physically connected to the original values in the multidimensional space (i.e., the distances between the points in the final map are not linear combinations of the initial distances). The goal is simply to maintain objects with similar properties close to each other while, at the same time, separating those with very different properties.

\begin{figure}
    \centering
    \includegraphics[width=1\linewidth]{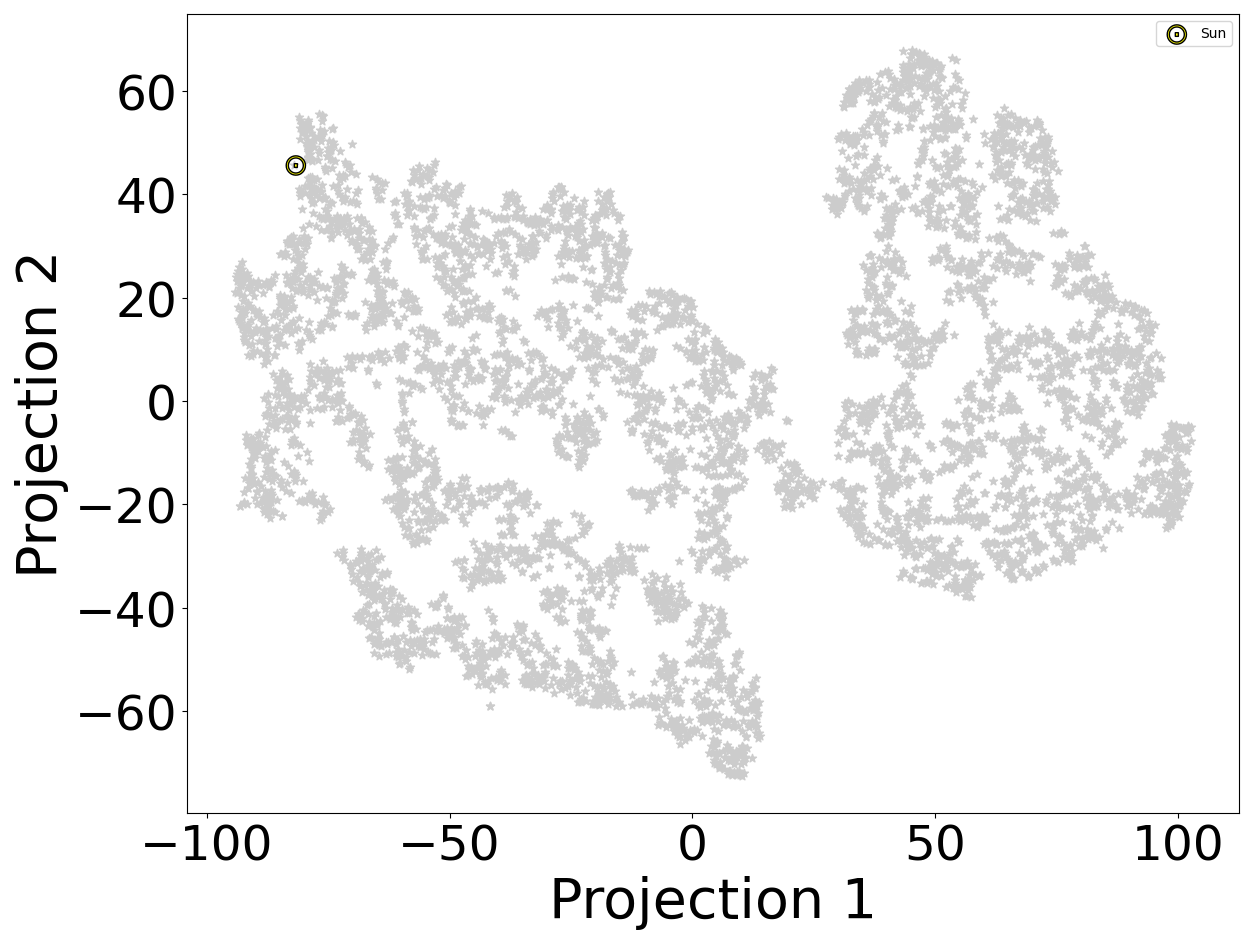}
    \caption{Reduced 2D t-SNE map for our sample of metal-poor stars. The projected dimensions themselves are not connected to the physical quantities used in the analysis. This is one example from the 50 realisations of the 2D maps that were created in our analysis.}%
    \label{fig:tsnemap}
\end{figure}

An advantage of t-SNE is that its efficiency does not depend on the density of the points, since it automatically adjusts how clumpy the projection will be. This is summarised in the perplexity parameter, which can be thought of as the average number of neighbours that one data point has. For optimal use of t-SNE, \citet{wattenberg2016} recommends exploring the perplexity parameter and how it affects the reduction in dimensionality. After performing these tests, we set the perplexity at 35.

An example of a 2D map obtained for our sample of metal-poor stars is presented in Fig.~\ref{fig:tsnemap}. The Sun has been included in the analysis to help with the visualisation. On this map, one can immediately notice structures at different levels. Stars that are clumped closely together around the Sun can be expected to show properties of disc stars (i.e., prograde circular orbits close to the Galactic plane). Metal-poor halo stars will be positioned in another region of the map. Indeed, the two main islands that can be visually identified in this figure seem to be related to a division between disc and halo stars (despite our metallicity cut, some important contamination by disc stars remained). 

It is important to note here that the position of the points and the density of the regions on the map change each time the algorithm is run. t-SNE preserves neither the global structure nor the density structure. The individual position of each point or the shape of each clump is not important. In the t-SNE results, the important factor is not how dense an area is on the final map, but the local structures that remain in the final map. As we stressed above, the goal is to keep similar data points together and to drive dissimilar data points away from each other. One other way to put this is, as t-SNE tries to agglomerate similar points together, the distance loses meaning. Two close points are similar, but the distance is not a direct measure of similarity or dissimilarity. The density and sizes of the clumps are adjusted mainly as a function of the perplexity parameter. To take some of the variation introduced by t-SNE into account, we performed 50 different realisations of the 2D map of our sample. We use these realisations to derive some statistics of how the stellar populations can be divided.

Due to the above facts, for the next step of the analysis, we decided not to use clustering methods based on density \citep[such as DBSCAN and its variations like HDBSCAN, see chapter 4 of][]{wattenberg2016} as they could in principle produce misleading results. Another problem with DBSCAN and its variations is shown in \citet{Ou2023}. As these authors discuss, HDBSCAN results are unstable when the points are resampled by their uncertainties. We actually performed a few tests using HDBSCAN but finally decided to apply a different approach to separate the structures seen on the t-SNE map. That is what we will describe next.

\begin{figure}
    \centering
    \includegraphics[width=1\linewidth]{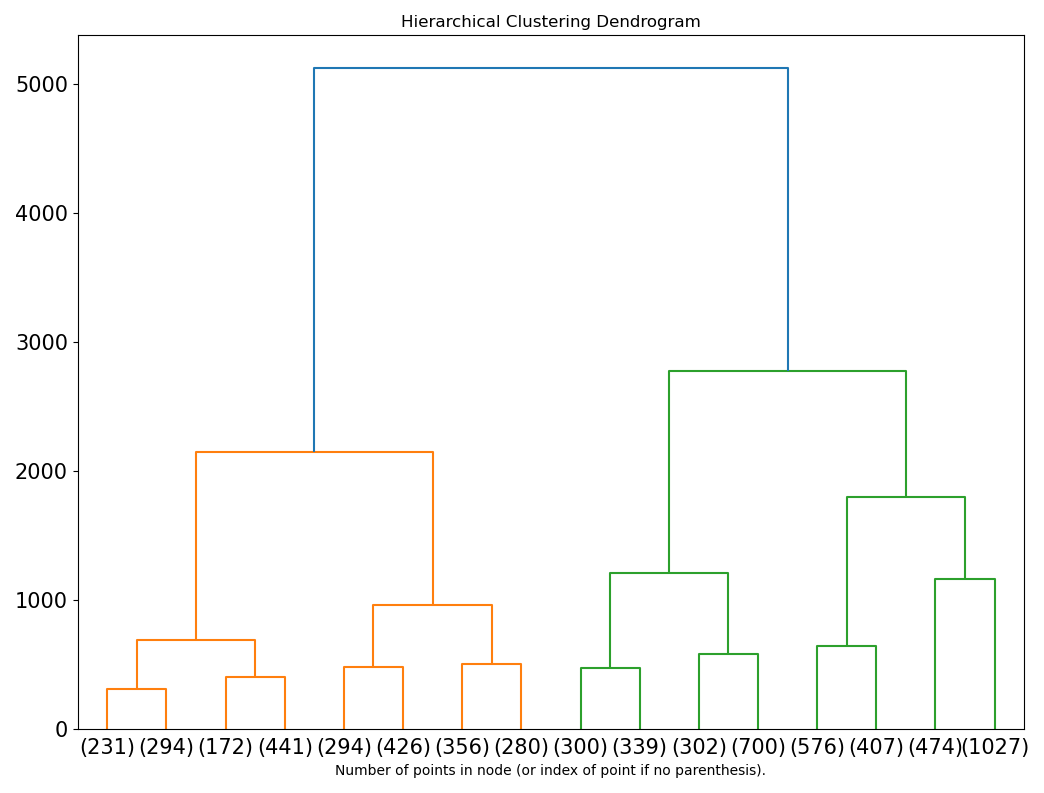}
    \caption{Dendogram of Ward hierarchical clustering for one of the 50 t-SNE realizations. In orange are the clusters that can be associated to the halo and in green the clusters with stars belonging to the disc.}
    \label{fig:dendogramhward}
\end{figure}

\subsection{Hierarchical Clustering}\label{sec:clustering}

To separate the groups (or clusters) in the t-SNE map, we used hierarchical clustering.  Hierarchical Aglomerative Clustering (HAC) tries to cluster objects by a distance, starting from the closest points and progressively including the farther ones, until all points have been grouped together. During this process, the minimum distance needed to identify a group is recorded. Even though, as we highlighted above, distances have no physical meaning on a t-SNE map, in our tests we found this to be the method that could best separate local structures for further exploration.

Several different metrics can be used to determine the type of distance to be used in HAC, from Euclidean to Mahalanobis\footnote{A distance, in multi-dimensional space, measured with respect to a centroid (such as the mean) of a distribution of points.}, for example. The method proposed by \citet{Ward1963}, which is the one we use here, consists of passing the Euclidean distances between the points through the Lance-Williams formula. This is done to compute the distances between each pair of clusters using an unstructured linkage criterion, which are then used to decide which clusters to merge next (i.e., to determine which are the closest clusters). In our case, a structured Ward hierarchical clustering with a k-Nearest Neighbour linkage was used\footnote{An example of structured and unstructured Ward's hierarchical clustering can be seen in the {\sf scikit-learn} documentation in \href{https://scikit-learn.org/stable/auto_examples/cluster/plot_ward_structured_vs_unstructured.html}{here}.}. We decided to go for this approach, as it better preserves the structure between the data points. The method was implemented using the {\sf Python} module {\sf scikit-learn} \citep{pedregosa2011}.\par

To evaluate the number of clusters in which to divide the data, we used a dendogram generated from the HAC (Fig.~\ref{fig:dendogramhward}). The vertical axis of this diagram shows the Euclidean distances between the clusters. The horizontal axis separates the number of groups that were generated at each distance. Here, we show the groups up to the third level of division. The first-level division appears to recover a separation between what seems to be halo- and disc-dominated populations. We note here that deciding which is the optimal number of groups in which the sample should be divided is subjective, since the method itself does not provide a metric that can be used to judge the goodness of a subdivision with respect to another.

\begin{figure}
    \centering
    \includegraphics[width=1\linewidth]{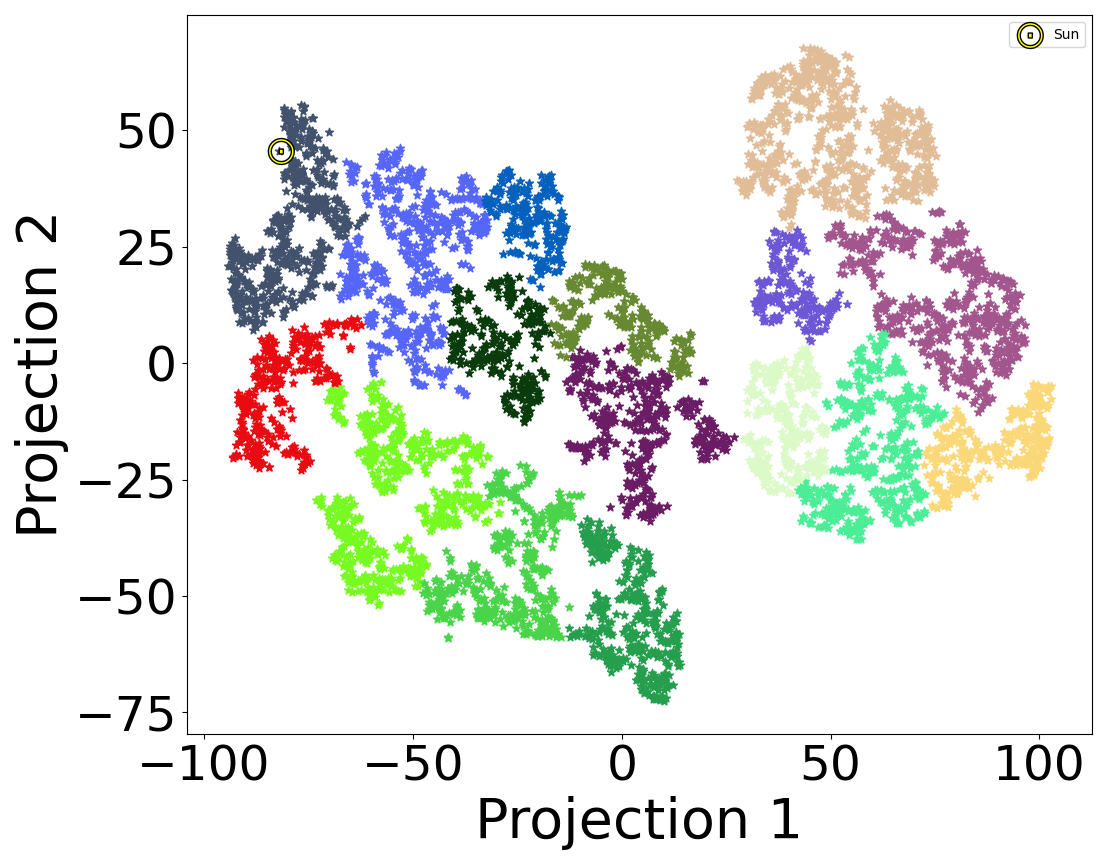}
    \caption{The same example projection of Fig.~\ref{fig:tsnemap} with the clusters produced by agglomerative clustering.}
    \label{fig:tsnemapwithclusters}
\end{figure}

\begin{figure}
    \centering
    \includegraphics[width=1\linewidth]{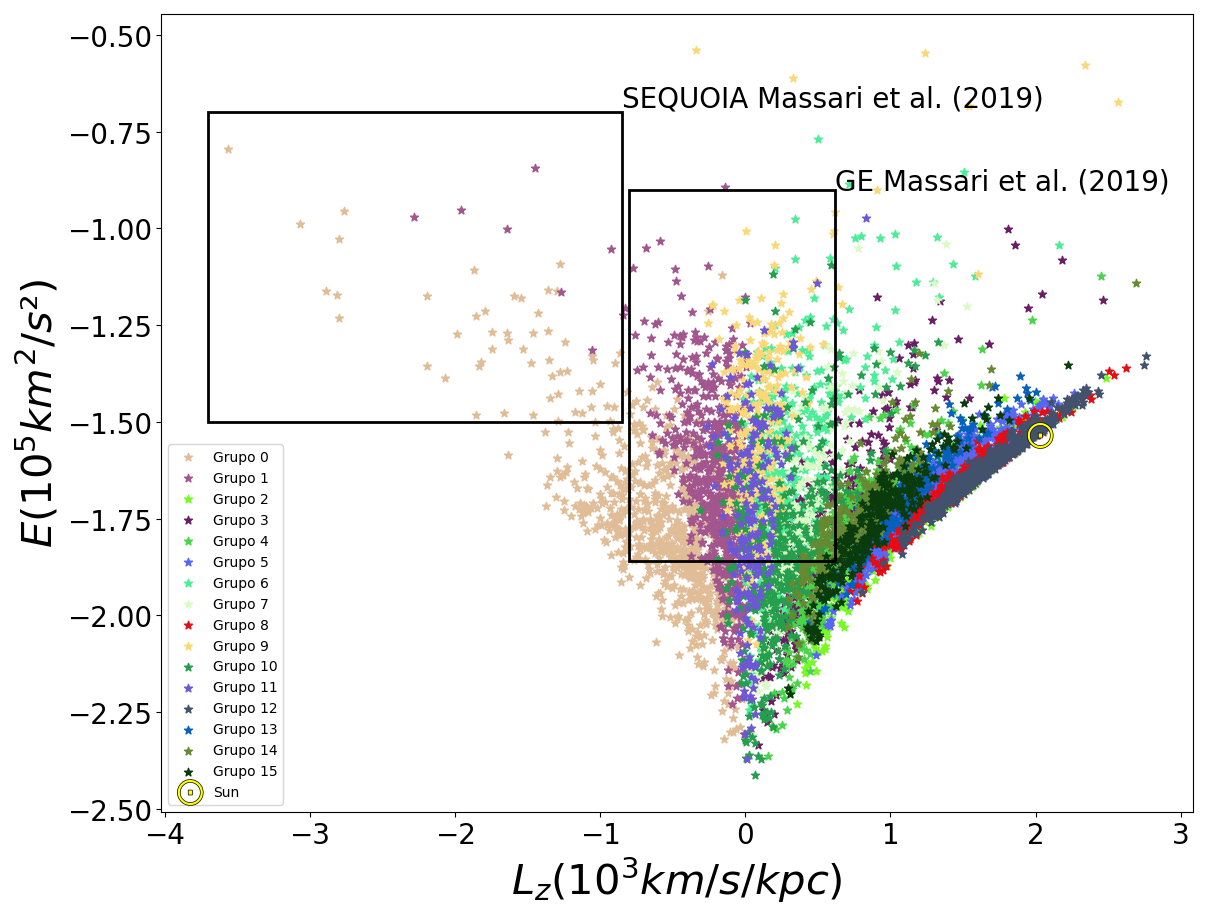}
    \caption{Lindblad diagram for the projection of Fig.~\ref{fig:tsnemap} with the clusters produced by agglomerative clustering with the same colors as Fig.~\ref{fig:tsnemapwithclusters}. The boxes correspond to the selection criteria for stars belonging to Sequoia and \Gaia-Enceladus defined in \citet{Massari2019}.}
    \label{fig:Lindbladwithclusters}
\end{figure}

We decided to explore a division that selects 16 clusters, as it seems to divide the t-SNE map with enough granularity for an exploratory study. An example of the distribution of the 16 groups on the t-SNE map is shown in Fig.~\ref{fig:tsnemapwithclusters}. The different colours in this plot are used to show how the clustering by HAC separates the clumps of stars. The Lindblad diagram showing the same divisions is shown in Fig.~\ref{fig:Lindbladwithclusters}. At first glance, it seems that the method divided the stars following some pattern in the \En\space and \Lz~diagram. Nevertheless, it is possible to see that there are stars of several clusters occupying the same \En\space by \Lz\space region (particularly in the disc region). Therefore, it is apparent that the separation we obtained is not simply a matter of cutting the sample at certain values of \Lz. 

For the purposes of this paper, we will focus on the discussion of the clusters identified in the halo part of the 2D t-SNE map (the ``island'' to the right of Figs.\ \ref{fig:tsnemap} and \ref{fig:tsnemapwithclusters}). Clusters belonging to this island were separated and selected in each of the 50 t-SNE projections. This exercise was useful in giving an idea of the probabilities that each star is associated with a certain structure on the t-SNE map. However, it is not the case that in all 50 t-SNE maps, the halo island is divided into six groups (as is the case for the map shown in Fig.\ \ref{fig:tsnemapwithclusters}). In our analysis, the number of clusters found on the halo island ranged from four to seven. 

\section{Discussion}\label{sec:results}

The accreted stellar population that is now generally recognised as GE was established as the result of a major merger in the works of \citet{Helmi2018} and \citet{Belokurov2018} and thanks to \Gaia~DR2 data \citep[][]{GaiaCollab2018a,GaiaCollab2018b}. In fact, data from \Gaia \space and spectroscopic surveys had stimulated several other efforts aimed at better characterising Galactic halo stars that were in progress in parallel \citep{Deason2018,FernandezAlvar2018,Hayes2018,Haywood2018}. However, one should note that even before that, hints of this population had been found due to the low abundance of $\alpha$ elements, retrograde motions, peculiar kinematics, or eccentric orbits of its stars \citep[e.g.][]{Carney1996,Majewski1996,ChibaBeers2000,Gratton2003}. That such stars could have an origin in the accretion of a dwarf galaxy had already been suggested \citep[e.g.][]{Gilmore2002, Brook2003}.

Several authors have investigated different ways to select stars with a high probability of belonging to the GE merger \citep[e.g.][]{Massari2019, Naidu2020, Feuillet2021, Buder2022, Carrillo2023}. It is worth to mention that \citet{Massari2019} selection was developed for studying globular clusters, not for selecting field stars as is the case for the other references. We start to explore our sample trying to define which of the halo groups most likely corresponds to what has been defined in the literature as the GE merger. For that, we chose among the groups identified in the t-SNE map the one in which the distribution of stars shows close to zero angular momentum and radial motion, characteristics generally attributed to this population. That is the main decision in our exploratory study where some prior knowledge was used. This selection was repeated in each of the 50 projections performed with t-SNE. Even if the number of clusters on the halo island changed (from a minimum of four to a maximum of seven), we could always define the probable GE-dominated group as the cluster with stars that have an average angular momentum close to zero and a radial motion. The number of stars included in this cluster varied from projection to projection. Therefore, we computed statistics from these selections. What we finally identify as our GE-dominated population is made up of 317 stars that appeared in this group on at least 80\% of the t-SNE maps. 

\begin{figure}
    \centering
    \includegraphics[width=1\linewidth]{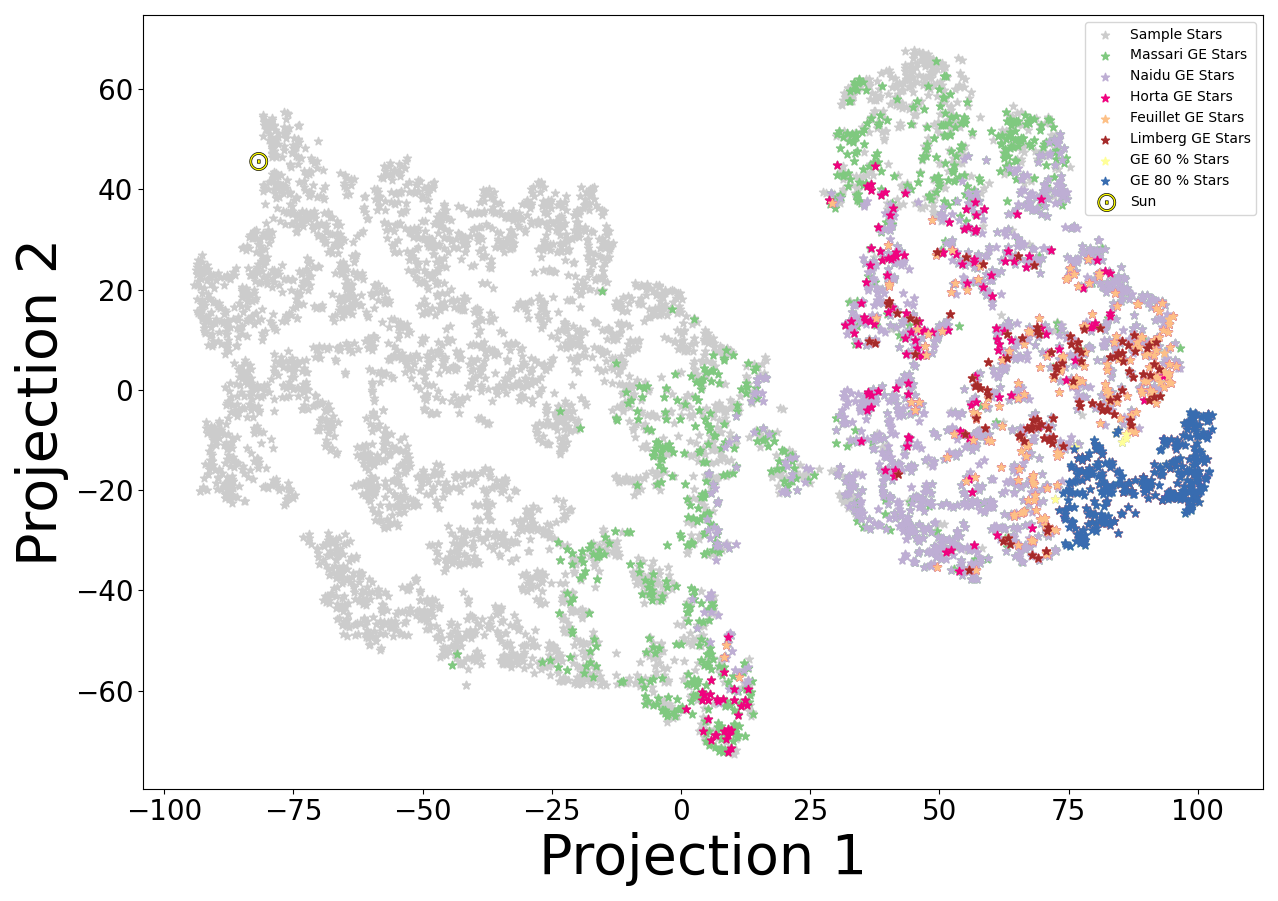}
    \caption{t-SNE map comparing the selection of \Gaia-Enceladus stars in the literature with ours. Our selection of stars with a probability of at least 80\% of belonging to GE is shown in blue. The stars that would be selected if we relax the criterion to 60\% are shown in yellow. The stars in green, purple, pink, salmon, and brown were selected according to the works of \citet{Massari2019}, \citet{Naidu2020}, \citet{Horta2023}, \citet{Feuillet2021}, and \citet{Limberg2022}, respectively.}
    \label{fig:tsne_comp_lit}
\end{figure}

\begin{figure*}
    \centering
    \includegraphics[width=.45\linewidth]{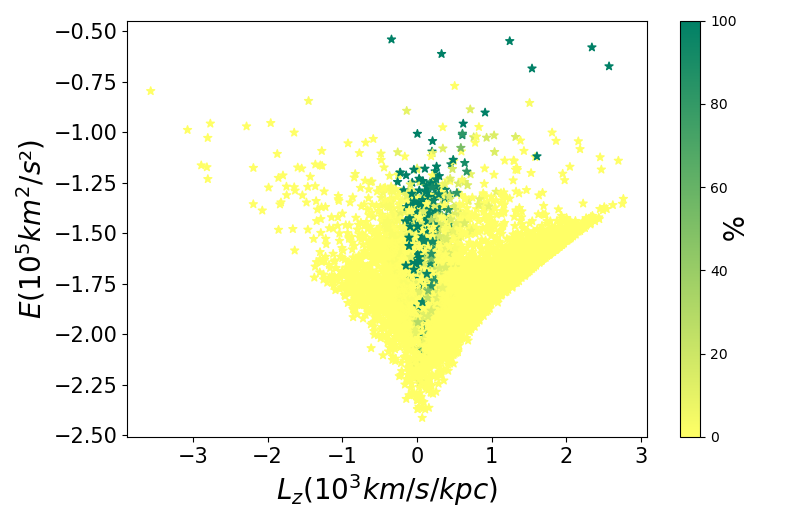}
    \includegraphics[width=.45\linewidth]{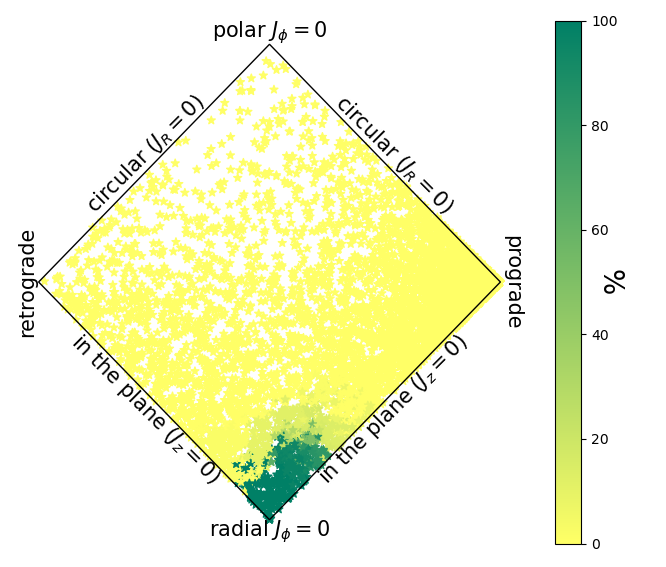}
    \caption{Diagrams showing how frequently each star of our sample was selected in the group we associate with the \Gaia-Enceladus merger. \textit{Left} - Lindblad diagram (\Lz\space by \En). \textit{Right} - the action map. Labels indicate the type of orbit that can be found in each region of the map.}
    \label{fig:GE_percentage}
\end{figure*}

\begin{figure*}
    \centering
    \includegraphics[width=.45\linewidth]{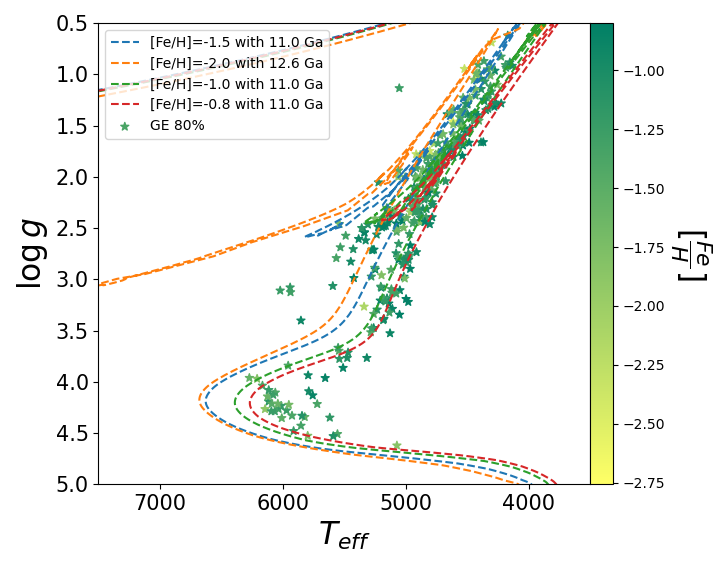}
    \includegraphics[width=.45\linewidth, height=.35\linewidth]{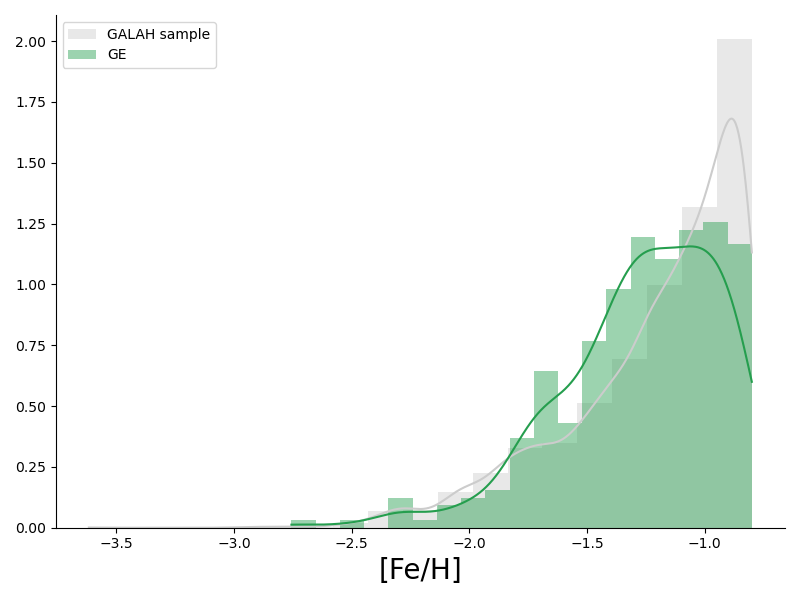}    \caption{\textit{Left} - Kiel Diagram of the GE-dominated group with MIST isochrones for \FeH \space = $-$0.8, $-$1.0 and $-$1.5~dex with an age of 11.0~Ga and \FeH \space = $-$2.0~dex with an age of 12.6~Ga. \textit{Right} - Normalised metallicity histogram comparing the GE-dominated group and the entire GALAH sample with \FeH$~\leq-0.8$~dex.}
    \label{fig:GE_met_isochrones}
\end{figure*}

Before discussing the properties of the GE-dominated group, as an exercise, we compared our selection of GE stars with those selected using the criteria adopted by the different authors. We selected stars from our sample of 6\,618 metal-poor stars using the criteria of \citet{Massari2019}, \citet{Naidu2020}, \citet{Feuillet2021}, \citet{Limberg2022}, and \citet{Horta2023}. These multiple selections are compared in an example t-SNE map in Fig.~\ref{fig:tsne_comp_lit}. The colours are over-plotted on top of each other, but the samples overlap, like a Matryoshka doll. Interestingly, some of the GE selections even enter the island which we generically associate with the stellar populations of the disc. Our own selection, adopting a 80\% (or even a 60\%) probability cut, appears to be the most restrictive, returning a sample concentrated in a corner of the map. Most importantly, the figure shows that at least a part of our GE sample would also be selected using many of the other criteria. However, it seems important to note here that for all criteria in the literature, there are stars on the t-SNE map that neighbour probable GE members, but that were not selected by them as part of GE. This, of course, reflects the fact that in the various works mentioned above, properties different from those we used to build the t-SNE map were used to select GE stars. The question of how to select the purest possible sample of GE stars remains open. We refer to \citet{Buder2022} for an extensive discussion on this topic using observational data and to \citet{Lane2022} for a discussion using simulated data \citep[see also][]{Carrillo2023}. These last two references indeed find that using actions can return a sample of accreted stars with high purity, at the expense of completeness. It would be interesting to check, using the same simulations, the impact on purity and completeness of using actions and chemistry together for a selection of accreted stars. In the following sections, we explore in detail the stellar content of the groups defined in our analysis.

\begin{figure*}
    \centering
    \includegraphics[width=.45\linewidth]{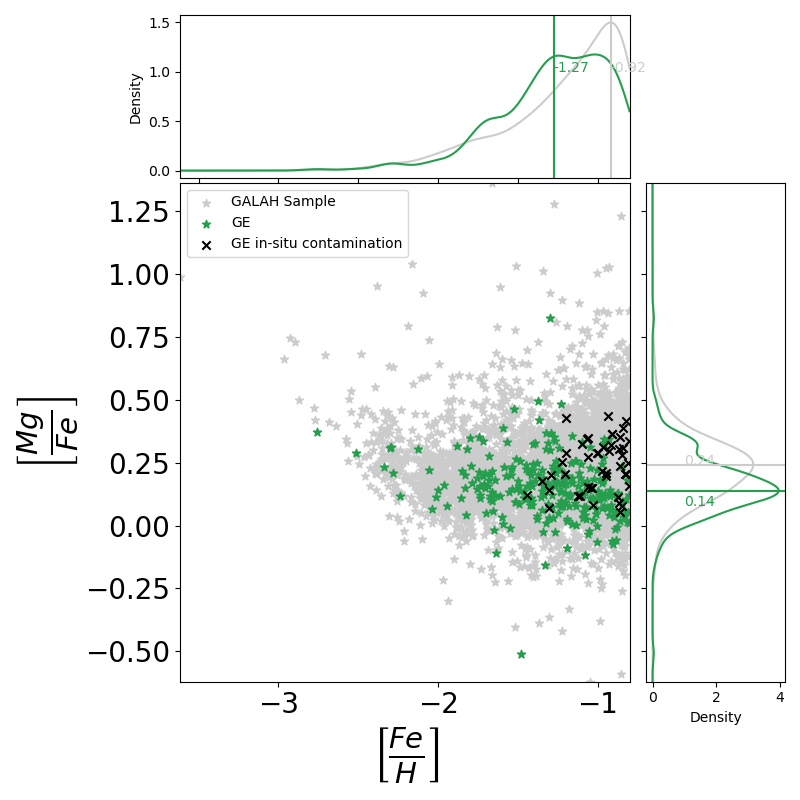}
    \includegraphics[width=.45\linewidth]{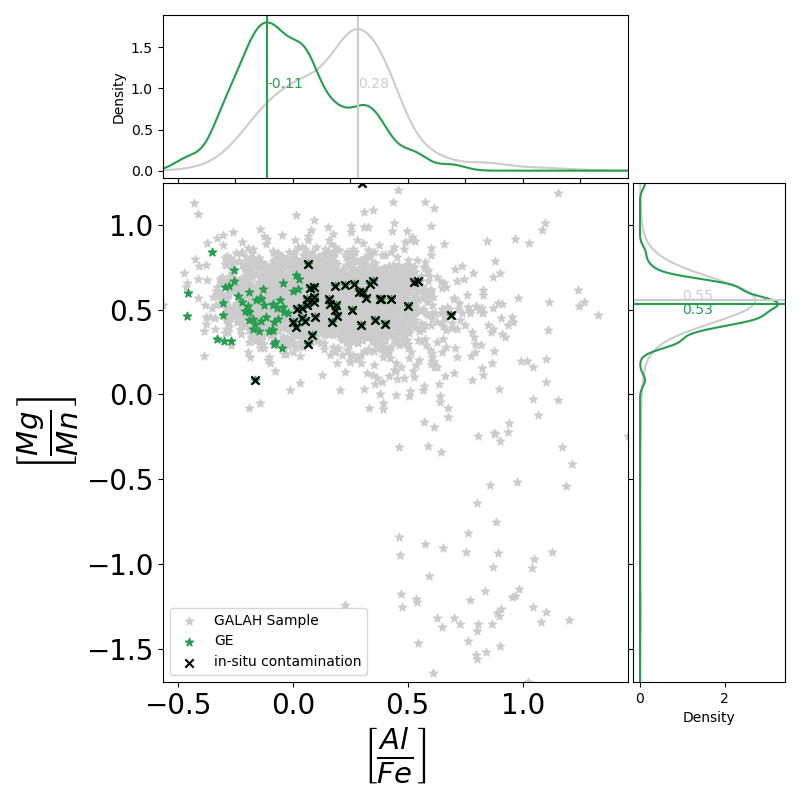}    
    \caption{In gray the stars selected with [Fe/H] $\leq$ $-$0.8 are shown. The stars of the GE-dominated cluster are shown in other colours (347 stars in the left panel; 94 stars in the right panel). Black crosses are (43) stars with high [Al/Fe] ratios and thus are likely of in situ origin. Green star symbols are objects with low [Al/Fe] ratios or without Mn and/or Al abundances of good quality. The latter group is likely dominated by stars of accreted origin. \textit{Left} - The \MgFe~ by \FeH \space diagram. \textit{Right} - The \MgMn~ by \AlFe \space diagram. }
    \label{fig:GE_descriptive}
\end{figure*}

\subsection{General properties of the GE-dominated group}

The stars selected as part of the GE-dominated cluster are depicted in dynamical diagrams in Fig.~\ref{fig:GE_percentage}. In the Lindblad diagram (left panel of the figure), the selected stars form a plume around L$_z$ = 0 that grows wider the less bound the stars become. A certain preference for stars with slightly prograde motions is apparent. This is confirmed on the action map (right panel). The stars occupy the corner of radial orbits with an asymmetry toward prograde motions. This in itself already suggests that we might be looking at a mixture of accreted and (heated) thick disc stars \citep[see e.g. the discussion on the action map in][]{Lane2022}.

A Kiel diagram of the stars is shown in the left panel of Fig.~\ref{fig:GE_met_isochrones}. For illustration, the plot includes isochrones from the MESA Isochrones and Stellar Tracks (MIST) database \citep{MIST1,MIST2}. The age-metallicity combinations of the isochrones were chosen to reproduce the GE age-metallicity relationship derived in \citet{GiribaldiSmiljanic23}. There is a general agreement between the parameters and the isochrones, showing at least a qualitative agreement with what is expected for a population dominated by GE. The exception is the region of the turn-off, where the stars have cooler temperatures than expected. This problem with the $T_{\rm eff}$ values for turn-off stars was already shown in \citet{GiribaldiSmiljanic23}, where the GALAH values were found to be too cool by 200-300 K, on average, compared with the $T_{\rm eff}$ values obtained using the infrared flux method \citep{Casagrande2021}, which in turn agree with the accurate $T_{\rm eff}$ scale obtained in \citet{Giribaldi2021}.

A histogram with the metallicity distribution for these stars is presented in the right panel of Fig.~\ref{fig:GE_met_isochrones}. The mean metallicity value of our sample is \FeH \space = $-$1.29~dex (median of \FeH \space = $-$1.26~dex). This value is very similar to the mean metallicity of GE stars derived in other works, which was found to mostly be between $-$1.15 and $-$1.3 \citep{Hayes2018,Mackereth2019,Matsuno2019,Das2020,Feuillet2021}. Our GE-dominated sample contains a good number of stars with metallicity as low as \FeH \space = $-$2.0, with a few objects reaching \FeH \space = $-$2.7. Other works have shown that the metal-poor tail of GE extends to metallicities below [Fe/H] = $-$3.0 \citep[e.g.][]{Das2020, Naidu2020, Monty2020, Bonifacio2021, Cordoni2021, Kielty2021, GiribaldiSmiljanic23}. However, several other works concentrate the discussion mostly on low-$\alpha$ stars that can be easily identified when \FeH $\geq$ \space $-$1.5 \citep[e.g.][]{Montalban2021,Myeong2022} and miss the metal-poor tail that we include here.

\subsection{Accreted and in situ stars in the GE-dominated group}
\label{sec:AccretedandGE}

Figure \ref{fig:GE_descriptive} depicts the GE dominated group in two chemical diagrams now commonly used to separate accreted stars from in situ formed stars, \MgFe$\times$\FeH\space and \MgMn$\times$\AlFe\space \citep[see][]{Hawkins2015, Das2020}. We note here that the number of stars with good Al and Mn abundances in this group (94) is much smaller than the number of stars with good Mg abundances (327). Lower sequences of the [$\alpha$/Fe] and [Al/Fe] ratios as a function of [Fe/H] are generally seen in stars of dwarf galaxies when compared to Milky Way stars \citep{Tolstoy2003, Venn2004, Kirby2010, Vargas2013, Hasselquist2021}. Because of that, such characteristics are also expected to be seen in accreted stars.

The GE-dominated group includes only stars of high [Mg/Mn] (i.e., rich in $\alpha$ elements, which are supernovae (SNe) type II products, and poor in Mn, which is mainly a SNe type Ia product). However, they span a range in [Al/Fe] ratios, from $-$0.5 to $>$ +0.5, indicating a possible mixture of in situ and accreted stars. Aluminium is produced mainly in non-explosive hydrogen burning \citep{Arnould1999} and the evolution of [Al/Fe] with metallicity can be explained by Al being mainly returned to the ISM by SNe II with little contribution from SNe Ia \citep{Kobayashi2020}. As stars in dwarf galaxies do not show high values of [Al/Fe] \citep{Hasselquist2021}, such stars can be interpreted here as probably belonging to a heated population originating from the Galactic (proto) thick disc. In Fig. \ref{fig:GE_descriptive}, we separate the possible in situ and accreted stars using the definitions proposed by \citet{Das2020}. Of the 94 stars shown in the right panel of Fig. \ref{fig:GE_descriptive}, 43 (45. 7\%) seem to have an in situ origin. In fact, when looking at \MgFe$\times$\FeH\space (left panel of Fig. \ref{fig:GE_descriptive}), we see that the stars with high [Al/Fe] mostly have a high [Mg/Fe] ratio, which is closer to expectations of in situ populations. The presence of this in-situ contamination is consistent with what is seen on the action map (Fig. \ref{fig:GE_percentage}), as discussed above.

Our selection of probable accreted GE stars displays a sequence with lower [Mg/Fe] values relative to the rest of the metal-poor sample, even for metallicities down to [Fe/H] = $-$2.0 (despite the visible large scatter in both groups of stars). We see a difference of about 0.1 dex, between the GE stars and the remaining sample, with the accreted GE sample having a mean \MgFe \space of 0.14(12)~dex. The [Mg/Fe] knee is not fully obvious to the eye (because of the metallicity cut in our sample), but there is a hint of its presence between [Fe/H] = $-$1.2 and $-$1.0. This agrees with, for example, \citet{Mackereth2019} and \citet{Horta2023} who found the knee around \FeH \space $\sim$ $-$1.2~dex using APOGEE data (but see also \citeauthor{Monty2020} \citeyear{Monty2020}, who found the GE knee at lower metallicity, [Fe/H] $\sim$ $-$1.6, using a different sample of stars).

\begin{figure*}
    \centering
    \includegraphics[width=.45\linewidth]{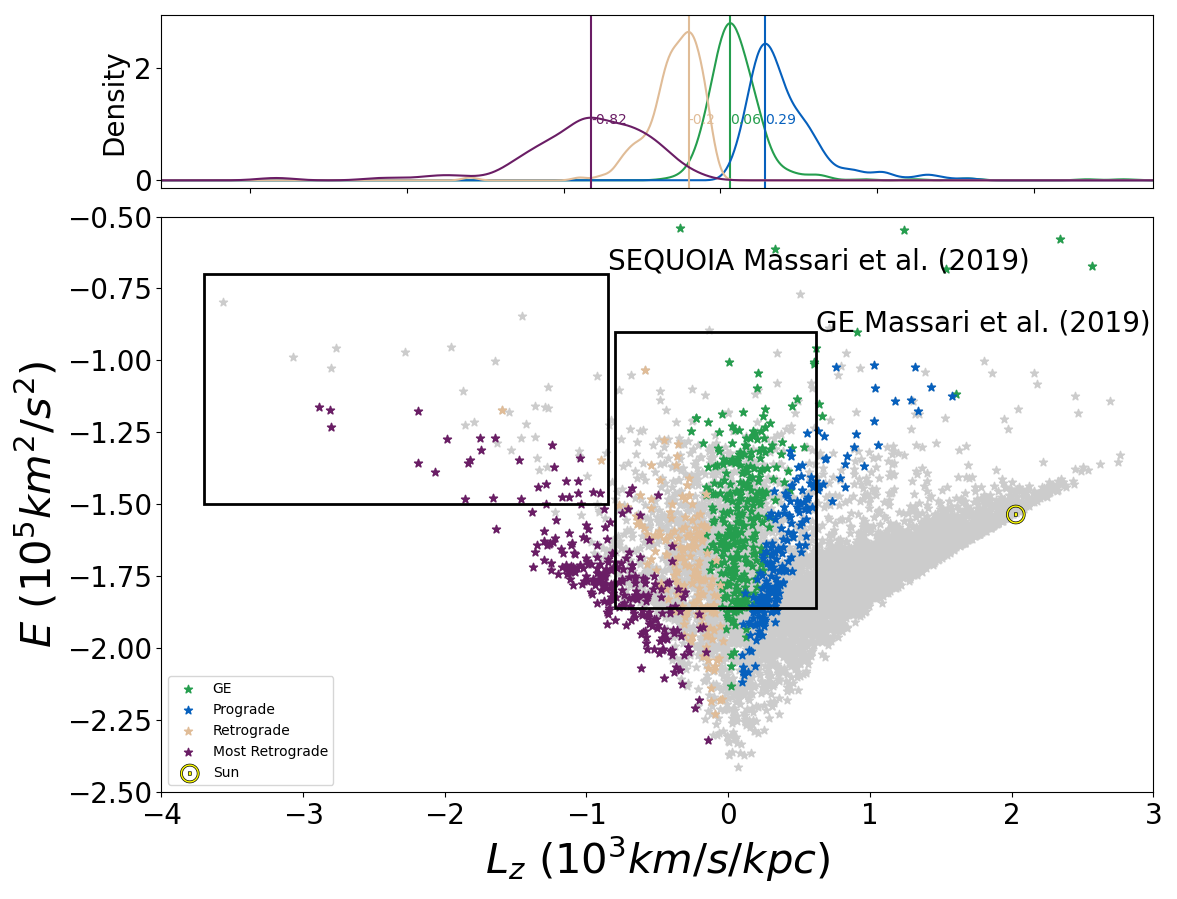}
    \includegraphics[width=.45\linewidth]{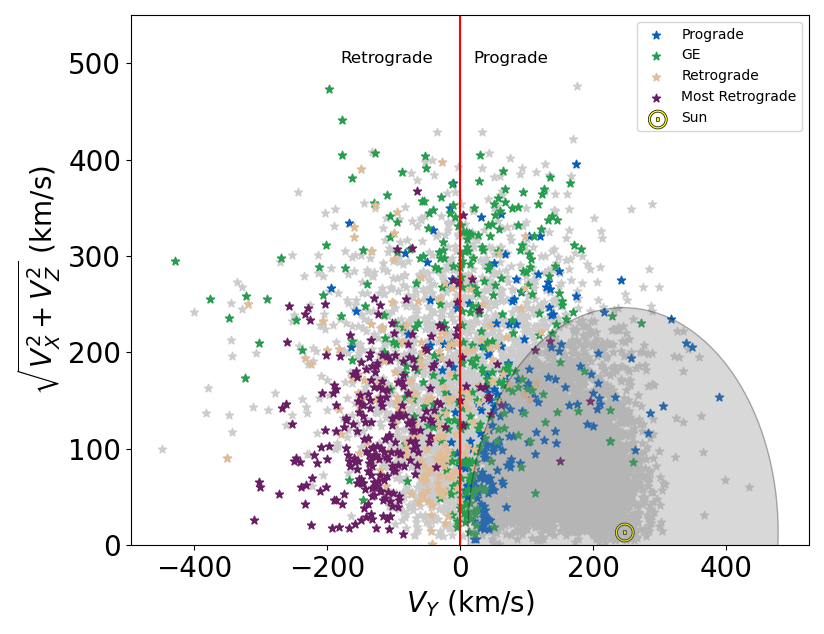}
    \includegraphics[width=.45\linewidth]{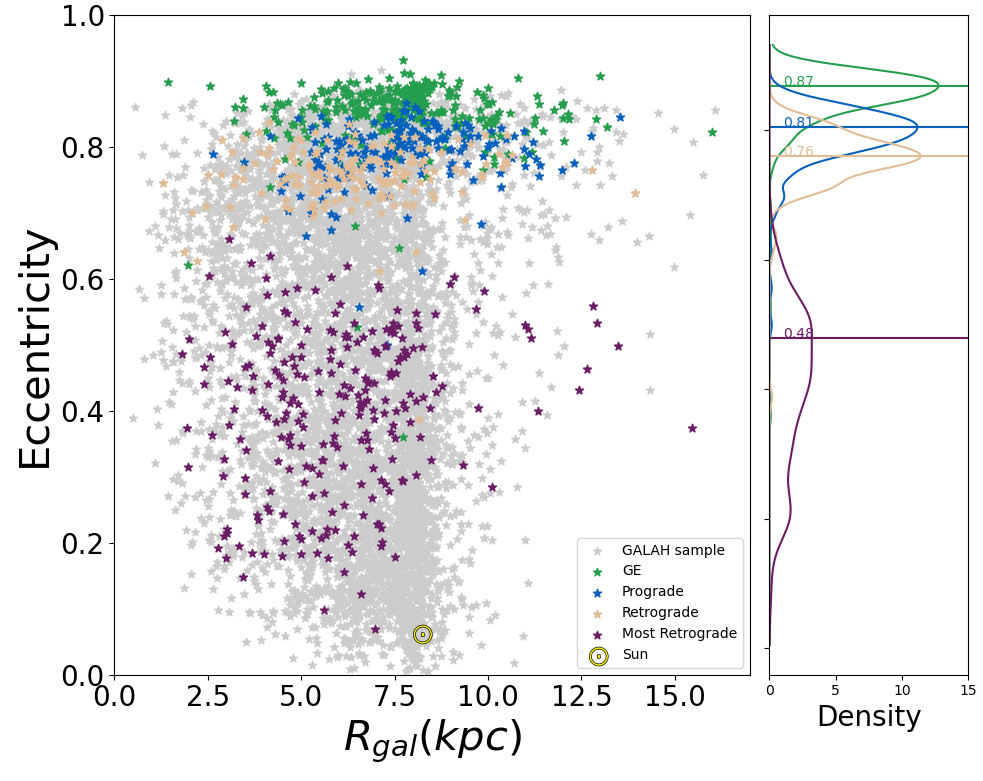}
    \includegraphics[width=.45\linewidth]{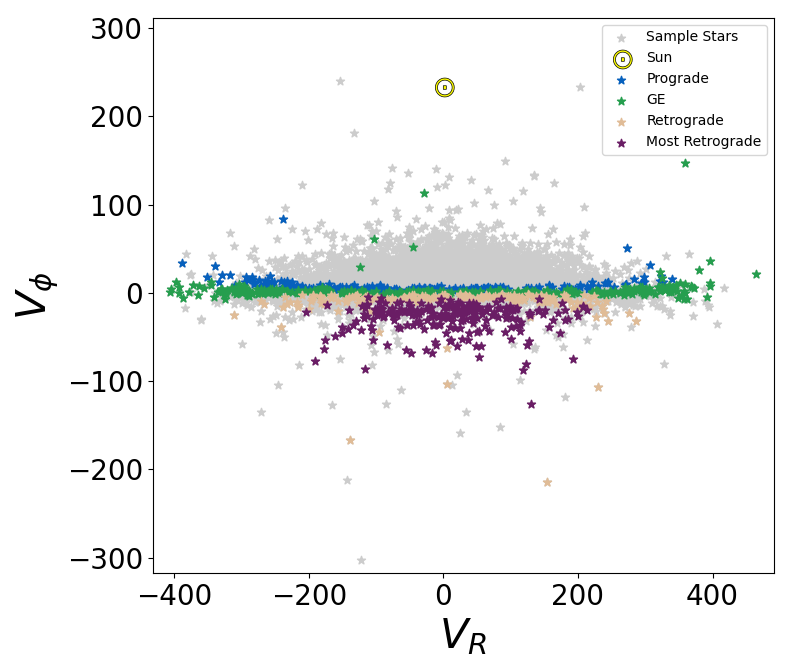}
    \includegraphics[width=.45\linewidth]{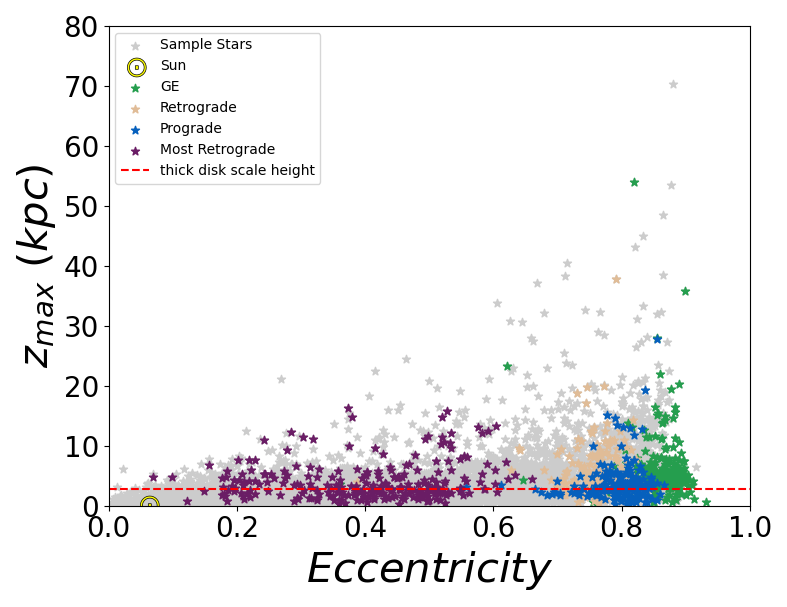}
    \includegraphics[width=.45\linewidth]{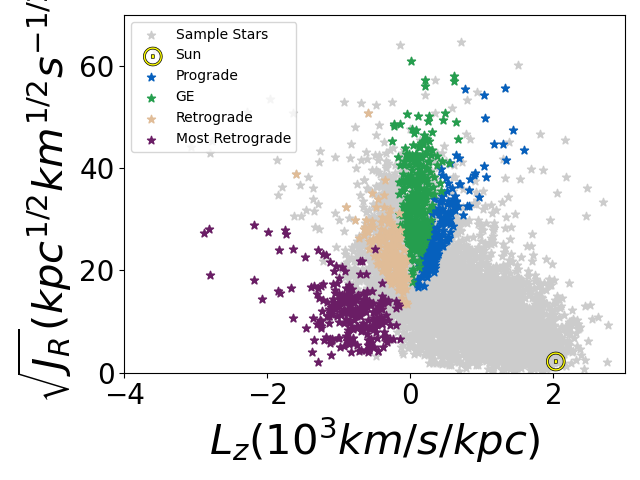}
    \caption{In all panels, the GE is shown in light green, the prograde halo group in blue, the retrograde group in light brown, and the most retrograde group in violet.\textit{Top Left} - The Lindblad diagram (\Lz$\times$\En).  \textit{Top Right} - The Toomre Diagram. \textit{Middle Left} - Eccentricity as a function of the galactocentric distance ($\mathrm{R_{gal}}$). \textit{Middle Right} - Cylindrical velocity components, \Vphi~as a function of the radial component \Vr. \textit{Bottom Left} - Maximum distance from the Galactic (\zmax) plane as a function of eccentricity. The dashed line at \zmax = 2~kpc indicates the scale height of the thick disc \citep{Li2018}. \textit{Bottom Right} - Radial action ($\sqrt{J_R}$) as a function of the orbital angular momentum $L_Z$.}
    \label{fig:GE_comp_withothergroups}
\end{figure*}

\begin{figure}
    \centering
    \includegraphics[width=\linewidth]{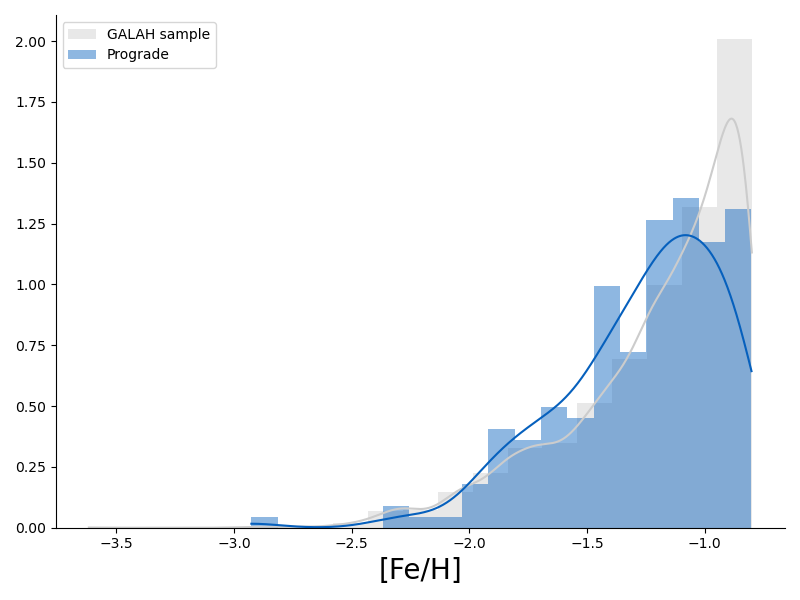}\\
    \includegraphics[width=\linewidth]{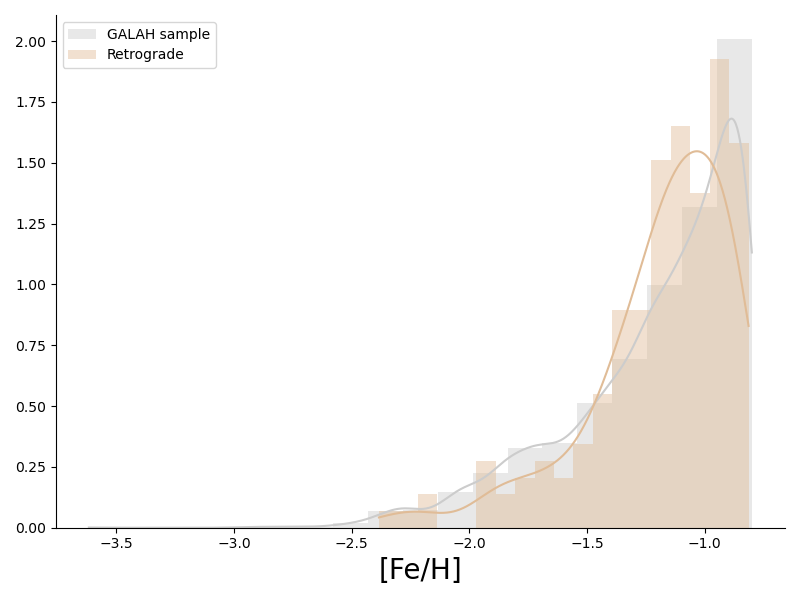}\\
        \includegraphics[width=\linewidth]{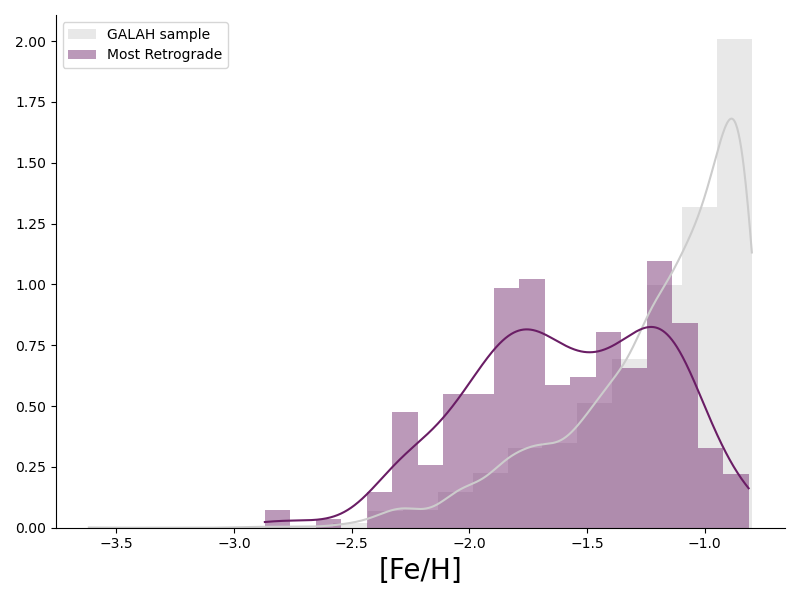}\\
    \caption{Histograms with the metallicity distribution of the three other clusters defined in the analysis: The prograde cluster in the top panel, the retrograde cluster in the middle, and the most retrograde cluster in the bottom.}
    \label{fig:other_groups_feh}
\end{figure}

In Fig. \ref{fig:GE_descriptive} we can see that some of the stars labelled as in situ contaminants in the GE-dominated cluster also display low [Mg/Fe] ratios. This dispersion can be attributed to the errors in the abundances, creating a large scatter in the points. In any case, this clearly shows that in this sample a simple cut in the diagram \MgFe$\times$\FeH\space is not enough to isolate stars of accreted origin. Moreover, we are left to wonder whether the in situ contamination abruptly ends at the straight cut apparent in the right panel of Fig. \ref{fig:GE_descriptive}. In fact, that stars in dwarf galaxies display low values of [Al/Fe] says nothing about the behaviour of this ratio in metal-poor stars of the Milky Way. Chemical evolution models seem to show that low [Al/Fe] values are a general feature of metal-poor stars and could also be present in stars formed in the Milky Way \citep{Kobayashi2020}. Therefore, it is not clear what the true level of in situ contamination is, even in the stars shown in green in the right panel of Fig. \ref{fig:GE_descriptive}, which we assume to be most likely of accreted origin. To advance in this discussion and understand if there are chemical differences between the in situ and accreted stars with [Fe/H] $\lesssim$ $-$1.2, we clearly need: i) stellar abundances of higher precision and ii) guidance from chemical evolution models. At the moment, we are forced to conclude that even a combination of chemical and dynamical parameters is not able to fully separate a pure sample of accreted origin, at least in the corner of the parameter space where we selected our GE-dominated cluster. Before discussing the abundances of other chemical elements, we first turn our attention to other groups of halo stars separated in the t-SNE maps.

\subsection{Other halo groups}

To continue exploring our sample of metal-poor stars, we defined three additional clusters from the t-SNE maps. What we shall henceforth call the prograde cluster is defined as the one that has a mean positive $L_z$ value just higher than the GE cluster. The (intermediate) retrograde group is the one that has a mean negative value of $L_z$ that is just lower than the one of the GE cluster. The most retrograde group is the one with the most extreme negative mean value of $L_z$. The stars assigned to these groups are those that appear in the same cluster in at least 80\% of the 50 t-SNE maps. The GE-dominated cluster is the largest with 327 stars. The most retrograde, prograde, and retrograde clusters have 253, 198 and 176 stars, respectively.

In Fig. \ref{fig:GE_comp_withothergroups} the kinematic and dynamic properties of all four groups of stars are compared in several different diagrams. The average properties of each group are summarised in Table \ref{tab:parameters}. In the Linblad and Toomre diagrams (top row of Fig. \ref{fig:GE_comp_withothergroups}), the most retrograde group is in the same region as the Thamnos substructure \citep{ Koppelman2019}, although it also shows a tail toward very retrograde loosely bound stars (top left in the figure) entering the region identified as the Sequoia merger \citep[see][]{Myeong2018}. We do not attempt to separate both substructures here, as the probable Sequoia tail is poorly populated. In both the Lindblad and Toomre diagrams, the other three clusters (at least partially) fall into regions that other authors might use to define GE \citep{Helmi2018, Massari2019}. The prograde cluster, however, has most of its stars in a region of the Toomre diagram that would be used to define a kinematic thick disc. These stars would end up excluded from most discussions of accreted stars. As the numbers in Table \ref{tab:parameters} show, there is a decrease in the mean orbital energy when moving from the prograde to the most retrograde group (although the dispersion is always high).

The average eccentricity of our GE-dominated cluster is remarkably high, with a mean value of 0.86 (left panel in the middle row of Fig. \ref{fig:GE_comp_withothergroups}). Both prograde and retrograde clusters also have high average eccentricity values, 0.80 and 0.76, respectively. High eccentricity is one of the characteristics often used to separate samples of GE stars \citep[e.g.][]{Naidu2020}. Typical values for this separation are 0.7 or 0.8. Therefore, such a cut could end up including many of the stars in the prograde and retrograde clusters as part of GE. However, we remark that eccentricity is often not used by itself, and only after some sample cleaning is used to remove disc stars and possibly other structures. It is interesting that the prograde cluster tends to have a somewhat higher mean eccentricity than the retrograde one. This could indicate a higher contamination from accreted stars or be caused by the heating up of the disc caused by mergers. In the velocity plot (right panel, middle row), it is possible to see that the GE-dominated cluster is indeed the one that dominates in the regions of extreme radial motions; $|V_{R}|$ $\geq$ 200 km s$^{-1}$ (and is so by construction). The prograde and retrograde clusters share a distribution around $V_{\phi}$ $\sim$ 0 km s$^{-1}$, but the tails at extreme $V_{R}$ values are less prominent (in particular for the prograde cluster).

\begin{table*}[htp!]
\caption{Chemo-dynamical parameters for the halo groups.}
\label{tab:parameters}
\centering
\footnotesize
\begin{threeparttable}
\begin{tabular}{lcccc}
\hline\hline
Parameter & GE & Prograde & Retrograde &  Most Retrograde \\
\hline
\Lz ($10^3$~km s$^{-1}$ kpc$^{-1}$) & 78(274) & 341(255) & -260(185) & -821(428)\\
\En ($10^5$~km$^2$ s$^{-2}$) & -1.7065(2358) & -1.5473(2518) & -1.7290(1963) & -1.7602(1925)\\
\Jr & 1080(566) & 661(484) & 523(275) & 145(148) \\
\Jz & 57(485) & 59(56)& 174(169) & 175(242)\\
Eccentricity & 0.86(5) & 0.80(5) & 0.76(5) & 0.42(12)\\
\FeH & -1.23(35) &-1.20(34) &-1.12(31) &-1.57(41) \\
\MgFe & 0.16(14) & 0.18(12) & 0.15(14) & 0.20(13) \\
Number of stars & 317 & 198 & 176 & 253 \\
\hline
[O/Fe] & 0.55(22)&0.59(23)&0.56(23)&0.70(25)\\
Number of stars &271&169&145&166\\
\hline
[Na/Fe]&-0.17(17)& -0.13(20)& -0.20(19)& -0.11(24)\\
Number of stars &282&176&163&198\\
\hline
[Al/Fe]&-0.03(23)& 0.09(24)& -0.02(21)& 0.15(30)\\
Number of stars &94&78&84&66\\
\hline
[Si/Fe]&0.17(15)& 0.21(16)& 0.16(18)& 0.25(17)\\
Number of stars &297&184&172&228\\
\hline
[K/Fe]&0.11(15)& 0.12(15)& 0.12(14)& 0.14(15)\\
Number of stars &306&191&172&251\\
\hline
[Ca/Fe]&0.23(12)& 0.23(12)& 0.23(13)& 0.26(15)\\
Number of stars &306&192&171&245\\
\hline
[Sc/Fe]&0.08(11)& 0.10(10)& 0.06(10)& 0.08(11)\\
Number of stars &296&185&175&248\\
\hline
[Ti/Fe]&0.20(21)& 0.23(22)& 0.18(22)& 0.24(22)\\
Number of stars &280&163&164&212\\
\hline
[V/Fe]&0.05(38)& 0.09(46)& 0.05(33)& 0.03(43)\\
Number of stars &85&50&62&66\\
\hline
[Cr/Fe]&-0.14(16)& -0.15(16)& -0.17(15)& -0.14(19)\\
Number of stars &273&167&159&201\\
\hline
[Mn/Fe]&-0.34(13)& -0.35(15)& -0.37(14)& -0.40(15)\\
Number of stars &308&190&171&245\\
\hline
[Co/Fe]&-0.08(54)& -0.01(61)& -0.10(33)& -0.04(41)\\
Number of stars &103&56&82&79\\
\hline
[Ni/Fe]&-0.16(13)& -0.14(15)& -0.14(19)& -0.14(16)\\
Number of stars &217&126&150&176\\
\hline
[Cu/Fe]&-0.51(17)& -0.47(25)& -0.48(19)& -0.46(20)\\
Number of stars &178&109&129&120\\
\hline
[Zn/Fe]&0.16(19)& 0.18(20)& 0.18(21)& 0.20(17)\\
Number of stars &308&192&173&245\\
\hline
[Rb/Fe]&1.03(05)& 1.12(40)& 0.68(52)& 0.80(29)\\
Number of stars &2&3&9&7\\
\hline
[Sr/Fe]&1.36(29)& 1.02(16)& 1.29(36)& 1.14(27)\\
Number of stars &10&5&7&5\\
\hline
[Y/Fe]&0.07(33)& 0.13(32)& 0.12(29)& 0.06(36)\\
Number of stars &301&185&170&245\\
\hline
[Zr/Fe]&0.37(54)& 0.42(45)& 0.36(42)& 0.46(58)\\
Number of stars &86&49&67&71\\
\hline
[Mo/Fe]&0.74(45)& 0.56(28)& 0.51(66)& 0.87(51)\\
Number of stars &5&2&6&11\\
\hline
[Ru/Fe]&0.55(60)& 0.48(38)& 0.60(26)& 0.88(39)\\
Number of stars &10&7&10&12\\
\hline
[Ba/Fe]&0.21(42)& 0.23(41)& 0.39(36)& 0.19(40)\\
Number of stars &313&196&174&252\\
\hline
[La/Fe]&0.31(37)& 0.37(40)& 0.26(32)& 0.34(31)\\
Number of stars &187&106&134&140\\
\hline
[Ce/Fe]&-0.10(35)& -0.09(34)& -0.18(27)& -0.09(38)\\
Number of stars &76&42&53&33\\
\hline
[Nd/Fe]&0.53(29)& 0.51(24)& 0.49(23)& 0.47(24)\\
Number of stars &230&118&145&173\\
\hline
[Sm/Fe]&0.25(43)& 0.17(34)& 0.16(32)& 0.12(29)\\
Number of stars &156&84&115&114\\
\hline
[Eu/Fe]&0.49(19)& 0.48(27)& 0.46(15)& 0.37(23)\\
Number of stars &137&74&98&62\\
\hline
\hline
\end{tabular}
\begin{tablenotes}
\item{} \textbf{Notes.} The values for the parameters of this table are medians. Numbers in parentheses show the standard deviation for the value, not the standard deviation of the mean. The number of stars with the best quality flag (flag\_X\_fe=0) is shown under the parameter. If for a certain element only a single star with good quality abundance was left, no standard deviation is shown. Abundances of Li and C, affected by stellar evolution in giants, were excluded.
\end{tablenotes}
\end{threeparttable}
\end{table*} 

In terms of eccentricity and velocities, the most retrograde cluster is very different from the others. Eccentricity values vary from about 0.2 to 0.6 (with an average of 0.42). \citet{Myeong2019} found Sequoia stars to have typical eccentricity values around 0.6. This again suggests that the number of Sequoia stars in our very retrograde cluster is small. Thamnos, on the other hand, was found to have lower eccentricity values than GE \citep{Koppelman2020}, with values below $\sim$ 0.65 \citep{Kordopatis2020, Horta2023}. It seems likely that this cluster is dominated by Thamnos stars. However, our most retrograde cluster has a smaller secondary peak in its eccentricity distribution around 0.2, with a few stars with smaller values. A very retrograde group of stars with thick disc chemistry was reported by \citet{Koppelman2019}, but was not discussed in detail. Those stars could be the same as the one in this peak with small eccentricity values. They are not necessarily part of the disc, but could of the inner halo with motions confined to the disc region.   

The bottom row of Fig. \ref{fig:GE_comp_withothergroups} shows plots of $Z_{\rm max}$ as a function of eccentricity (left) and of the square root of the radial action as a function of angular momentum (right). In the latter plot, the clusters are all well separated, which simply reflects the way they were defined in the first place. Our GE-dominated cluster matches the regions that \citet{Feuillet2021} and \citet{Matsuno2021} use to find GE stars. The most retrograde cluster also matches the position of Sequoia stars in \citet{Feuillet2021} but, as we discussed above, this cluster is mostly dominated by stars from Thamnos. In the former plot, we can see that most of the sample, in the four groups, tends to be inside the inner halo with orbits that take the stars out to at most 10 kpc from the Galactic plane. \citet{Li2018} reports a scale height of the thick disc component of the Galaxy of around 2.76~kpc. Interestingly, the stars of the most retrograde group are the ones that are concentrated mainly on typical disc heights, while the others show a tendency of higher dispersion. Stars from the GE-dominated cluster have a clearer tail to higher \zmax\space distances than the stars of the prograde and retrograde clusters. There are very few stars with $Z_{\rm max}$ $\gtrsim$ 20 kpc, which are most likely part of the outer halo \citep{Carollo2007, Carollo2010}. The number of these outer halo stars tends to increase for higher values of eccentricity.

\begin{figure*}[ht!]
    \centering
    \includegraphics[width=.45\linewidth]{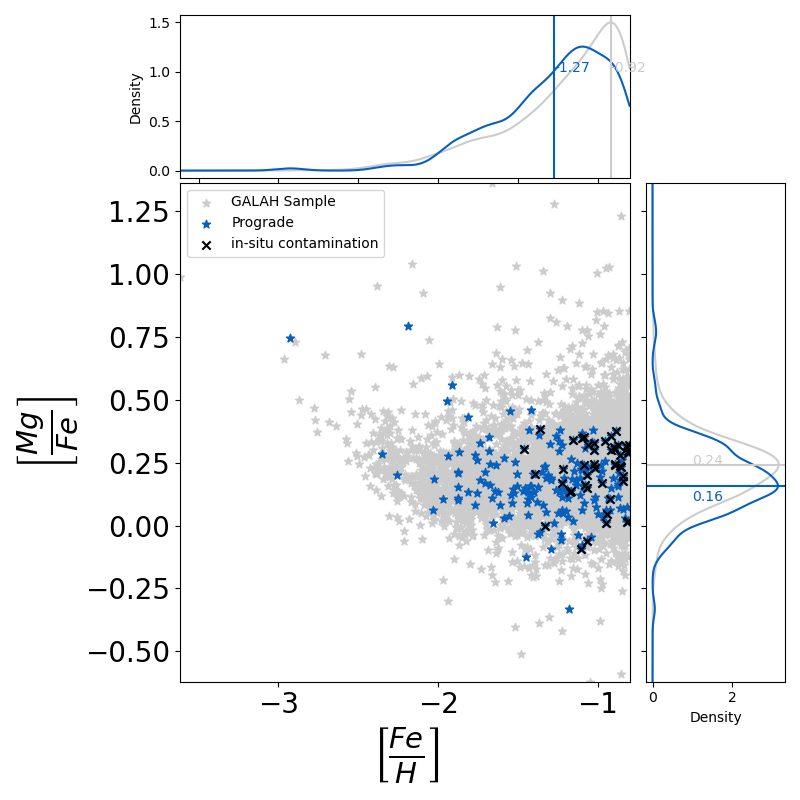}
    \includegraphics[width=.45\linewidth]{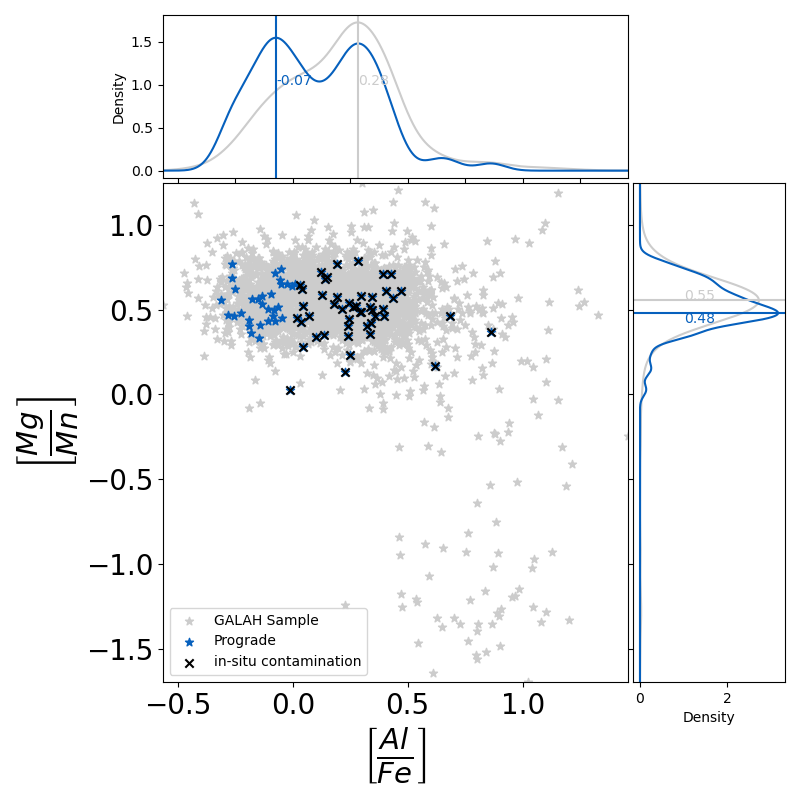}\\
    \caption{ Black crosses are (47) stars with high values of \AlFe\space of probable in situ origin. The blue star symbols are objects with a low \AlFe\space ratio or without values for the Al and/or Mn abundances.\textit{Left} - The \MgFe~ by \FeH\space diagram highlighting 198 stars belonging to the prograde group. \textit{Right} - The \MgMn~ by \AlFe\space diagram highlighting 78 stars belonging to the prograde group.}
    \label{fig:prograde_descriptive}
\end{figure*}

\begin{figure*}
    \centering
    \includegraphics[width=.45\linewidth]{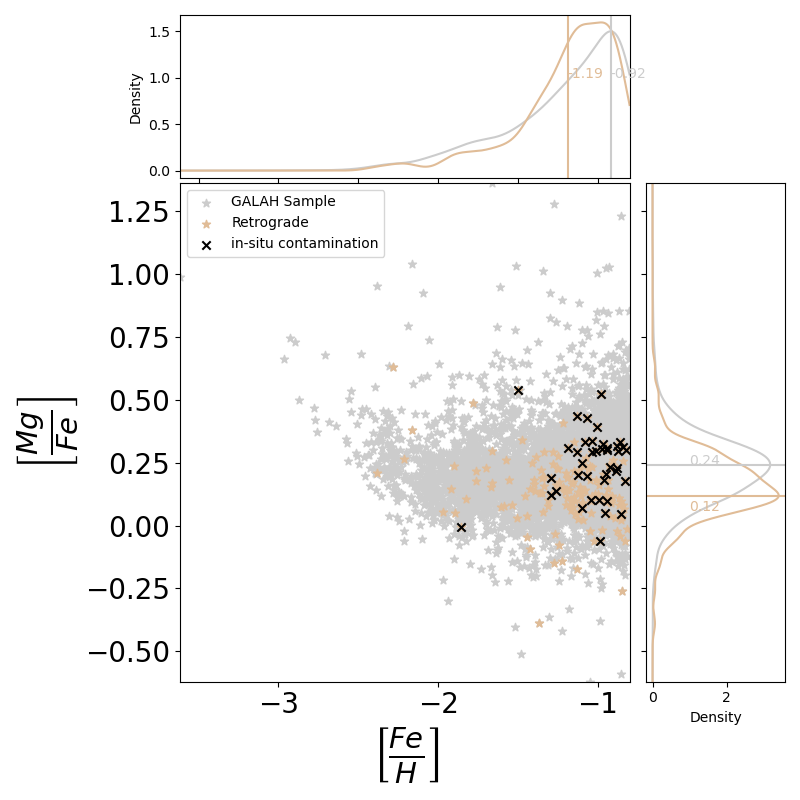}
    \includegraphics[width=.45\linewidth]{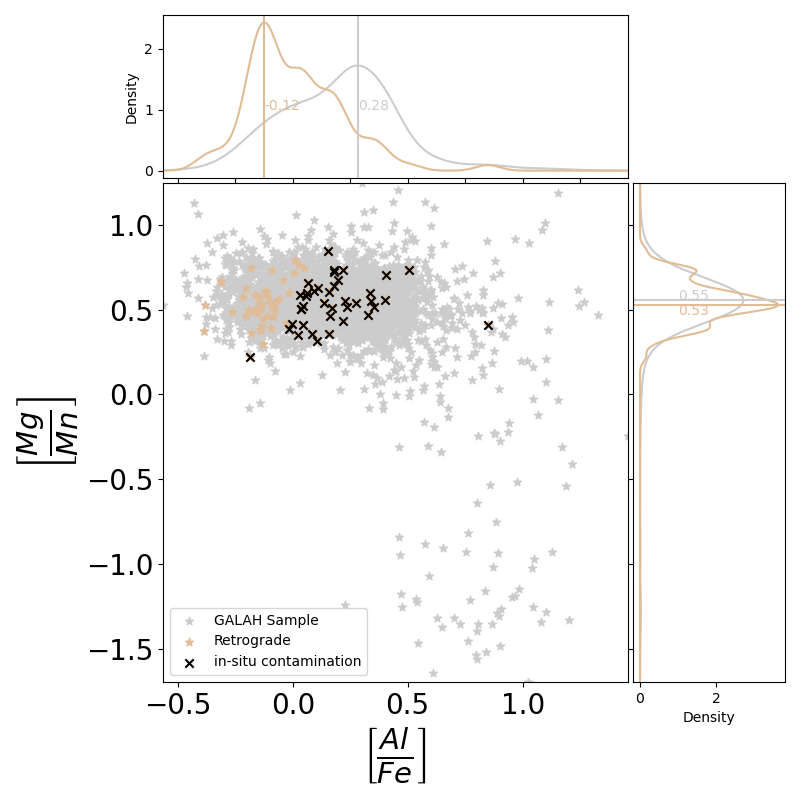}\\
    \caption{Black crosses are (39) stars with high values of \AlFe\space of probable in situ origin. The light brown star symbols are objects with a low \AlFe\space ratio or without values for the Al and/or Mn abundances.\textit{Left} - The \MgFe~ by \FeH\space diagram highlighting 176 stars belonging to the retrograde group. \textit{Right} - The \MgMn~ by \AlFe\space diagram highlighting 84 stars belonging to the retrograde group.}
    \label{fig:retrograde_descriptive}
\end{figure*}
\begin{figure*}
    \centering
    \includegraphics[width=.45\linewidth]{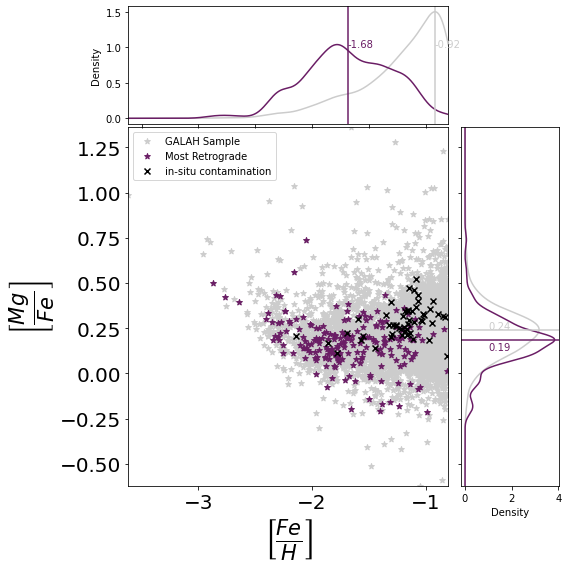}
    \includegraphics[width=.45\linewidth]{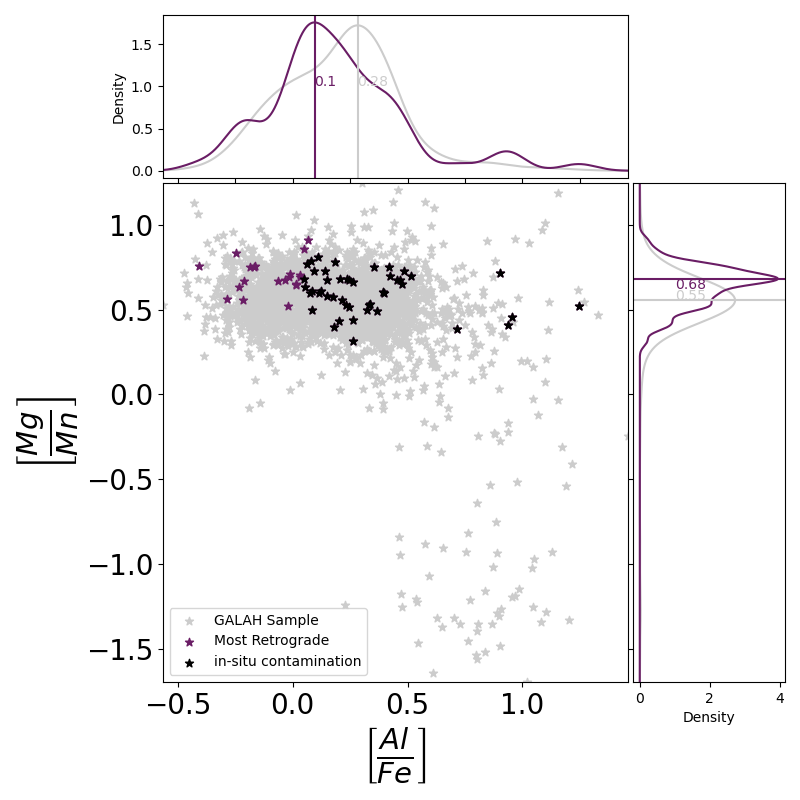}\\
    \caption{Black crosses are (46) stars with high values of \AlFe\space of probable in situ origin. The violet star symbols are objects with a low \AlFe\space ratio or without values for the Al and/or Mn abundances.\textit{Left} - The \MgFe~ by \FeH\space diagram highlighting 253 stars belonging to the most retorgrade group. \textit{Right} - The \MgMn~ by \AlFe\space diagram highlighting 66 stars belonging to the most retrograde group. }
    \label{fig:thule_descriptive}
\end{figure*}

The metallicity distributions of the three additional clusters are shown in Fig. \ref{fig:other_groups_feh}. The similarity between the GE, retrograde, and prograde clusters is also visible here (see the top row of Fig. \ref{fig:other_groups_feh} compared to the left panel of Fig. \ref{fig:GE_met_isochrones}). The one cluster that seems to have distinct properties is the very retrograde one (bottom panel in Fig. \ref{fig:other_groups_feh}), which is clearly more populated in metal-poor stars \citep[as also discussed elsewhere, e.g.,][]{Kordopatis2020, Horta2023}. Let us now turn our attention back to the chemical abundances to investigate whether there are more similarities (or differences) among these clusters.

\subsection{Accreted and in situ stars in the other halo groups}

Figures~\ref{fig:prograde_descriptive}, \ref{fig:retrograde_descriptive}, \ref{fig:thule_descriptive} show the diagrams of \MgFe\space versus \FeH\space and \MgMn\space versus \AlFe\space for the prograde, intermediate retrograde, and most retrograde groups, respectively. Again, we note that requiring good abundances of Mn and Al decreases the size of the samples. For the prograde group, we found that 47 of the 78 stars ($\sim$ 60\%) with Al and Mn abundances are probably of in situ origin due to their high values of [Al/Fe] \citep{Das2020}. For the intermediate retrograde group, 39 of 84 stars ($\sim$46\%) are possibly in situ. For the most retrograde group, 46 of 66 stars ($\sim$70\%) seem to probably be of in situ origin. 

Although the number of in situ stars should probably decrease for groups that are increasingly more retrograde, we find that the most retrograde group is the one with the largest fractional in situ contamination. Nevertheless, it is hard to judge the significance of this finding, as this is also affected by the data quality and the capability of detecting and analysing the Al lines in these stars. In any case, we do clearly find that the in situ contamination is important in all dynamical parameter space, accounting for about 50\% or more stars with measured Al and Mn abundances. This in situ contamination is probably what was identified in \citet{GiribaldiSmiljanic23} and named Erebus, an old in situ component affected by the chaotic conditions in the early Galaxy. As discussed in \citet{GiribaldiSmiljanic23}, there is probably some relation between Erebus and the Aurora component identified by \citet{BelokurovKravtsov2022}. 

In all cases, the in situ contamination concentrates in the region of higher metallicity values (\FeH $\gtrsim$ $-$1.4). This is probably related to the limit of detectability of the Al and/or Mn lines. The in situ stars also tend to have values of [Mg/Fe] that are higher than for the others, although the scatter is always high. The most retrograde group is the one with stars that on average have higher values of \MgFe, a feature that is probably related to its lower metallicity and the difficulty in separating accreted and in situ stars using Mg abundances at this regime. Chemical differences at the lower metallicity end are very unclear; Al and Mn abundances are mostly missing, and Mg values seem to overlap. In the sections below, we investigate whether other abundances give additional information on possible chemical differences between the groups.

\subsection{Europium and barium abundances}

\citet{Aguado2021} analysed nine stars from GE and Sequoia, with [Fe/H] $<$ $-$1.4, and found that they are enhanced in r-process elements with \BaEu\space $\sim$ $-$0.6 dex and [Eu/Fe] = 0.6--0.7. \citet{Matsuno2021}, using DR3 of GALAH, also found enrichment in r-process elements in GE stars, although not at the same level as \citet{Aguado2021}, with \BaEu $\sim$ 0.0 and \EuFe $\sim$ 0.5. New observations reported by \citet{Naidu2022} also show an Eu enhancement in GE stars, with \BaEu $\sim$ $-$0.45 (but the [Eu/Fe] ratios are not given). A globular cluster enriched with r-processes elements (NGC 1261; [Fe/H] = $-$1.26) has also been associated with \Gaia -Enceladus \citep{KochHansen2021}, further supporting the idea that this system contained an environment enriched by the r-process.

\begin{figure*}
    \centering
    \includegraphics[width=.45\linewidth]{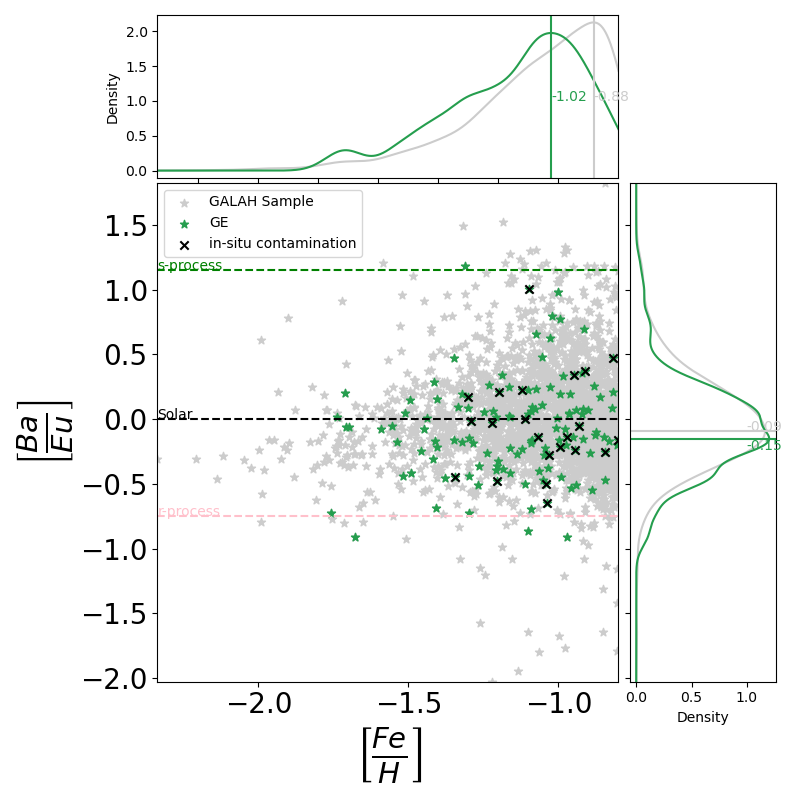}
    \includegraphics[width=.45\linewidth]{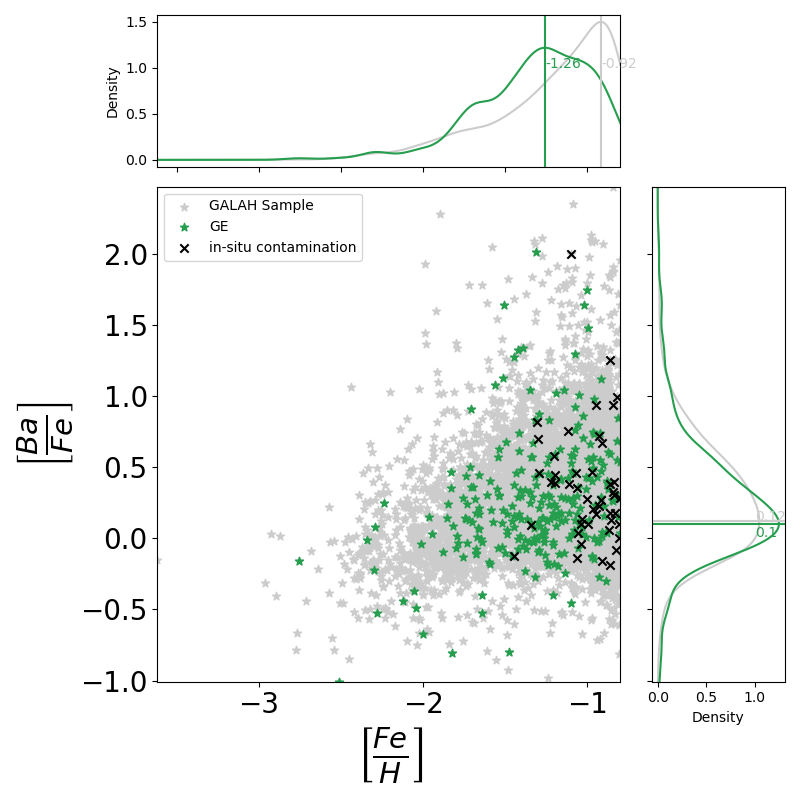}\\
    \includegraphics[width=.45\linewidth]{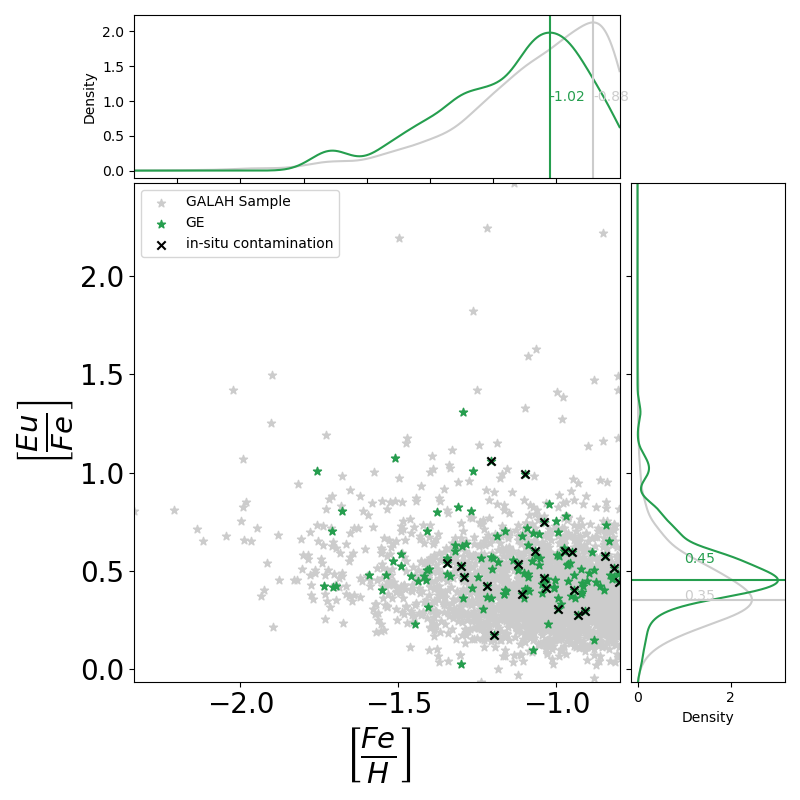}
    \includegraphics[width=.45\linewidth]{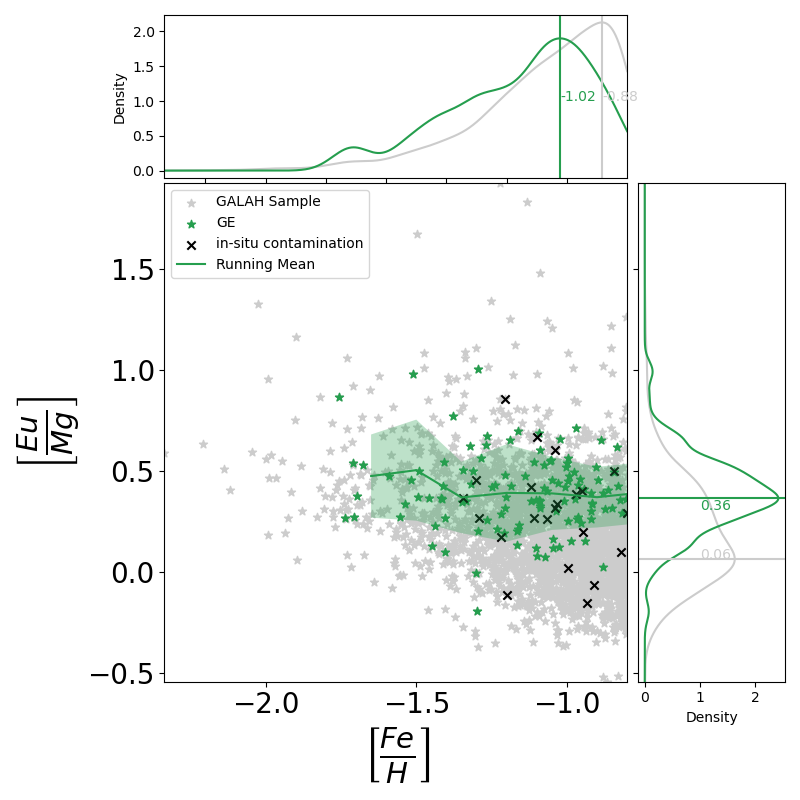}
    \caption{Diagrams involving abundances of Ba and Eu for the GE-dominated group. In grey, the whole sample of metal-poor stars is shown. Black crosses are stars of the GE-dominated group that were found to be of in situ origin with high values of [Al/Fe]. The green symbols are the remaining stars of the group. \textit{Top Left} - \BaEu~ as a function of \FeH. Dashed lines indicate the solar ratio, and values for pure contributions for the s- and r-processes as calculated by \citet{Bisterzo2014}. The horizontal lines on the KDE plot mark the peak values. \textit{Top Right} - \EuBa~ as a function of \EuFe. The limits in [Eu/Fe] for defining the r-I and r-II stars proposed by \citet{Beers2005} are indicated. \textit{Bottom Left} - \EuFe~ as a function of \FeH. \textit{Bottom Right} - \EuMg~ as a function of \FeH. The green line and the green stripe show the running mean and the scatter, respectively.}
    \label{fig:GE_barium_europium}
\end{figure*}

In the panels of Fig.~\ref{fig:GE_barium_europium}, we present the neutron capture ratios for our GE-dominated group, compared to the abundance distribution of the entire sample. For this discussion, we exclude from the group those in situ stars with high values of [Al/Fe]. The mean values of \BaFe\space and \EuFe\space of our GE are 0.26(42)~dex and 0.52(18)~dex, respectively, which implies that the GE-dominated group is enriched in elements of the r-process. This qualitatively agrees with the conclusions of previous works. However, we do not find that GE is as heavily enriched as found in \citet{Aguado2021} and find it more enriched than reported by \citet{Matsuno2021}. The overall \BaEu~ratio is $-$0.26, but increases to $-$0.15 if we consider only stars that have both Ba and Eu abundances (top left panel of Fig.~\ref{fig:GE_barium_europium}).

We can also restrict the investigation to only those stars that have [Al/Fe] $<$ 0 (the numbers above include stars where Al and Mn were not detected). In this case, \EuFe\space remains similar, 0.48(15)~dex but the \BaFe\space increases to 0.45(37). With this, we recover the same result as \citet{Matsuno2021}, \BaEu\space $\sim$ 0.0. As these stars are at the high metallicity end of the sample, what we detect is an increase in Ba produced by the s-process as the metallicity increases. Only when we look at lower metallicities do the GE stars show evidence of a purer r-process enrichment. This agrees with the findings of \citet{Aguado2021}, whose sample was of lower-metallicity stars.

\begin{figure*}
    \centering
    \includegraphics[width=.45\linewidth]{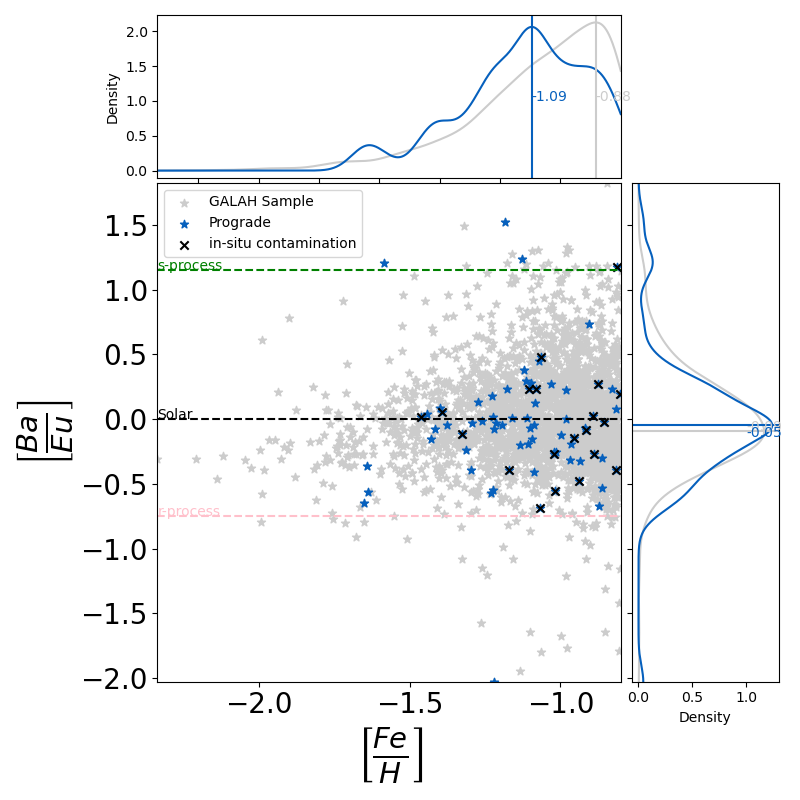}
    \includegraphics[width=.45\linewidth]{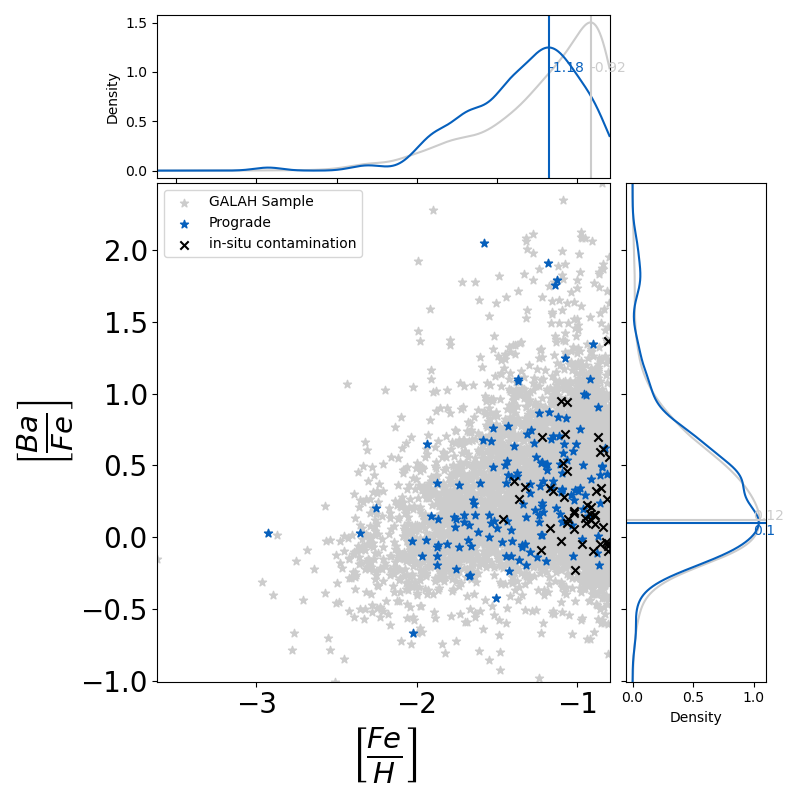}\\
    \includegraphics[width=.45\linewidth]{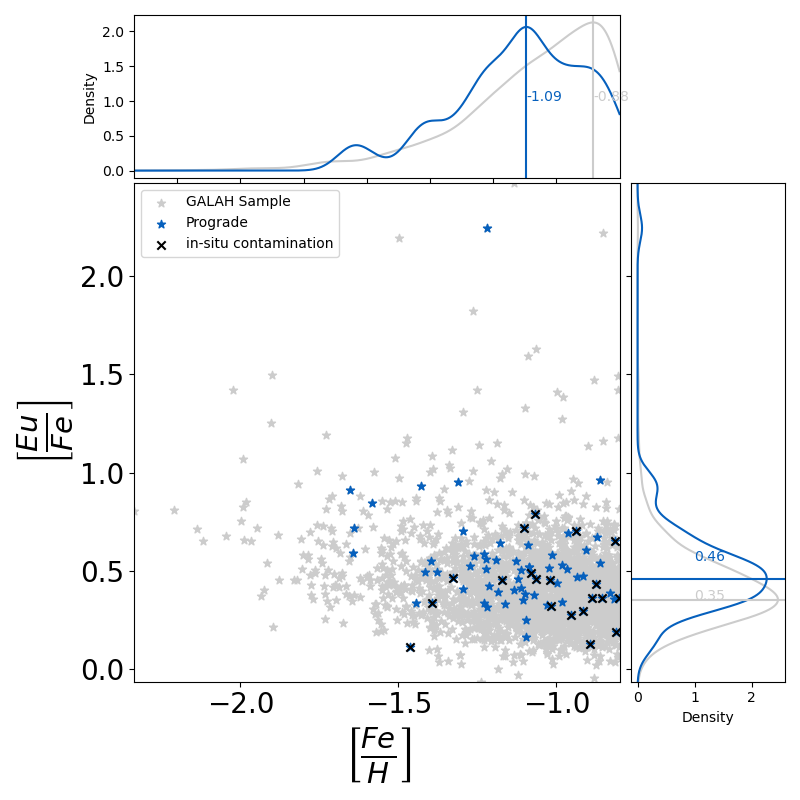}
    \includegraphics[width=.45\linewidth]{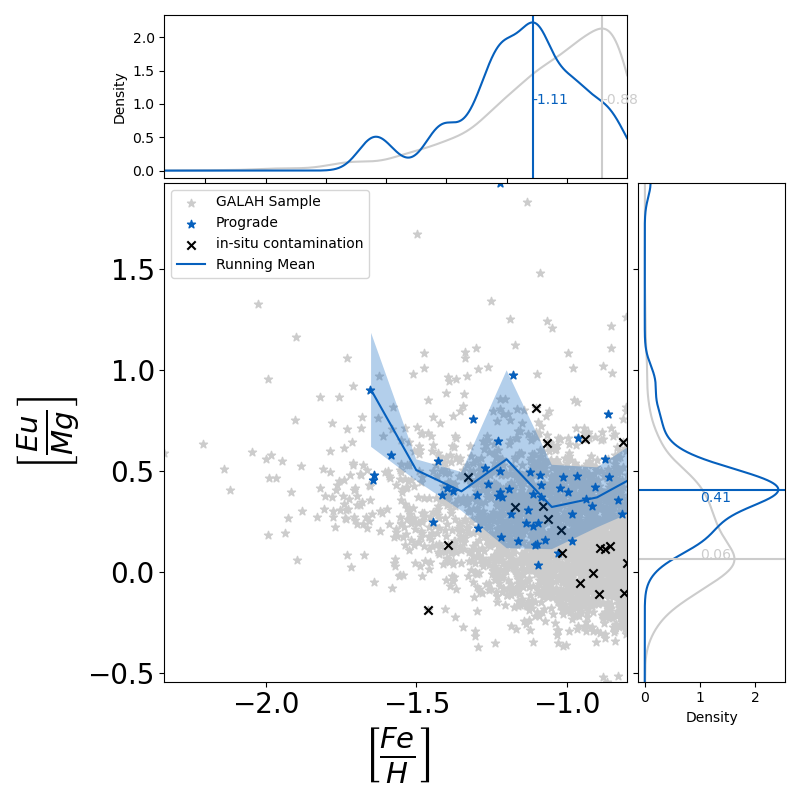}
    \caption{Same as Fig. \ref{fig:GE_barium_europium} but for the prograde group. The stars shown in blue have low [Al/Fe] or lack Al and/or Mn abundances.} 
    \label{fig:progrado_barium_europium}
\end{figure*}

\begin{figure*}
    \centering
    \includegraphics[width=.45\linewidth]{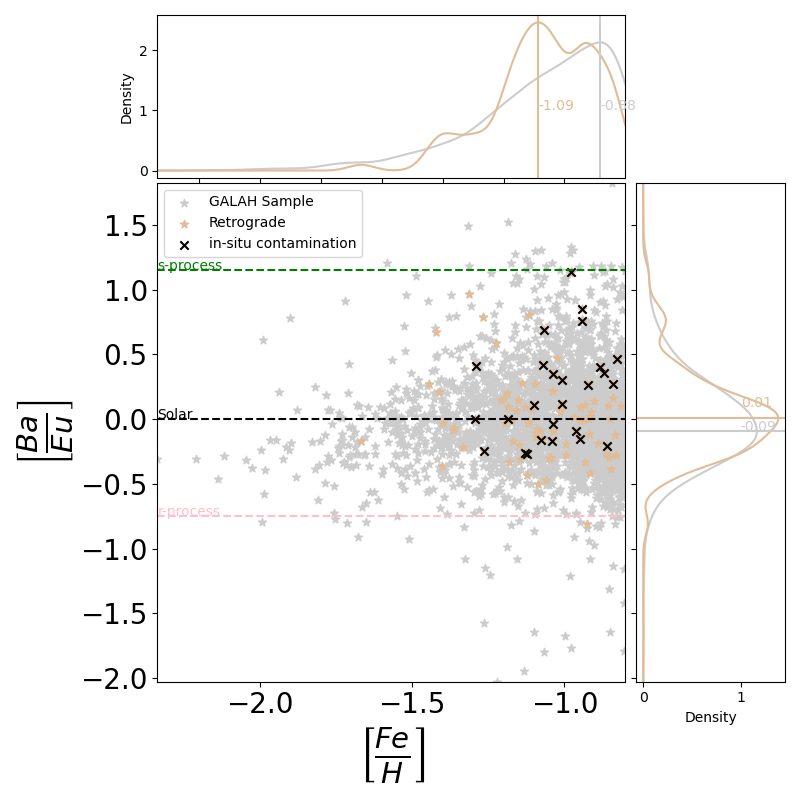}
    \includegraphics[width=.45\linewidth]{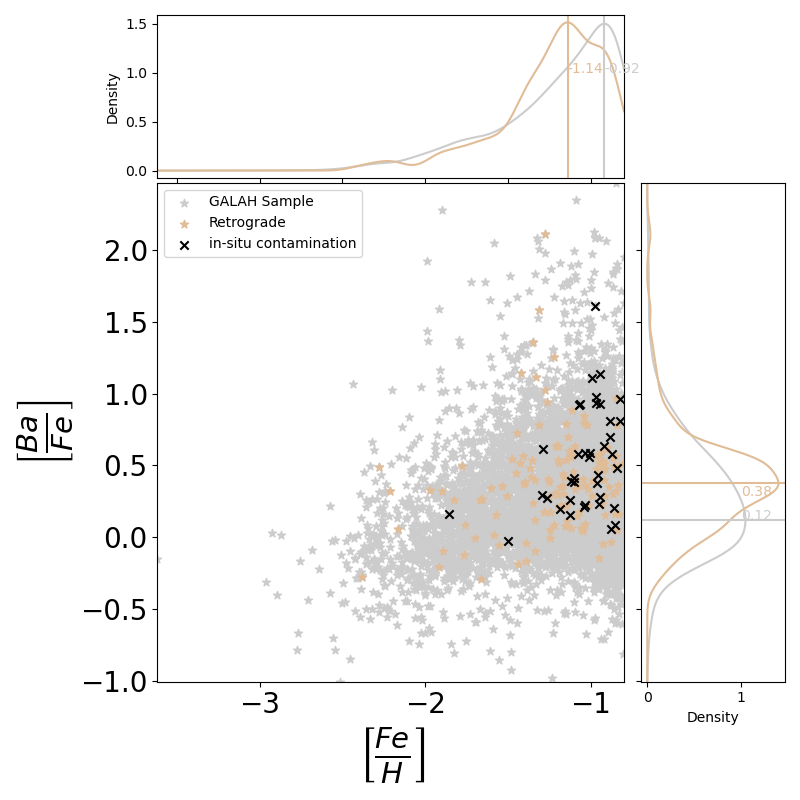}\\
    \includegraphics[width=.45\linewidth]{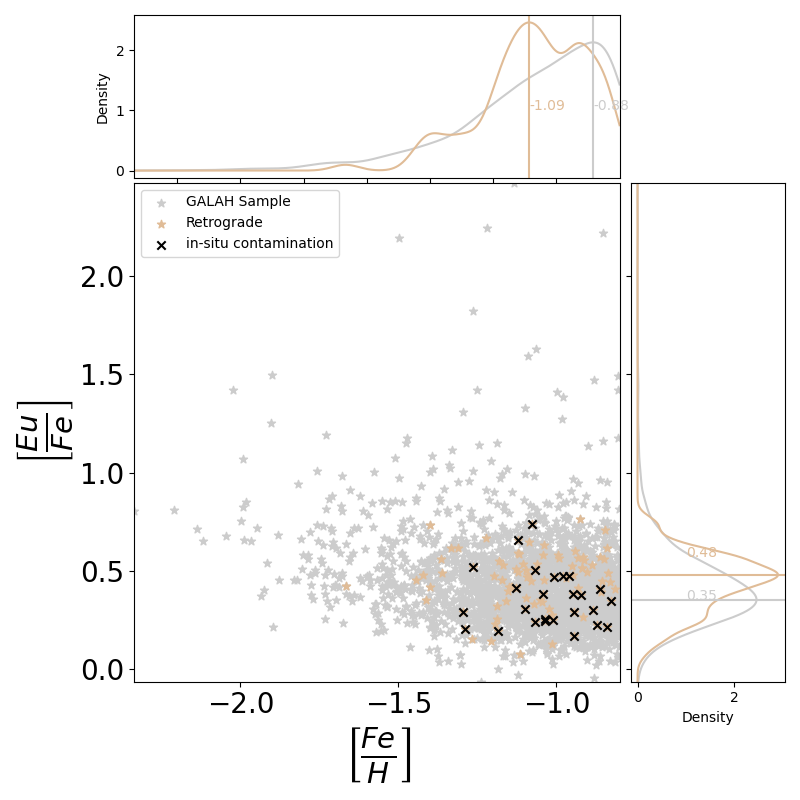}
    \includegraphics[width=.45\linewidth]{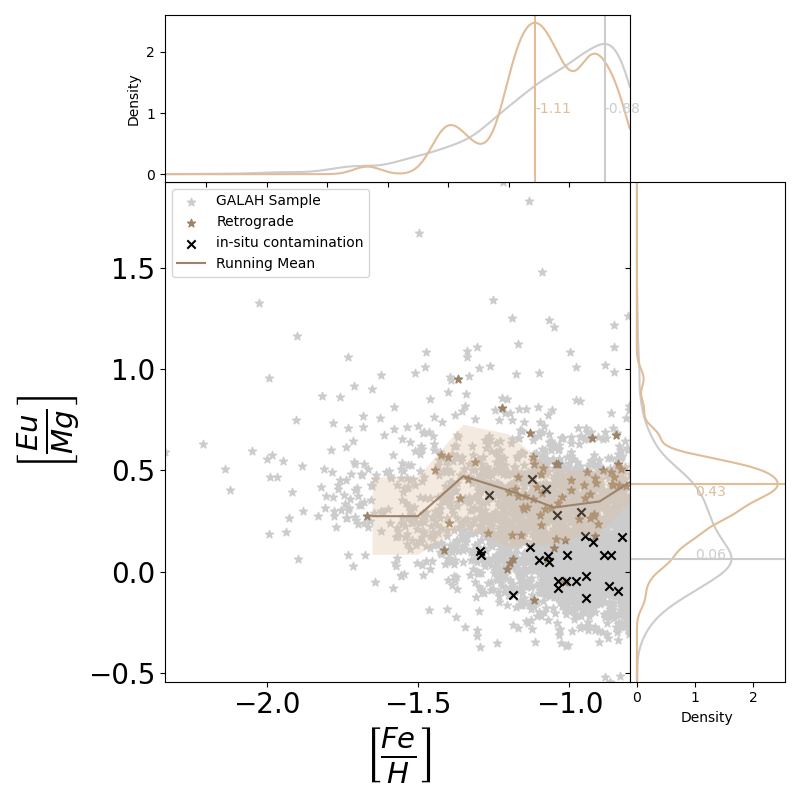}
    \caption{Same as Fig. \ref{fig:GE_barium_europium} but for the retrograde group. The stars shown in light brown have low [Al/Fe] or lack Al and/or Mn abundances.}
    \label{fig:retrogrado_barium_europium}
\end{figure*}

\begin{figure*}
    \centering
    \includegraphics[width=.45\linewidth]{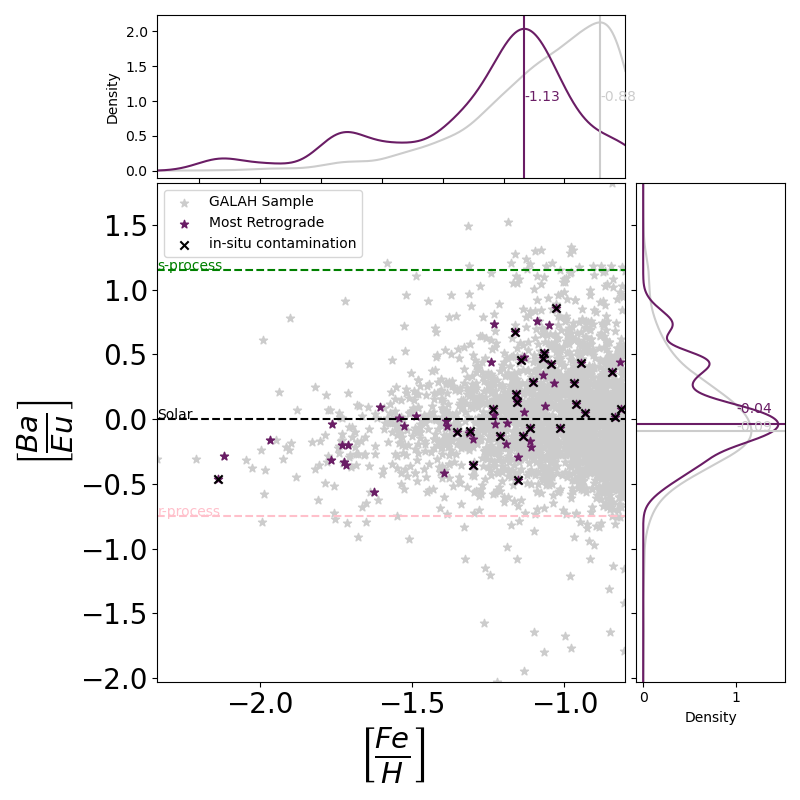}
    \includegraphics[width=.45\linewidth]{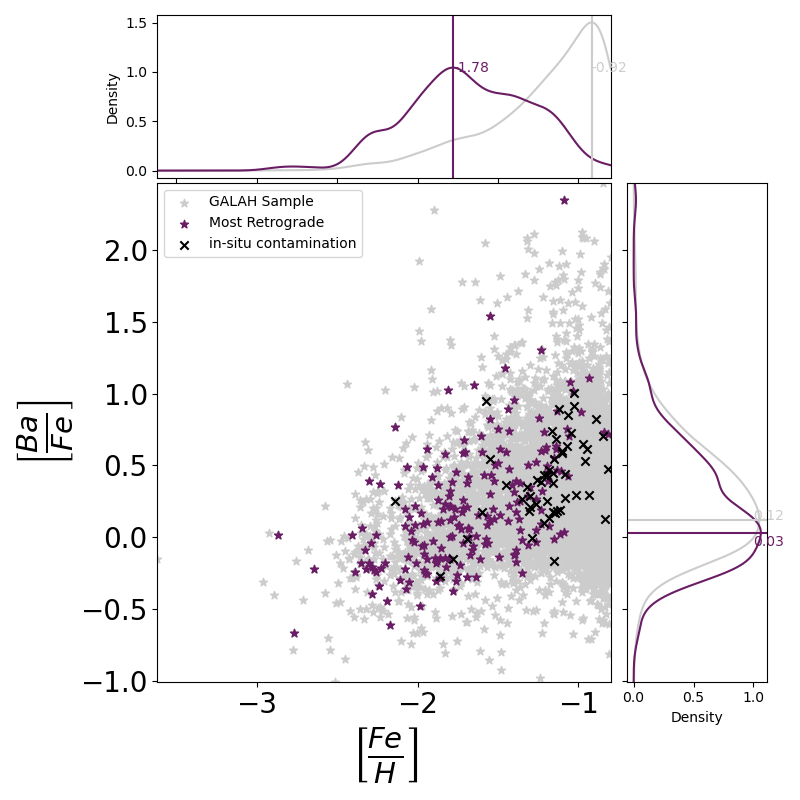}\\
    \includegraphics[width=.45\linewidth]{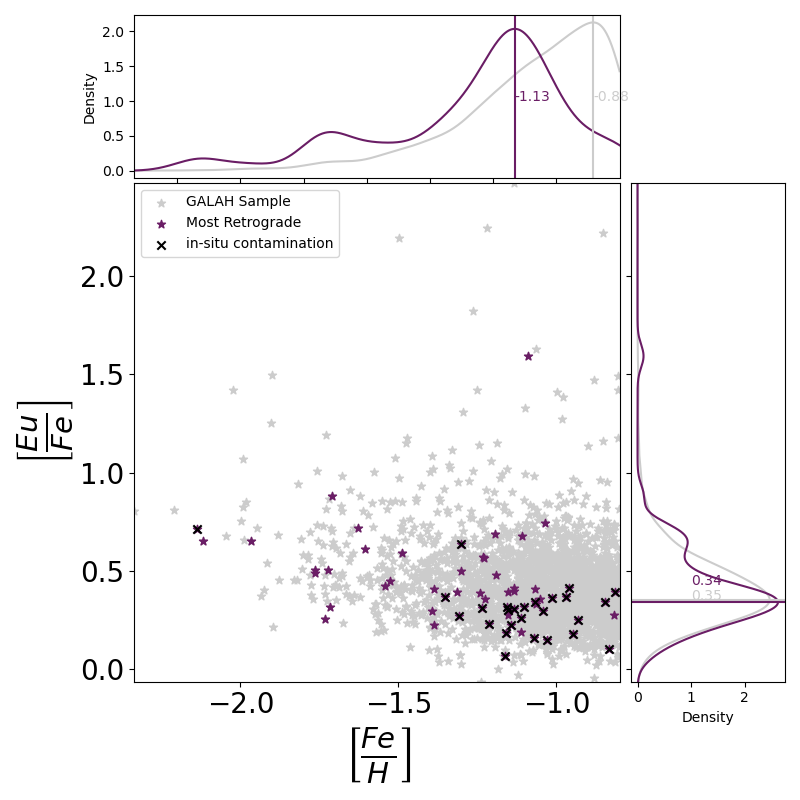}
    \includegraphics[width=.45\linewidth]{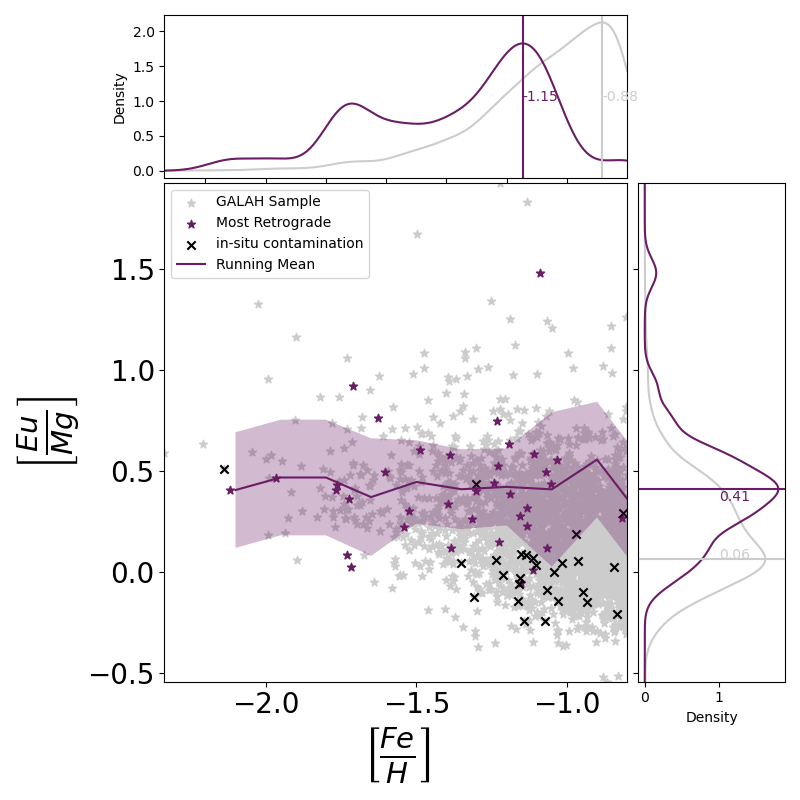}
    \caption{Same as Fig. \ref{fig:GE_barium_europium} but for the most retrograde group. The stars shown in violet have low [Al/Fe] or lack Al and/or Mn abundances.} 
    \label{fig:thule_barium_europium}
\end{figure*}

Similar behaviour is seen in the other halo groups, except for the intermediate retrograde one (see Figs. \ref{fig:progrado_barium_europium}, \ref{fig:retrogrado_barium_europium}, and \ref{fig:thule_barium_europium}). If we simply remove from the groups the stars that we think are in situ because of their high [Al/Fe] ratios, we obtain \EuFe\space = 0.55(29) and \BaFe\space = 0.33(44), \EuFe\space = 0.47(14) and \BaFe\space = 0.39(35), \EuFe\space = 0.50(25) and \BaFe\space = 0.20(41) for the prograde, retrograde, and most retrograde groups, respectively. The numbers change to \EuFe\space = 0.49(18) and \BaFe\space = 0.54(46), \EuFe\space = 0.45(15) and \BaFe\space = 0.39(21), \EuFe\space = 0.42(18) and \BaFe\space = 0.49(31) for the prograde, retrograde and most retrograde groups, respectively, if we focus on stars with low [Al/Fe] ratios only. Therefore, signs of an increased s-process contribution are also present in the prograde and most retrograde samples. For the intermediate retrograde sample, the different behaviour does not mean that an s-process contribution is not present. Instead, it reflects that this contribution of the s-process is already important at lower metallicities \citep[see, e.g., the models of ][]{Kobayashi2020}. This is supported by the increasing trend of [Ba/Fe] as a function of [Fe/H] seen in all groups (top right panels in Figs.~\ref{fig:GE_barium_europium}, \ref{fig:progrado_barium_europium}, \ref{fig:retrogrado_barium_europium} and \ref{fig:thule_barium_europium}).

An interesting indicator to look at is the \EuMg\space ratio. This ratio reflects the correlation between the production sites of the two elements. A flat trend of \EuMg\space versus \FeH\space shows that Eu and Mg are produced at the same rate. A positive slope indicates that there is an additional source of Eu that does not contribute much Mg. \citet{Asa2020} using this indicator concluded that Eu abundances in ultra-faint dwarf (UFD) galaxies likely come from two sources. One results in quick Eu enrichment and produces Mg at the same time (possibly SNe II). The other is a delayed contributor that produces more Eu than Mg, enriching the medium in r-process elements only (possibly neutron star mergers).

The \EuMg\space ratios are shown as a function of \FeH\space in the bottom right panels of Figs.~\ref{fig:GE_barium_europium}, \ref{fig:progrado_barium_europium}, \ref{fig:retrogrado_barium_europium}, and \ref{fig:thule_barium_europium}. In the GE, the intermediate retrograde and the slightly prograde groups, we detect a flat trend of \EuMg$\sim$+0.4~dex, indicating that these elements come continuously from the same source. Sometimes, particularly at the most metal-poor regime, there is one star or another that is very Eu rich, driving some variation in the running mean. Nevertheless, for most of the sample, the mean [Eu/Mg] seems constant. Although the mean seems flat, the most retrograde group appears to separate into two populations with distinct \EuMg\space values. A metal-poor part with \EuMg$\sim +$0.4~dex, similar to what is seen in the other groups, and another with \EuMg$\sim$ 0~dex and \FeH$\geq$ $-$1.5~dex. This is also seen in the [Eu/Fe] versus [Fe/H] plot (bottom left of Fig. \ref{fig:thule_barium_europium}). There is one group of stars that seems to remain Eu rich in the whole metallicity range, but another shows a decreasing [Eu/Fe] ratio with increasing metallicity. This latter population with \EuMg$\sim$ 0~dex and \FeH$\geq$ $-$1.5~dex, seems to match the behaviour of the in situ population better. 

That we see a flat \EuMg\space ratio among stars that are likely accreted, in all of our groups, means that in these stars Eu and Mg have similar rates of production. This implies that there is no clear sign of a source that only produces r-process elements, such as neutron star mergers (NSMs). This conclusion goes in contrary to the findings of other works, where such a contribution was needed \citep{Naidu2022}. We do see that the \EuMg\space ratio of GE stars becomes different from that of in situ stars at metallicities around \FeH\space = $\sim$ $-$1.3, but without change in its \EuMg\space value. Assuming the age-metallicity relationship of GE derived by \citet{GiribaldiSmiljanic23}, this metallicity corresponds to ages around 11.0--10.5 Ga. This means that even after $\sim$2.0-2.5 Ga of star formation history, the nucleosynthetic contribution from NSM did not become significant, implying perhaps longer delays for the contribution of this type of source.

\subsection{Other chemical abundances in the halo groups}\label{sec:ks_abun}

\begin{figure*}
    \centering
    \includegraphics[width=1\linewidth]{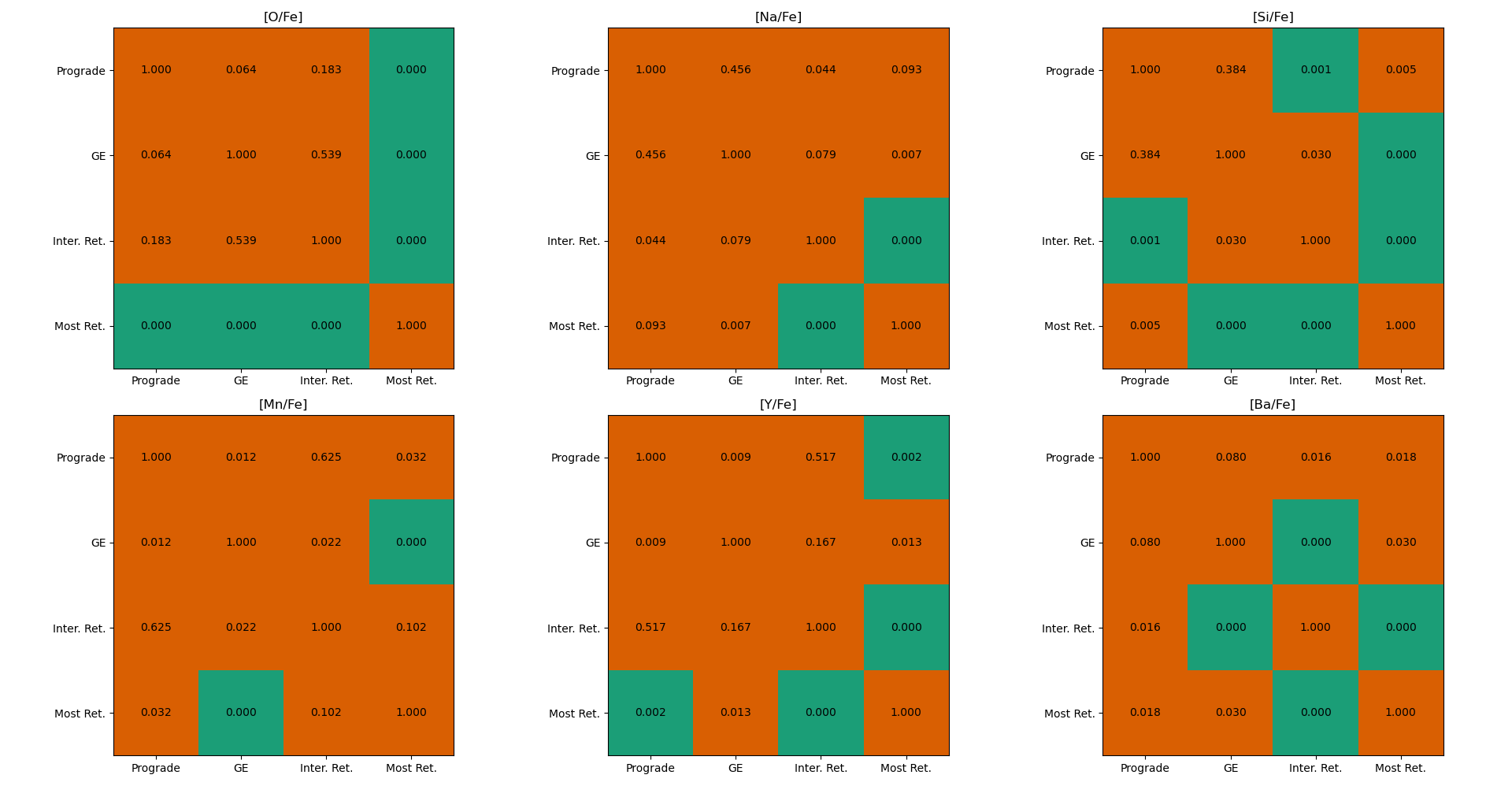}
    \caption{Results of the KS test applied to the chemical abundances of all four halo groups defined in this work. These results concern the case where the in situ contamination that has [Al/Fe] $>$ 0.0 was removed from all groups. Brown-coloured regions indicate that the null hypothesis was not rejected (i.e., there is no difference between the groups). The green regions indicate that the null hypothesis was rejected (there is a difference between the groups). Only elements where some difference was detected are shown. For all other elements, the KS test indicates no difference between the groups.}
    \label{fig:KStest}
\end{figure*}

Finally, we tested for chemical differences between halo groups using all other elements available in the GALAH DR3 catalogue. Distributions of abundances were compared using the Kolmogorov-Smirnov (KS) test. We adopted the traditional value of $p=0.05$ Bonferroni corrected for the number of comparisons we make, thus $p=0.002$ for considering the distributions as different (rejecting the hypothesis that they are the same). As in the discussion above, this comparison was done for two cases; first, removing only the confirmed in situ contamination that has [Al/Fe] $>$ 0.0 (and leaving inside the sample the stars without good Al abundances), and second, restricting the sample to only those objects with good Al abundances and [Al/Fe] $<$ 0.0. part of the results for the first case are shown in Fig.~\ref{fig:KStest} in the form of correlation matrices. 

Essentially, we find that only the most retrograde group appears to be chemically distinct from the other groups. It differs from all other groups in the distributions of [O/Fe]. In addition, it differs from the prograde group in [Y/Fe]; from the GE-dominated group in [Si/Fe] and \MnFe; and from the intermediate retrograde group in [Si/Fe], [Y/Fe] and \BaFe. Otherwise, we also detect a difference in [Si/Fe] between the prograde and intermediate retrograde groups and in [Ba/Fe] between the GE-dominated and intermediate retrograde groups. For the other elements available in GALAH DR3, we found no difference between the groups.  

If we concentrate instead on those stars with good Al abundances and [Al/Fe] $<$ 0.0, all differences disappear except for the [Mn/Fe] ratio of the most retrograde group, which becomes different from all other groups. However, one should keep in mind that sometimes the number of stars that are compared can be small, depending on the element being tested. In summary, from comparisons of both subsamples, we find that the most retrograde group is the only one with clear signs of being chemically different from the others. The abundances of the prograde, GE-dominated, and intermediate retrograde groups are very similar.

In addition, we separated from each group two samples, one containing only dwarfs and the other only giants. We repeated the abundance comparisons with KS tests only among samples of the same stellar types. As discussed in more detail in Appendix \ref{sec:appendix}, Fig. \ref{fig:appendix_ks_giants}, abundance differences appear only in the sample of giants. Also, in this case, the differences appear only for the most retrograde group. For dwarfs, all abundances are similar. Although this seems to indicate the existence of systematic effects between dwarfs and giants in the GALAH results, the differences seen in the giants support the discussion above.

We interpret these results, together with the information that was gathered in the previous sections, as signs that the accreted component part of the prograde, GE-dominated, and retrograde groups is the same. Therefore, our exploration seems to show no significant signs of accretion events other than GE in the region of the parameter space occupied by the three groups; at least in the sample we selected for analysis here. This conclusion could be different if, for example, the sample was selected to extend to higher metallicity values \citep[see e.g.][]{Myeong2022}. 

However, the most retrograde group shows signs of being chemically distinct from the others. We interpret this result as showing that the accreted stars present in this group possibly have a different origin. This could be attributed to a different accretion event, for example Thamnos and/or Sequoia \citep{Myeong2018, Koppelman2019} or to a separate region of the GE progenitor, where the chemical enrichment history was distinct \citep{Koppelman2020, Amarante2022}. In that sense, we call attention to the discussion in \citet{Amarante2022}, where it is argued on the basis of simulations that Sequoia could be explained as the metal-poor part of the same merger that gave origin to GE. \citet{Amarante2022} show that accreted stars that end up with very retrograde, high energy orbits tend to be more metal-poor, with a multi-peak metallicity distribution, more $\alpha$-rich, to have a larger spread in abundances, and to have less eccentric orbits. Although the stars in our most retrograde group are perhaps not in the same high-energy orbits, the remaining characteristics seen in the \citet{Amarante2022} simulations agree with what we detect.

Finally, it is also important to note that uncertainties in the abundances might still be affecting our capability of detecting chemical differences among the groups. Therefore, additional studies that derive high-quality abundances are needed to confirm or refute our findings. 

\section{Conclusions}\label{sec:conclusions}

We performed a comprehensive analysis of the chemodynamic properties of metal-poor stars ([Fe/H] $<$ $-$0.8) present in the DR3 of the GALAH survey \citep{Buder2021}. We used unsupervised machine learning methods, t-SNE and HAC, on a parameter space made of two dynamic (\Jx\space and \Jy) and two chemical ([Fe/H] and [Mg/Fe]) quantities to separate the stars into groups of similar properties. We do not use cuts in the parameter space, but attempt to let the sample itself tell how it should be best divided. This analysis was repeated 50 times to estimate the probability that the stars belong to the groups that were identified. On the basis of this analysis, we defined four groups of halo stars. 

The first group was selected as the one in which the stars show an angular momentum and radial action distribution similar to what has been defined as the \emph{Gaia}-Enceladus merger. We refer to this as the GE-dominated group. Two other groups were defined to be slightly more prograde and slightly more retrograde than the GE one. The final group was selected to be composed of stars that show the most extreme retrograde motions in the sample. We then performed a detailed exploratory study of the dynamical and chemical properties of the groups, trying to be as independent of predefined concepts as possible. The following conclusions can be drawn from this exploration.

\begin{itemize}
    \item Our GE-dominated group has a mean metallicity of [Fe/H] = $-$1.29 and extends to [Fe/H] $\sim$ $-$2.7. The metallicity distribution of the prograde and retrograde groups is very similar. On the other hand, the most retrograde group has a larger fraction of more metal-poor stars;
    \item The GE-dominated group contains stars with the most eccentric orbits in the sample (median eccentricity equal to 0.86);
    \item The prograde and retrograde groups also contain stars of very eccentric orbits, but with medians of lower values, 0.80 and 0.76, respectively;
    \item The most retrograde group has a completely different eccentricity distribution, between 0.2 and 0.6 with a mean of 0.42. Structures identified in the same regime of dynamic properties (Thamnos and Sequoia) have already been reported to have eccentricities of about 0.6 or lower;
    \item Using abundances of Mg, Mn, and Al, we find an important in-situ contamination in all four groups. We identify this contamination as the Erebus component defined in \citet{GiribaldiSmiljanic23}, which could also be related to the Aurora defined by \citet{BelokurovKravtsov2022}. At the high-metallicity end ($>$ $-$1.4), chemistry can tentatively be used to separate accreted stars from in situ ones. However, there is important scatter that sometimes blurs the division;
    \item At lower metallicities, it is less clear that chemistry (Mg, Al, and Mn) can separate in situ and accreted stars. However, abundances of Al and Mn are not available for a large number of stars in the catalogue. We note that chemical evolution models suggest that low-metallicity in situ stars can also have low Al abundances \citep{Kobayashi2020}, causing additional difficulties in separating stars of different origins.
    \item Because the chemodynamic separation between accreted and in situ stars becomes difficult, it is possible that unrecognised in situ contamination affects samples selected as being of possible low-metallicity accreted stars. This might have important consequences for works that attempt to infer the properties of the original systems that merged with the Galaxy.
    \item Regarding neutron capture elements, we find accreted stars from GE to be r-process rich at the low metallicity end of our sample. However, the abundance of Ba increases with metallicity, indicating that an increasing contribution of the s-process is present. The same is seen in accreted stars of all groups.
    \item The most retrograde group of stars seems to contain stars from two populations with different trends in [Eu/Fe] as a function of [Fe/H]. One population of stars remains Eu rich in the whole metallicity range, just as for the accreted stars in the other three halo groups. However, the other shows a decreasing [Eu/Fe] ratio with increasing metallicity. We interpret the latter as part of an in situ contamination.
    \item The \EuFe~ ratios and in particular the flat trend of \EuMg~ with [Fe/H] indicate that the accreted stars were enriched by a source that produces Mg and r-process elements simultaneously (e.g. CCSNe). We do not see signs of major contribution from a delayed source, such as NSM, that produces only r-process elements.
    \item We tested for differences in the abundances of all other elements available in the catalogue. We find that the accreted stars in the prograde, GE-dominated, and retrograde groups, at this regime of metallicity, do not differ among themselves. Essentially, only the stars of the most retrograde group show signs of different chemical abundances in some elements.
    \item On the basis of the observations above, we conclude that the current sample shows signs of at most two distinct accretion events. One that we can associate to GE with stars distributed in the region of the parameter space occupied by three of our groups. The second, which can be associated to Thamnos and/or Sequoia, is present only in the most retrograde group of stars. Nevertheless, we call attention to the discussion in \citet{Amarante2022}, which suggested that one single merger could explain the accreted stars in all four of our groups. However, we note here that the sample we analysed does not cover regions of the Galaxy where other substructures are found, as for example, Heracles \citep{Horta2021} in the inner Galaxy, the Cetus Polar Stream \citep[at about 34 kpc from the Sun,][]{Newberg2009}, or the Saggitarius stream \citep[beyond 20 kpc of the Sun,][]{Majewski2003}.

\end{itemize}

Finally, we point out that to advance in the characterisation and separation of stars belonging to different merger events, chemical abundances of high quality are needed. Moreover, we believe that it is important to take into account the guidance offered by chemical evolution models when this kind of separation is attempted.

\begin{acknowledgements}

The authors thank the anonymous referee for the quick and constructive feedback. ARS and RS acknowledge the support of the Polish National Science Centre (NCN) through the project 2018/31/B/ST9/01469. ARS would like to thank RS for the kind mentorship through these difficult times. ARS thanks Riano Giribaldi, Maria Luiza Linhares Dantas, and H\'elio Perottoni for the discussions. This research has made use of the SIMBAD database, operated at CDS, Strasbourg, France, and the NASA Astrophysics Data System. This work presents results from the European Space Agency (ESA) space mission Gaia. Gaia data are being processed by the Gaia Data Processing and Analysis Consortium (DPAC). Funding for the DPAC is provided by national institutions, in particular, the institutions participating in the Gaia MultiLateral Agreement (MLA). The Gaia mission website is \url{https://www.cosmos.esa.int/gaia}. The Gaia archive website is \url{https://archives.esac.esa.int/gaia}. This work also used GALAH survey data, their website is \url{http://galah-survey.org/}. This research made use of Astropy\footnote{\url{http://www.astropy.org}} a community-developed core Python package for Astronomy \citep{astropy:2013, astropy:2018}. This research used Mesa Isochrones and Stellar Tracks (MIST)\footnote{\url{waps.cfa.harvard.edu/MIST/}}\citep{MIST1,MIST2}.

\end{acknowledgements}
    
\bibliographystyle{aa} 
\bibliography{references} 

\begin{thebibliography}{151}
\expandafter\ifx\csname natexlab\endcsname\relax\def\natexlab#1{#1}\fi

\bibitem[{{Aguado} {et~al.}(2021){Aguado}, {Belokurov}, {Myeong}, {Evans},
  {Kobayashi}, {Sbordone}, {Chanam{\'e}}, {Navarrete}, \&
  {Koposov}}]{Aguado2021}
{Aguado}, D.~S., {Belokurov}, V., {Myeong}, G.~C., {et~al.} 2021, \apjl, 908,
  L8

\bibitem[{{Amarante} {et~al.}(2020){Amarante}, {Beraldo e Silva}, {Debattista},
  \& {Smith}}]{Amarante2020}
{Amarante}, J. A.~S., {Beraldo e Silva}, L., {Debattista}, V.~P., \& {Smith},
  M.~C. 2020, \apjl, 891, L30

\bibitem[{{Amarante} {et~al.}(2022){Amarante}, {Debattista}, {Beraldo e Silva},
  {Laporte}, \& {Deg}}]{Amarante2022}
{Amarante}, J. A.~S., {Debattista}, V.~P., {Beraldo e Silva}, L., {Laporte}, C.
  F.~P., \& {Deg}, N. 2022, \apj, 937, 12

\bibitem[{{Anders} {et~al.}(2018){Anders}, {Chiappini}, {Santiago},
  {Matijevi{\v{c}}}, {Queiroz}, {Steinmetz}, \& {Guiglion}}]{Anders2018}
{Anders}, F., {Chiappini}, C., {Santiago}, B.~X., {et~al.} 2018, \aap, 619,
  A125

\bibitem[{{Arnould} {et~al.}(1999){Arnould}, {Goriely}, \&
  {Jorissen}}]{Arnould1999}
{Arnould}, M., {Goriely}, S., \& {Jorissen}, A. 1999, \aap, 347, 572

\bibitem[{{Astropy Collaboration} {et~al.}(2018){Astropy Collaboration},
  {Price-Whelan}, {Sip{\H{o}}cz}, {G{\"u}nther}, {Lim}, {Crawford}, {Conseil},
  {Shupe}, {Craig}, {Dencheva}, {Ginsburg}, {Vand erPlas}, {Bradley},
  {P{\'e}rez-Su{\'a}rez}, {de Val-Borro}, {Aldcroft}, {Cruz}, {Robitaille},
  {Tollerud}, {Ardelean}, {Babej}, {Bach}, {Bachetti}, {Bakanov}, {Bamford},
  {Barentsen}, {Barmby}, {Baumbach}, {Berry}, {Biscani}, {Boquien}, {Bostroem},
  {Bouma}, {Brammer}, {Bray}, {Breytenbach}, {Buddelmeijer}, {Burke},
  {Calderone}, {Cano Rodr{\'\i}guez}, {Cara}, {Cardoso}, {Cheedella}, {Copin},
  {Corrales}, {Crichton}, {D'Avella}, {Deil}, {Depagne}, {Dietrich}, {Donath},
  {Droettboom}, {Earl}, {Erben}, {Fabbro}, {Ferreira}, {Finethy}, {Fox},
  {Garrison}, {Gibbons}, {Goldstein}, {Gommers}, {Greco}, {Greenfield},
  {Groener}, {Grollier}, {Hagen}, {Hirst}, {Homeier}, {Horton}, {Hosseinzadeh},
  {Hu}, {Hunkeler}, {Ivezi{\'c}}, {Jain}, {Jenness}, {Kanarek}, {Kendrew},
  {Kern}, {Kerzendorf}, {Khvalko}, {King}, {Kirkby}, {Kulkarni}, {Kumar},
  {Lee}, {Lenz}, {Littlefair}, {Ma}, {Macleod}, {Mastropietro}, {McCully},
  {Montagnac}, {Morris}, {Mueller}, {Mumford}, {Muna}, {Murphy}, {Nelson},
  {Nguyen}, {Ninan}, {N{\"o}the}, {Ogaz}, {Oh}, {Parejko}, {Parley}, {Pascual},
  {Patil}, {Patil}, {Plunkett}, {Prochaska}, {Rastogi}, {Reddy Janga},
  {Sabater}, {Sakurikar}, {Seifert}, {Sherbert}, {Sherwood-Taylor}, {Shih},
  {Sick}, {Silbiger}, {Singanamalla}, {Singer}, {Sladen}, {Sooley},
  {Sornarajah}, {Streicher}, {Teuben}, {Thomas}, {Tremblay}, {Turner},
  {Terr{\'o}n}, {van Kerkwijk}, {de la Vega}, {Watkins}, {Weaver}, {Whitmore},
  {Woillez}, {Zabalza}, \& {Astropy Contributors}}]{astropy:2018}
{Astropy Collaboration}, {Price-Whelan}, A.~M., {Sip{\H{o}}cz}, B.~M., {et~al.}
  2018, \aj, 156, 123

\bibitem[{{Astropy Collaboration} {et~al.}(2013){Astropy Collaboration},
  {Robitaille}, {Tollerud}, {Greenfield}, {Droettboom}, {Bray}, {Aldcroft},
  {Davis}, {Ginsburg}, {Price-Whelan}, {Kerzendorf}, {Conley}, {Crighton},
  {Barbary}, {Muna}, {Ferguson}, {Grollier}, {Parikh}, {Nair}, {Unther},
  {Deil}, {Woillez}, {Conseil}, {Kramer}, {Turner}, {Singer}, {Fox}, {Weaver},
  {Zabalza}, {Edwards}, {Azalee Bostroem}, {Burke}, {Casey}, {Crawford},
  {Dencheva}, {Ely}, {Jenness}, {Labrie}, {Lim}, {Pierfederici}, {Pontzen},
  {Ptak}, {Refsdal}, {Servillat}, \& {Streicher}}]{astropy:2013}
{Astropy Collaboration}, {Robitaille}, T.~P., {Tollerud}, E.~J., {et~al.} 2013,
  \aap, 558, A33

\bibitem[{{Bailer-Jones} {et~al.}(2021){Bailer-Jones}, {Rybizki}, {Fouesneau},
  {Demleitner}, \& {Andrae}}]{Bailer-Jones2021}
{Bailer-Jones}, C.~A.~L., {Rybizki}, J., {Fouesneau}, M., {Demleitner}, M., \&
  {Andrae}, R. 2021, \aj, 161, 147

\bibitem[{{Balbinot} {et~al.}(2023){Balbinot}, {Helmi}, {Callingham},
  {Matsuno}, {Dodd}, \& {Ruiz-Lara}}]{Balbinot2023}
{Balbinot}, E., {Helmi}, A., {Callingham}, T., {et~al.} 2023, arXiv e-prints,
  arXiv:2306.02756

\bibitem[{{Barb{\'a}} {et~al.}(2019){Barb{\'a}}, {Minniti}, {Geisler},
  {Alonso-Garc{\'\i}a}, {Hempel}, {Monachesi}, {Arias}, \&
  {G{\'o}mez}}]{Barba2019}
{Barb{\'a}}, R.~H., {Minniti}, D., {Geisler}, D., {et~al.} 2019, \apjl, 870,
  L24

\bibitem[{{Beers} \& {Christlieb}(2005)}]{Beers2005}
{Beers}, T.~C. \& {Christlieb}, N. 2005, \araa, 43, 531

\bibitem[{{Beers} {et~al.}(2014){Beers}, {Norris}, {Placco}, {Lee}, {Rossi},
  {Carollo}, \& {Masseron}}]{Beers2014}
{Beers}, T.~C., {Norris}, J.~E., {Placco}, V.~M., {et~al.} 2014, \apj, 794, 58

\bibitem[{{Belokurov} {et~al.}(2018){Belokurov}, {Erkal}, {Evans}, {Koposov},
  \& {Deason}}]{Belokurov2018}
{Belokurov}, V., {Erkal}, D., {Evans}, N.~W., {Koposov}, S.~E., \& {Deason},
  A.~J. 2018, \mnras, 478, 611

\bibitem[{{Belokurov} \& {Kravtsov}(2022)}]{BelokurovKravtsov2022}
{Belokurov}, V. \& {Kravtsov}, A. 2022, \mnras, 514, 689

\bibitem[{{Belokurov} {et~al.}(2020){Belokurov}, {Sanders}, {Fattahi}, {Smith},
  {Deason}, {Evans}, \& {Grand}}]{Belokurov2020}
{Belokurov}, V., {Sanders}, J.~L., {Fattahi}, A., {et~al.} 2020, \mnras, 494,
  3880

\bibitem[{{Belokurov} {et~al.}(2006){Belokurov}, {Zucker}, {Evans}, {Gilmore},
  {Vidrih}, {Bramich}, {Newberg}, {Wyse}, {Irwin}, {Fellhauer}, {Hewett},
  {Walton}, {Wilkinson}, {Cole}, {Yanny}, {Rockosi}, {Beers}, {Bell},
  {Brinkmann}, {Ivezi{\'c}}, \& {Lupton}}]{Belokurov2006}
{Belokurov}, V., {Zucker}, D.~B., {Evans}, N.~W., {et~al.} 2006, \apjl, 642,
  L137

\bibitem[{{Bensby} {et~al.}(2005){Bensby}, {Feltzing}, {Lundstr{\"o}m}, \&
  {Ilyin}}]{Bensby2005}
{Bensby}, T., {Feltzing}, S., {Lundstr{\"o}m}, I., \& {Ilyin}, I. 2005, \aap,
  433, 185

\bibitem[{{Bisterzo} {et~al.}(2014){Bisterzo}, {Travaglio}, {Gallino},
  {Wiescher}, \& {K{\"a}ppeler}}]{Bisterzo2014}
{Bisterzo}, S., {Travaglio}, C., {Gallino}, R., {Wiescher}, M., \&
  {K{\"a}ppeler}, F. 2014, \apj, 787, 10

\bibitem[{{Bland-Hawthorn} \& {Freeman}(2014)}]{BlandHawthorn2014}
{Bland-Hawthorn}, J. \& {Freeman}, K. 2014, Saas-Fee Advanced Course, 37, 1

\bibitem[{{Bonaca} {et~al.}(2020){Bonaca}, {Conroy}, {Cargile}, {Naidu},
  {Johnson}, {Zaritsky}, {Ting}, {Caldwell}, {Han}, \& {van
  Dokkum}}]{Bonaca2020}
{Bonaca}, A., {Conroy}, C., {Cargile}, P.~A., {et~al.} 2020, \apjl, 897, L18

\bibitem[{{Bonifacio} {et~al.}(2021){Bonifacio}, {Monaco}, {Salvadori},
  {Caffau}, {Spite}, {Sbordone}, {Spite}, {Ludwig}, {Di Matteo}, {Haywood},
  {Fran{\c{c}}ois}, {Koch-Hansen}, {Christlieb}, \& {Zaggia}}]{Bonifacio2021}
{Bonifacio}, P., {Monaco}, L., {Salvadori}, S., {et~al.} 2021, \aap, 651, A79

\bibitem[{{Borre} {et~al.}(2022){Borre}, {Aguirre B{\o}rsen-Koch}, {Helmi},
  {Koppelman}, {Nielsen}, {R{\o}rsted}, {Stello}, {Stokholm}, {Winther},
  {Davies}, {Hon}, {Kruijssen}, {Laporte}, {Reyes}, \& {Yu}}]{Borre22}
{Borre}, C.~C., {Aguirre B{\o}rsen-Koch}, V., {Helmi}, A., {et~al.} 2022,
  \mnras, 514, 2527

\bibitem[{{Bovy}(2015)}]{Bovy2015}
{Bovy}, J. 2015, \apjs, 216, 29

\bibitem[{{Brook} {et~al.}(2003){Brook}, {Kawata}, {Gibson}, \&
  {Flynn}}]{Brook2003}
{Brook}, C.~B., {Kawata}, D., {Gibson}, B.~K., \& {Flynn}, C. 2003, \apjl, 585,
  L125

\bibitem[{{Buder} {et~al.}(2022){Buder}, {Lind}, {Ness}, {Feuillet}, {Horta},
  {Monty}, {Buck}, {Nordlander}, {Bland-Hawthorn}, {Casey}, {de Silva},
  {D'Orazi}, {Freeman}, {Hayden}, {Kos}, {Martell}, {Lewis}, {Lin},
  {Schlesinger}, {Sharma}, {Simpson}, {Stello}, {Zucker}, {Zwitter},
  {Ciuc{\u{a}}}, {Horner}, {Kobayashi}, {Ting}, {Wyse}, \& {Wyse}}]{Buder2022}
{Buder}, S., {Lind}, K., {Ness}, M.~K., {et~al.} 2022, \mnras, 510, 2407

\bibitem[{{Buder} {et~al.}(2021){Buder}, {Sharma}, {Kos}, {Amarsi},
  {Nordlander}, {Lind}, {Martell}, {Asplund}, {Bland-Hawthorn}, {Casey}, {de
  Silva}, {D'Orazi}, {Freeman}, {Hayden}, {Lewis}, {Lin}, {Schlesinger},
  {Simpson}, {Stello}, {Zucker}, {Zwitter}, {Beeson}, {Buck}, {Casagrande},
  {Clark}, {{\v{C}}otar}, {da Costa}, {de Grijs}, {Feuillet}, {Horner},
  {Kafle}, {Khanna}, {Kobayashi}, {Liu}, {Montet}, {Nandakumar}, {Nataf},
  {Ness}, {Spina}, {Tepper-Garc{\'\i}a}, {Ting}, {Traven},
  {Vogrin{\v{c}}i{\v{c}}}, {Wittenmyer}, {Wyse}, {{\v{Z}}erjal},
  {{\v{Z}}erjal}, \& {Galah Collaboration}}]{Buder2021}
{Buder}, S., {Sharma}, S., {Kos}, J., {et~al.} 2021, \mnras, 506, 150

\bibitem[{{Bullock} \& {Johnston}(2005)}]{BullockJohnston2005}
{Bullock}, J.~S. \& {Johnston}, K.~V. 2005, \apj, 635, 931

\bibitem[{{Carney} {et~al.}(1996){Carney}, {Laird}, {Latham}, \&
  {Aguilar}}]{Carney1996}
{Carney}, B.~W., {Laird}, J.~B., {Latham}, D.~W., \& {Aguilar}, L.~A. 1996,
  \aj, 112, 668

\bibitem[{{Carollo} {et~al.}(2010){Carollo}, {Beers}, {Chiba}, {Norris},
  {Freeman}, {Lee}, {Ivezi{\'c}}, {Rockosi}, \& {Yanny}}]{Carollo2010}
{Carollo}, D., {Beers}, T.~C., {Chiba}, M., {et~al.} 2010, \apj, 712, 692

\bibitem[{{Carollo} {et~al.}(2007){Carollo}, {Beers}, {Lee}, {Chiba}, {Norris},
  {Wilhelm}, {Sivarani}, {Marsteller}, {Munn}, {Bailer-Jones}, {Fiorentin}, \&
  {York}}]{Carollo2007}
{Carollo}, D., {Beers}, T.~C., {Lee}, Y.~S., {et~al.} 2007, \nat, 450, 1020

\bibitem[{{Carollo} \& {Chiba}(2021)}]{Carollo2021}
{Carollo}, D. \& {Chiba}, M. 2021, \apj, 908, 191

\bibitem[{{Carrillo} {et~al.}(2023){Carrillo}, {Deason}, {Fattahi},
  {Callingham}, \& {Grand}}]{Carrillo2023}
{Carrillo}, A., {Deason}, A.~J., {Fattahi}, A., {Callingham}, T.~M., \&
  {Grand}, R. J.~J. 2023, arXiv e-prints, arXiv:2306.00770

\bibitem[{{Casagrande} {et~al.}(2021){Casagrande}, {Lin}, {Rains}, {Liu},
  {Buder}, {Horner}, {Asplund}, {Lewis}, {Martell}, {Nordlander}, {Stello},
  {Ting}, {Wittenmyer}, {Bland-Hawthorn}, {Casey}, {De Silva}, {D'Orazi},
  {Freeman}, {Hayden}, {Kos}, {Lind}, {Schlesinger}, {Sharma}, {Simpson},
  {Zucker}, \& {Zwitter}}]{Casagrande2021}
{Casagrande}, L., {Lin}, J., {Rains}, A.~D., {et~al.} 2021, \mnras, 507, 2684

\bibitem[{{Chiba} \& {Beers}(2000)}]{ChibaBeers2000}
{Chiba}, M. \& {Beers}, T.~C. 2000, \aj, 119, 2843

\bibitem[{{Choi} {et~al.}(2016){Choi}, {Dotter}, {Conroy}, {Cantiello},
  {Paxton}, \& {Johnson}}]{MIST2}
{Choi}, J., {Dotter}, A., {Conroy}, C., {et~al.} 2016, \apj, 823, 102

\bibitem[{{Chun} {et~al.}(2020){Chun}, {Lee}, \& {Lim}}]{Chun2020}
{Chun}, S.-H., {Lee}, J.-J., \& {Lim}, D. 2020, \apj, 900, 146

\bibitem[{{Cordoni} {et~al.}(2021){Cordoni}, {Da Costa}, {Yong}, {Mackey},
  {Marino}, {Monty}, {Nordlander}, {Norris}, {Asplund}, {Bessell}, {Casey},
  {Frebel}, {Lind}, {Murphy}, {Schmidt}, {Gao}, {Xylakis-Dornbusch}, {Amarsi},
  \& {Milone}}]{Cordoni2021}
{Cordoni}, G., {Da Costa}, G.~S., {Yong}, D., {et~al.} 2021, \mnras, 503, 2539

\bibitem[{{Das} {et~al.}(2020){Das}, {Hawkins}, \& {Jofr{\'e}}}]{Das2020}
{Das}, P., {Hawkins}, K., \& {Jofr{\'e}}, P. 2020, \mnras, 493, 5195

\bibitem[{{De Silva} {et~al.}(2015){De Silva}, {Freeman}, {Bland-Hawthorn},
  {Martell}, {de Boer}, {Asplund}, {Keller}, {Sharma}, {Zucker}, {Zwitter},
  {Anguiano}, {Bacigalupo}, {Bayliss}, {Beavis}, {Bergemann}, {Campbell},
  {Cannon}, {Carollo}, {Casagrande}, {Casey}, {Da Costa}, {D'Orazi}, {Dotter},
  {Duong}, {Heger}, {Ireland}, {Kafle}, {Kos}, {Lattanzio}, {Lewis}, {Lin},
  {Lind}, {Munari}, {Nataf}, {O'Toole}, {Parker}, {Reid}, {Schlesinger},
  {Sheinis}, {Simpson}, {Stello}, {Ting}, {Traven}, {Watson}, {Wittenmyer},
  {Yong}, \& {{\v{Z}}erjal}}]{deSilva2015}
{De Silva}, G.~M., {Freeman}, K.~C., {Bland-Hawthorn}, J., {et~al.} 2015,
  \mnras, 449, 2604

\bibitem[{{Deason} {et~al.}(2018){Deason}, {Belokurov}, {Koposov}, \&
  {Lancaster}}]{Deason2018}
{Deason}, A.~J., {Belokurov}, V., {Koposov}, S.~E., \& {Lancaster}, L. 2018,
  \apjl, 862, L1

\bibitem[{{Di Matteo} {et~al.}(2019){Di Matteo}, {Haywood}, {Lehnert}, {Katz},
  {Khoperskov}, {Snaith}, {G{\'o}mez}, \& {Robichon}}]{DiMatteo2019}
{Di Matteo}, P., {Haywood}, M., {Lehnert}, M.~D., {et~al.} 2019, \aap, 632, A4

\bibitem[{{Dodd} {et~al.}(2023){Dodd}, {Callingham}, {Helmi}, {Matsuno},
  {Ruiz-Lara}, {Balbinot}, \& {L{\"o}vdal}}]{Dodd2023}
{Dodd}, E., {Callingham}, T.~M., {Helmi}, A., {et~al.} 2023, \aap, 670, L2

\bibitem[{{Donlon} {et~al.}(2022){Donlon}, {Newberg}, {Kim}, \&
  {L{\'e}pine}}]{Donlon2022}
{Donlon}, Thomas, I., {Newberg}, H.~J., {Kim}, B., \& {L{\'e}pine}, S. 2022,
  \apjl, 932, L16

\bibitem[{{Donlon} {et~al.}(2020){Donlon}, {Newberg}, {Sanderson}, \&
  {Widrow}}]{Donlon2020}
{Donlon}, Thomas, I., {Newberg}, H.~J., {Sanderson}, R., \& {Widrow}, L.~M.
  2020, \apj, 902, 119

\bibitem[{{Dotter}(2016)}]{MIST1}
{Dotter}, A. 2016, \apjs, 222, 8

\bibitem[{{Eggen} {et~al.}(1962){Eggen}, {Lynden-Bell}, \&
  {Sandage}}]{Eggen1962}
{Eggen}, O.~J., {Lynden-Bell}, D., \& {Sandage}, A.~R. 1962, \apj, 136, 748

\bibitem[{{Fern{\'a}ndez-Alvar} {et~al.}(2018){Fern{\'a}ndez-Alvar}, {Carigi},
  {Schuster}, {Hayes}, {{\'A}vila-Vergara}, {Majewski}, {Allende Prieto},
  {Beers}, {S{\'a}nchez}, {Zamora}, {Garc{\'\i}a-Hern{\'a}ndez}, {Tang},
  {Fern{\'a}ndez-Trincado}, {Tissera}, {Geisler}, \&
  {Villanova}}]{FernandezAlvar2018}
{Fern{\'a}ndez-Alvar}, E., {Carigi}, L., {Schuster}, W.~J., {et~al.} 2018,
  \apj, 852, 50

\bibitem[{{Feuillet} {et~al.}(2020){Feuillet}, {Feltzing}, {Sahlholdt}, \&
  {Casagrande}}]{Feuillet2020}
{Feuillet}, D.~K., {Feltzing}, S., {Sahlholdt}, C.~L., \& {Casagrande}, L.
  2020, \mnras, 497, 109

\bibitem[{{Feuillet} {et~al.}(2021){Feuillet}, {Sahlholdt}, {Feltzing}, \&
  {Casagrande}}]{Feuillet2021}
{Feuillet}, D.~K., {Sahlholdt}, C.~L., {Feltzing}, S., \& {Casagrande}, L.
  2021, \mnras

\bibitem[{{Forbes}(2020)}]{Forbes2020}
{Forbes}, D.~A. 2020, \mnras, 493, 847

\bibitem[{{Freeman} \& {Bland-Hawthorn}(2002)}]{Ken2002}
{Freeman}, K. \& {Bland-Hawthorn}, J. 2002, \araa, 40, 487

\bibitem[{{Fuhrmann}(2008)}]{Fuhrmann2008}
{Fuhrmann}, K. 2008, \mnras, 384, 173

\bibitem[{{Gaia Collaboration} {et~al.}(2018{\natexlab{a}}){Gaia
  Collaboration}, {Babusiaux}, {van Leeuwen}, {Barstow}, {Jordi}, {Vallenari},
  {Bossini}, {Bressan}, {Cantat-Gaudin}, {van Leeuwen}, {Brown}, {Prusti}, {de
  Bruijne}, {Bailer-Jones}, {Biermann}, {Evans}, {Eyer}, {Jansen}, {Klioner},
  {Lammers}, {Lindegren}, {Luri}, {Mignard}, {Panem}, {Pourbaix}, {Randich},
  {Sartoretti}, {Siddiqui}, {Soubiran}, {Walton}, {Arenou}, {Bastian},
  {Cropper}, {Drimmel}, {Katz}, {Lattanzi}, {Bakker}, {Cacciari},
  {Casta{\~n}eda}, {Chaoul}, {Cheek}, {De Angeli}, {Fabricius}, {Guerra},
  {Holl}, {Masana}, {Messineo}, {Mowlavi}, {Nienartowicz}, {Panuzzo},
  {Portell}, {Riello}, {Seabroke}, {Tanga}, {Th{\'e}venin}, {Gracia-Abril},
  {Comoretto}, {Garcia-Reinaldos}, {Teyssier}, {Altmann}, {Andrae}, {Audard},
  {Bellas-Velidis}, {Benson}, {Berthier}, {Blomme}, {Burgess}, {Busso},
  {Carry}, {Cellino}, {Clementini}, {Clotet}, {Creevey}, {Davidson}, {De
  Ridder}, {Delchambre}, {Dell'Oro}, {Ducourant},
  {Fern{\'a}ndez-Hern{\'a}ndez}, {Fouesneau}, {Fr{\'e}mat}, {Galluccio},
  {Garc{\'\i}a-Torres}, {Gonz{\'a}lez-N{\'u}{\~n}ez}, {Gonz{\'a}lez-Vidal},
  {Gosset}, {Guy}, {Halbwachs}, {Hambly}, {Harrison}, {Hern{\'a}ndez},
  {Hestroffer}, {Hodgkin}, {Hutton}, {Jasniewicz}, {Jean-Antoine-Piccolo},
  {Jordan}, {Korn}, {Krone-Martins}, {Lanzafame}, {Lebzelter}, {L{\"o}ffler},
  {Manteiga}, {Marrese}, {Mart{\'\i}n-Fleitas}, {Moitinho}, {Mora}, {Muinonen},
  {Osinde}, {Pancino}, {Pauwels}, {Petit}, {Recio-Blanco}, {Richards},
  {Rimoldini}, {Robin}, {Sarro}, {Siopis}, {Smith}, {Sozzetti}, {S{\"u}veges},
  {Torra}, {van Reeven}, {Abbas}, {Abreu Aramburu}, {Accart}, {Aerts},
  {Altavilla}, {{\'A}lvarez}, {Alvarez}, {Alves}, {Anderson}, {Andrei},
  {Anglada Varela}, {Antiche}, {Antoja}, {Arcay}, {Astraatmadja}, {Bach},
  {Baker}, {Balaguer-N{\'u}{\~n}ez}, {Balm}, {Barache}, {Barata}, {Barbato},
  {Barblan}, {Barklem}, {Barrado}, {Barros}, {Bartholom{\'e} Mu{\~n}oz},
  {Bassilana}, {Becciani}, {Bellazzini}, {Berihuete}, {Bertone}, {Bianchi},
  {Bienaym{\'e}}, {Blanco-Cuaresma}, {Boch}, {Boeche}, {Bombrun}, {Borrachero},
  {Bouquillon}, {Bourda}, {Bragaglia}, {Bramante}, {Breddels}, {Brouillet},
  {Br{\"u}semeister}, {Brugaletta}, {Bucciarelli}, {Burlacu}, {Busonero},
  {Butkevich}, {Buzzi}, {Caffau}, {Cancelliere}, {Cannizzaro}, {Carballo},
  {Carlucci}, {Carrasco}, {Casamiquela}, {Castellani}, {Castro-Ginard},
  {Charlot}, {Chemin}, {Chiavassa}, {Cocozza}, {Costigan}, {Cowell}, {Crifo},
  {Crosta}, {Crowley}, {Cuypers}, {Dafonte}, {Damerdji}, {Dapergolas}, {David},
  {David}, {de Laverny}, {De Luise}, {De March}, {de Martino}, {de Souza}, {de
  Torres}, {Debosscher}, {del Pozo}, {Delbo}, {Delgado}, {Delgado}, {Diakite},
  {Diener}, {Distefano}, {Dolding}, {Drazinos}, {Dur{\'a}n}, {Edvardsson},
  {Enke}, {Eriksson}, {Esquej}, {Eynard Bontemps}, {Fabre}, {Fabrizio},
  {Faigler}, {Falc{\~a}o}, {Farr{\`a}s Casas}, {Federici}, {Fedorets},
  {Fernique}, {Figueras}, {Filippi}, {Findeisen}, {Fonti}, {Fraile}, {Fraser},
  {Fr{\'e}zouls}, {Gai}, {Galleti}, {Garabato}, {Garc{\'\i}a-Sedano},
  {Garofalo}, {Garralda}, {Gavel}, {Gavras}, {Gerssen}, {Geyer}, {Giacobbe},
  {Gilmore}, {Girona}, {Giuffrida}, {Glass}, {Gomes}, {Granvik}, {Gueguen},
  {Guerrier}, {Guiraud}, {Guti{\'e}}, {Haigron}, {Hatzidimitriou}, {Hauser},
  {Haywood}, {Heiter}, {Helmi}, {Heu}, {Hilger}, {Hobbs}, {Hofmann}, {Holland},
  {Huckle}, {Hypki}, {Icardi}, {Jan{\ss}en}, {Jevardat de Fombelle}, {Jonker},
  {Juh{\'a}sz}, {Julbe}, {Karampelas}, {Kewley}, {Klar}, {Kochoska}, {Kohley},
  {Kolenberg}, {Kontizas}, {Kontizas}, {Koposov}, {Kordopatis},
  {Kostrzewa-Rutkowska}, {Koubsky}, {Lambert}, {Lanza}, {Lasne}, {Lavigne}, {Le
  Fustec}, {Le Poncin-Lafitte}, {Lebreton}, {Leccia}, {Leclerc},
  {Lecoeur-Taibi}, {Lenhardt}, {Leroux}, {Liao}, {Licata}, {Lindstr{\o}m},
  {Lister}, {Livanou}, {Lobel}, {L{\'o}pez}, {Managau}, {Mann}, {Mantelet},
  {Marchal}, {Marchant}, {Marconi}, {Marinoni}, {Marschalk{\'o}}, {Marshall},
  {Martino}, {Marton}, {Mary}, {Massari}, {Matijevi{\v{c}}}, {Mazeh},
  {McMillan}, {Messina}, {Michalik}, {Millar}, {Molina}, {Molinaro},
  {Moln{\'a}r}, {Montegriffo}, {Mor}, {Morbidelli}, {Morel}, {Morris},
  {Mulone}, {Muraveva}, {Musella}, {Nelemans}, {Nicastro}, {Noval},
  {O'Mullane}, {Ord{\'e}novic}, {Ord{\'o}{\~n}ez-Blanco}, {Osborne}, {Pagani},
  {Pagano}, {Pailler}, {Palacin}, {Palaversa}, {Panahi}, {Pawlak},
  {Piersimoni}, {Pineau}, {Plachy}, {Plum}, {Poggio}, {Poujoulet},
  {Pr{\v{s}}a}, {Pulone}, {Racero}, {Ragaini}, {Rambaux}, {Ramos-Lerate},
  {Regibo}, {Reyl{\'e}}, {Riclet}, {Ripepi}, {Riva}, {Rivard}, {Rixon},
  {Roegiers}, {Roelens}, {Romero-G{\'o}mez}, {Rowell}, {Royer}, {Ruiz-Dern},
  {Sadowski}, {Sagrist{\`a} Sell{\'e}s}, {Sahlmann}, {Salgado}, {Salguero},
  {Sanna}, {Santana-Ros}, {Sarasso}, {Savietto}, {Schultheis}, {Sciacca},
  {Segol}, {Segovia}, {S{\'e}gransan}, {Shih}, {Siltala}, {Silva}, {Smart},
  {Smith}, {Solano}, {Solitro}, {Sordo}, {Soria Nieto}, {Souchay}, {Spagna},
  {Spoto}, {Stampa}, {Steele}, {Steidelm{\"u}ller}, {Stephenson}, {Stoev},
  {Suess}, {Surdej}, {Szabados}, {Szegedi-Elek}, {Tapiador}, {Taris}, {Tauran},
  {Taylor}, {Teixeira}, {Terrett}, {Teyssandier}, {Thuillot}, {Titarenko},
  {Torra Clotet}, {Turon}, {Ulla}, {Utrilla}, {Uzzi}, {Vaillant}, {Valentini},
  {Valette}, {van Elteren}, {Van Hemelryck}, {Vaschetto}, {Vecchiato},
  {Veljanoski}, {Viala}, {Vicente}, {Vogt}, {von Essen}, {Voss}, {Votruba},
  {Voutsinas}, {Walmsley}, {Weiler}, {Wertz}, {Wevers}, {Wyrzykowski},
  {Yoldas}, {{\v{Z}}erjal}, {Ziaeepour}, {Zorec}, {Zschocke}, {Zucker},
  {Zurbach}, \& {Zwitter}}]{GaiaCollab2018b}
{Gaia Collaboration}, {Babusiaux}, C., {van Leeuwen}, F., {et~al.}
  2018{\natexlab{a}}, \aap, 616, A10

\bibitem[{{Gaia Collaboration} {et~al.}(2018{\natexlab{b}}){Gaia
  Collaboration}, {Brown}, {Vallenari}, {Prusti}, {de Bruijne}, {Babusiaux},
  {Bailer-Jones}, {Biermann}, {Evans}, {Eyer}, {Jansen}, {Jordi}, {Klioner},
  {Lammers}, {Lindegren}, {Luri}, {Mignard}, {Panem}, {Pourbaix}, {Randich},
  {Sartoretti}, {Siddiqui}, {Soubiran}, {van Leeuwen}, {Walton}, {Arenou},
  {Bastian}, {Cropper}, {Drimmel}, {Katz}, {Lattanzi}, {Bakker}, {Cacciari},
  {Casta{\~n}eda}, {Chaoul}, {Cheek}, {De Angeli}, {Fabricius}, {Guerra},
  {Holl}, {Masana}, {Messineo}, {Mowlavi}, {Nienartowicz}, {Panuzzo},
  {Portell}, {Riello}, {Seabroke}, {Tanga}, {Th{\'e}venin}, {Gracia-Abril},
  {Comoretto}, {Garcia-Reinaldos}, {Teyssier}, {Altmann}, {Andrae}, {Audard},
  {Bellas-Velidis}, {Benson}, {Berthier}, {Blomme}, {Burgess}, {Busso},
  {Carry}, {Cellino}, {Clementini}, {Clotet}, {Creevey}, {Davidson}, {De
  Ridder}, {Delchambre}, {Dell'Oro}, {Ducourant},
  {Fern{\'a}ndez-Hern{\'a}ndez}, {Fouesneau}, {Fr{\'e}mat}, {Galluccio},
  {Garc{\'\i}a-Torres}, {Gonz{\'a}lez-N{\'u}{\~n}ez}, {Gonz{\'a}lez-Vidal},
  {Gosset}, {Guy}, {Halbwachs}, {Hambly}, {Harrison}, {Hern{\'a}ndez},
  {Hestroffer}, {Hodgkin}, {Hutton}, {Jasniewicz}, {Jean-Antoine-Piccolo},
  {Jordan}, {Korn}, {Krone-Martins}, {Lanzafame}, {Lebzelter}, {L{\"o}ffler},
  {Manteiga}, {Marrese}, {Mart{\'\i}n-Fleitas}, {Moitinho}, {Mora}, {Muinonen},
  {Osinde}, {Pancino}, {Pauwels}, {Petit}, {Recio-Blanco}, {Richards},
  {Rimoldini}, {Robin}, {Sarro}, {Siopis}, {Smith}, {Sozzetti}, {S{\"u}veges},
  {Torra}, {van Reeven}, {Abbas}, {Abreu Aramburu}, {Accart}, {Aerts},
  {Altavilla}, {{\'A}lvarez}, {Alvarez}, {Alves}, {Anderson}, {Andrei},
  {Anglada Varela}, {Antiche}, {Antoja}, {Arcay}, {Astraatmadja}, {Bach},
  {Baker}, {Balaguer-N{\'u}{\~n}ez}, {Balm}, {Barache}, {Barata}, {Barbato},
  {Barblan}, {Barklem}, {Barrado}, {Barros}, {Barstow}, {Bartholom{\'e}
  Mu{\~n}oz}, {Bassilana}, {Becciani}, {Bellazzini}, {Berihuete}, {Bertone},
  {Bianchi}, {Bienaym{\'e}}, {Blanco-Cuaresma}, {Boch}, {Boeche}, {Bombrun},
  {Borrachero}, {Bossini}, {Bouquillon}, {Bourda}, {Bragaglia}, {Bramante},
  {Breddels}, {Bressan}, {Brouillet}, {Br{\"u}semeister}, {Brugaletta},
  {Bucciarelli}, {Burlacu}, {Busonero}, {Butkevich}, {Buzzi}, {Caffau},
  {Cancelliere}, {Cannizzaro}, {Cantat-Gaudin}, {Carballo}, {Carlucci},
  {Carrasco}, {Casamiquela}, {Castellani}, {Castro-Ginard}, {Charlot},
  {Chemin}, {Chiavassa}, {Cocozza}, {Costigan}, {Cowell}, {Crifo}, {Crosta},
  {Crowley}, {Cuypers}, {Dafonte}, {Damerdji}, {Dapergolas}, {David}, {David},
  {de Laverny}, {De Luise}, {De March}, {de Martino}, {de Souza}, {de Torres},
  {Debosscher}, {del Pozo}, {Delbo}, {Delgado}, {Delgado}, {Di Matteo},
  {Diakite}, {Diener}, {Distefano}, {Dolding}, {Drazinos}, {Dur{\'a}n},
  {Edvardsson}, {Enke}, {Eriksson}, {Esquej}, {Eynard Bontemps}, {Fabre},
  {Fabrizio}, {Faigler}, {Falc{\~a}o}, {Farr{\`a}s Casas}, {Federici},
  {Fedorets}, {Fernique}, {Figueras}, {Filippi}, {Findeisen}, {Fonti},
  {Fraile}, {Fraser}, {Fr{\'e}zouls}, {Gai}, {Galleti}, {Garabato},
  {Garc{\'\i}a-Sedano}, {Garofalo}, {Garralda}, {Gavel}, {Gavras}, {Gerssen},
  {Geyer}, {Giacobbe}, {Gilmore}, {Girona}, {Giuffrida}, {Glass}, {Gomes},
  {Granvik}, {Gueguen}, {Guerrier}, {Guiraud}, {Guti{\'e}rrez-S{\'a}nchez},
  {Haigron}, {Hatzidimitriou}, {Hauser}, {Haywood}, {Heiter}, {Helmi}, {Heu},
  {Hilger}, {Hobbs}, {Hofmann}, {Holland}, {Huckle}, {Hypki}, {Icardi},
  {Jan{\ss}en}, {Jevardat de Fombelle}, {Jonker}, {Juh{\'a}sz}, {Julbe},
  {Karampelas}, {Kewley}, {Klar}, {Kochoska}, {Kohley}, {Kolenberg},
  {Kontizas}, {Kontizas}, {Koposov}, {Kordopatis}, {Kostrzewa-Rutkowska},
  {Koubsky}, {Lambert}, {Lanza}, {Lasne}, {Lavigne}, {Le Fustec}, {Le
  Poncin-Lafitte}, {Lebreton}, {Leccia}, {Leclerc}, {Lecoeur-Taibi},
  {Lenhardt}, {Leroux}, {Liao}, {Licata}, {Lindstr{\o}m}, {Lister}, {Livanou},
  {Lobel}, {L{\'o}pez}, {Managau}, {Mann}, {Mantelet}, {Marchal}, {Marchant},
  {Marconi}, {Marinoni}, {Marschalk{\'o}}, {Marshall}, {Martino}, {Marton},
  {Mary}, {Massari}, {Matijevi{\v{c}}}, {Mazeh}, {McMillan}, {Messina},
  {Michalik}, {Millar}, {Molina}, {Molinaro}, {Moln{\'a}r}, {Montegriffo},
  {Mor}, {Morbidelli}, {Morel}, {Morris}, {Mulone}, {Muraveva}, {Musella},
  {Nelemans}, {Nicastro}, {Noval}, {O'Mullane}, {Ord{\'e}novic},
  {Ord{\'o}{\~n}ez-Blanco}, {Osborne}, {Pagani}, {Pagano}, {Pailler},
  {Palacin}, {Palaversa}, {Panahi}, {Pawlak}, {Piersimoni}, {Pineau}, {Plachy},
  {Plum}, {Poggio}, {Poujoulet}, {Pr{\v{s}}a}, {Pulone}, {Racero}, {Ragaini},
  {Rambaux}, {Ramos-Lerate}, {Regibo}, {Reyl{\'e}}, {Riclet}, {Ripepi}, {Riva},
  {Rivard}, {Rixon}, {Roegiers}, {Roelens}, {Romero-G{\'o}mez}, {Rowell},
  {Royer}, {Ruiz-Dern}, {Sadowski}, {Sagrist{\`a} Sell{\'e}s}, {Sahlmann},
  {Salgado}, {Salguero}, {Sanna}, {Santana-Ros}, {Sarasso}, {Savietto},
  {Schultheis}, {Sciacca}, {Segol}, {Segovia}, {S{\'e}gransan}, {Shih},
  {Siltala}, {Silva}, {Smart}, {Smith}, {Solano}, {Solitro}, {Sordo}, {Soria
  Nieto}, {Souchay}, {Spagna}, {Spoto}, {Stampa}, {Steele},
  {Steidelm{\"u}ller}, {Stephenson}, {Stoev}, {Suess}, {Surdej}, {Szabados},
  {Szegedi-Elek}, {Tapiador}, {Taris}, {Tauran}, {Taylor}, {Teixeira},
  {Terrett}, {Teyssandier}, {Thuillot}, {Titarenko}, {Torra Clotet}, {Turon},
  {Ulla}, {Utrilla}, {Uzzi}, {Vaillant}, {Valentini}, {Valette}, {van Elteren},
  {Van Hemelryck}, {van Leeuwen}, {Vaschetto}, {Vecchiato}, {Veljanoski},
  {Viala}, {Vicente}, {Vogt}, {von Essen}, {Voss}, {Votruba}, {Voutsinas},
  {Walmsley}, {Weiler}, {Wertz}, {Wevers}, {Wyrzykowski}, {Yoldas},
  {{\v{Z}}erjal}, {Ziaeepour}, {Zorec}, {Zschocke}, {Zucker}, {Zurbach}, \&
  {Zwitter}}]{GaiaCollab2018a}
{Gaia Collaboration}, {Brown}, A.~G.~A., {Vallenari}, A., {et~al.}
  2018{\natexlab{b}}, \aap, 616, A1

\bibitem[{{Gaia Collaboration} {et~al.}(2021){Gaia Collaboration}, {Brown},
  {Vallenari}, {Prusti}, {de Bruijne}, {Babusiaux}, {Biermann}, {Creevey},
  {Evans}, {Eyer}, {Hutton}, {Jansen}, {Jordi}, {Klioner}, {Lammers},
  {Lindegren}, {Luri}, {Mignard}, {Panem}, {Pourbaix}, {Randich}, {Sartoretti},
  {Soubiran}, {Walton}, {Arenou}, {Bailer-Jones}, {Bastian}, {Cropper},
  {Drimmel}, {Katz}, {Lattanzi}, {van Leeuwen}, {Bakker}, {Cacciari},
  {Casta{\~n}eda}, {De Angeli}, {Ducourant}, {Fabricius}, {Fouesneau},
  {Fr{\'e}mat}, {Guerra}, {Guerrier}, {Guiraud}, {Jean-Antoine Piccolo},
  {Masana}, {Messineo}, {Mowlavi}, {Nicolas}, {Nienartowicz}, {Pailler},
  {Panuzzo}, {Riclet}, {Roux}, {Seabroke}, {Sordo}, {Tanga}, {Th{\'e}venin},
  {Gracia-Abril}, {Portell}, {Teyssier}, {Altmann}, {Andrae}, {Bellas-Velidis},
  {Benson}, {Berthier}, {Blomme}, {Brugaletta}, {Burgess}, {Busso}, {Carry},
  {Cellino}, {Cheek}, {Clementini}, {Damerdji}, {Davidson}, {Delchambre},
  {Dell'Oro}, {Fern{\'a}ndez-Hern{\'a}ndez}, {Galluccio}, {Garc{\'\i}a-Lario},
  {Garcia-Reinaldos}, {Gonz{\'a}lez-N{\'u}{\~n}ez}, {Gosset}, {Haigron},
  {Halbwachs}, {Hambly}, {Harrison}, {Hatzidimitriou}, {Heiter},
  {Hern{\'a}ndez}, {Hestroffer}, {Hodgkin}, {Holl}, {Jan{\ss}en}, {Jevardat de
  Fombelle}, {Jordan}, {Krone-Martins}, {Lanzafame}, {L{\"o}ffler}, {Lorca},
  {Manteiga}, {Marchal}, {Marrese}, {Moitinho}, {Mora}, {Muinonen}, {Osborne},
  {Pancino}, {Pauwels}, {Petit}, {Recio-Blanco}, {Richards}, {Riello},
  {Rimoldini}, {Robin}, {Roegiers}, {Rybizki}, {Sarro}, {Siopis}, {Smith},
  {Sozzetti}, {Ulla}, {Utrilla}, {van Leeuwen}, {van Reeven}, {Abbas}, {Abreu
  Aramburu}, {Accart}, {Aerts}, {Aguado}, {Ajaj}, {Altavilla}, {{\'A}lvarez},
  {{\'A}lvarez Cid-Fuentes}, {Alves}, {Anderson}, {Anglada Varela}, {Antoja},
  {Audard}, {Baines}, {Baker}, {Balaguer-N{\'u}{\~n}ez}, {Balbinot}, {Balog},
  {Barache}, {Barbato}, {Barros}, {Barstow}, {Bartolom{\'e}}, {Bassilana},
  {Bauchet}, {Baudesson-Stella}, {Becciani}, {Bellazzini}, {Bernet}, {Bertone},
  {Bianchi}, {Blanco-Cuaresma}, {Boch}, {Bombrun}, {Bossini}, {Bouquillon},
  {Bragaglia}, {Bramante}, {Breedt}, {Bressan}, {Brouillet}, {Bucciarelli},
  {Burlacu}, {Busonero}, {Butkevich}, {Buzzi}, {Caffau}, {Cancelliere},
  {C{\'a}novas}, {Cantat-Gaudin}, {Carballo}, {Carlucci}, {Carnerero},
  {Carrasco}, {Casamiquela}, {Castellani}, {Castro-Ginard}, {Castro Sampol},
  {Chaoul}, {Charlot}, {Chemin}, {Chiavassa}, {Cioni}, {Comoretto}, {Cooper},
  {Cornez}, {Cowell}, {Crifo}, {Crosta}, {Crowley}, {Dafonte}, {Dapergolas},
  {David}, {David}, {de Laverny}, {De Luise}, {De March}, {De Ridder}, {de
  Souza}, {de Teodoro}, {de Torres}, {del Peloso}, {del Pozo}, {Delbo},
  {Delgado}, {Delgado}, {Delisle}, {Di Matteo}, {Diakite}, {Diener},
  {Distefano}, {Dolding}, {Eappachen}, {Edvardsson}, {Enke}, {Esquej}, {Fabre},
  {Fabrizio}, {Faigler}, {Fedorets}, {Fernique}, {Fienga}, {Figueras},
  {Fouron}, {Fragkoudi}, {Fraile}, {Franke}, {Gai}, {Garabato},
  {Garcia-Gutierrez}, {Garc{\'\i}a-Torres}, {Garofalo}, {Gavras}, {Gerlach},
  {Geyer}, {Giacobbe}, {Gilmore}, {Girona}, {Giuffrida}, {Gomel}, {Gomez},
  {Gonzalez-Santamaria}, {Gonz{\'a}lez-Vidal}, {Granvik},
  {Guti{\'e}rrez-S{\'a}nchez}, {Guy}, {Hauser}, {Haywood}, {Helmi}, {Hidalgo},
  {Hilger}, {H{\l}adczuk}, {Hobbs}, {Holland}, {Huckle}, {Jasniewicz},
  {Jonker}, {Juaristi Campillo}, {Julbe}, {Karbevska}, {Kervella}, {Khanna},
  {Kochoska}, {Kontizas}, {Kordopatis}, {Korn}, {Kostrzewa-Rutkowska},
  {Kruszy{\'n}ska}, {Lambert}, {Lanza}, {Lasne}, {Le Campion}, {Le Fustec},
  {Lebreton}, {Lebzelter}, {Leccia}, {Leclerc}, {Lecoeur-Taibi}, {Liao},
  {Licata}, {Lindstr{\o}m}, {Lister}, {Livanou}, {Lobel}, {Madrero Pardo},
  {Managau}, {Mann}, {Marchant}, {Marconi}, {Marcos Santos}, {Marinoni},
  {Marocco}, {Marshall}, {Martin Polo}, {Mart{\'\i}n-Fleitas}, {Masip},
  {Massari}, {Mastrobuono-Battisti}, {Mazeh}, {McMillan}, {Messina},
  {Michalik}, {Millar}, {Mints}, {Molina}, {Molinaro}, {Moln{\'a}r},
  {Montegriffo}, {Mor}, {Morbidelli}, {Morel}, {Morris}, {Mulone}, {Munoz},
  {Muraveva}, {Murphy}, {Musella}, {Noval}, {Ord{\'e}novic}, {Orr{\`u}},
  {Osinde}, {Pagani}, {Pagano}, {Palaversa}, {Palicio}, {Panahi}, {Pawlak},
  {Pe{\~n}alosa Esteller}, {Penttil{\"a}}, {Piersimoni}, {Pineau}, {Plachy},
  {Plum}, {Poggio}, {Poretti}, {Poujoulet}, {Pr{\v{s}}a}, {Pulone}, {Racero},
  {Ragaini}, {Rainer}, {Raiteri}, {Rambaux}, {Ramos}, {Ramos-Lerate}, {Re
  Fiorentin}, {Regibo}, {Reyl{\'e}}, {Ripepi}, {Riva}, {Rixon}, {Robichon},
  {Robin}, {Roelens}, {Rohrbasser}, {Romero-G{\'o}mez}, {Rowell}, {Royer},
  {Rybicki}, {Sadowski}, {Sagrist{\`a} Sell{\'e}s}, {Sahlmann}, {Salgado},
  {Salguero}, {Samaras}, {Sanchez Gimenez}, {Sanna}, {Santove{\~n}a},
  {Sarasso}, {Schultheis}, {Sciacca}, {Segol}, {Segovia}, {S{\'e}gransan},
  {Semeux}, {Shahaf}, {Siddiqui}, {Siebert}, {Siltala}, {Slezak}, {Smart},
  {Solano}, {Solitro}, {Souami}, {Souchay}, {Spagna}, {Spoto}, {Steele},
  {Steidelm{\"u}ller}, {Stephenson}, {S{\"u}veges}, {Szabados}, {Szegedi-Elek},
  {Taris}, {Tauran}, {Taylor}, {Teixeira}, {Thuillot}, {Tonello}, {Torra},
  {Torra}, {Turon}, {Unger}, {Vaillant}, {van Dillen}, {Vanel}, {Vecchiato},
  {Viala}, {Vicente}, {Voutsinas}, {Weiler}, {Wevers}, {Wyrzykowski}, {Yoldas},
  {Yvard}, {Zhao}, {Zorec}, {Zucker}, {Zurbach}, \& {Zwitter}}]{GaiaEDR3}
{Gaia Collaboration}, {Brown}, A.~G.~A., {Vallenari}, A., {et~al.} 2021, \aap,
  649, A1

\bibitem[{{Gaia Collaboration} {et~al.}(2016){Gaia Collaboration}, {Prusti},
  {de Bruijne}, {Brown}, {Vallenari}, {Babusiaux}, {Bailer-Jones}, {Bastian},
  {Biermann}, {Evans}, {Eyer}, {Jansen}, {Jordi}, {Klioner}, {Lammers},
  {Lindegren}, {Luri}, {Mignard}, {Milligan}, {Panem}, {Poinsignon},
  {Pourbaix}, {Randich}, {Sarri}, {Sartoretti}, {Siddiqui}, {Soubiran},
  {Valette}, {van Leeuwen}, {Walton}, {Aerts}, {Arenou}, {Cropper}, {Drimmel},
  {H{\o}g}, {Katz}, {Lattanzi}, {O'Mullane}, {Grebel}, {Holland}, {Huc},
  {Passot}, {Bramante}, {Cacciari}, {Casta{\~n}eda}, {Chaoul}, {Cheek}, {De
  Angeli}, {Fabricius}, {Guerra}, {Hern{\'a}ndez}, {Jean-Antoine-Piccolo},
  {Masana}, {Messineo}, {Mowlavi}, {Nienartowicz}, {Ord{\'o}{\~n}ez-Blanco},
  {Panuzzo}, {Portell}, {Richards}, {Riello}, {Seabroke}, {Tanga},
  {Th{\'e}venin}, {Torra}, {Els}, {Gracia-Abril}, {Comoretto},
  {Garcia-Reinaldos}, {Lock}, {Mercier}, {Altmann}, {Andrae}, {Astraatmadja},
  {Bellas-Velidis}, {Benson}, {Berthier}, {Blomme}, {Busso}, {Carry},
  {Cellino}, {Clementini}, {Cowell}, {Creevey}, {Cuypers}, {Davidson}, {De
  Ridder}, {de Torres}, {Delchambre}, {Dell'Oro}, {Ducourant}, {Fr{\'e}mat},
  {Garc{\'\i}a-Torres}, {Gosset}, {Halbwachs}, {Hambly}, {Harrison}, {Hauser},
  {Hestroffer}, {Hodgkin}, {Huckle}, {Hutton}, {Jasniewicz}, {Jordan},
  {Kontizas}, {Korn}, {Lanzafame}, {Manteiga}, {Moitinho}, {Muinonen},
  {Osinde}, {Pancino}, {Pauwels}, {Petit}, {Recio-Blanco}, {Robin}, {Sarro},
  {Siopis}, {Smith}, {Smith}, {Sozzetti}, {Thuillot}, {van Reeven}, {Viala},
  {Abbas}, {Abreu Aramburu}, {Accart}, {Aguado}, {Allan}, {Allasia},
  {Altavilla}, {{\'A}lvarez}, {Alves}, {Anderson}, {Andrei}, {Anglada Varela},
  {Antiche}, {Antoja}, {Ant{\'o}n}, {Arcay}, {Atzei}, {Ayache}, {Bach},
  {Baker}, {Balaguer-N{\'u}{\~n}ez}, {Barache}, {Barata}, {Barbier}, {Barblan},
  {Baroni}, {Barrado y Navascu{\'e}s}, {Barros}, {Barstow}, {Becciani},
  {Bellazzini}, {Bellei}, {Bello Garc{\'\i}a}, {Belokurov}, {Bendjoya},
  {Berihuete}, {Bianchi}, {Bienaym{\'e}}, {Billebaud}, {Blagorodnova},
  {Blanco-Cuaresma}, {Boch}, {Bombrun}, {Borrachero}, {Bouquillon}, {Bourda},
  {Bouy}, {Bragaglia}, {Breddels}, {Brouillet}, {Br{\"u}semeister},
  {Bucciarelli}, {Budnik}, {Burgess}, {Burgon}, {Burlacu}, {Busonero}, {Buzzi},
  {Caffau}, {Cambras}, {Campbell}, {Cancelliere}, {Cantat-Gaudin}, {Carlucci},
  {Carrasco}, {Castellani}, {Charlot}, {Charnas}, {Charvet}, {Chassat},
  {Chiavassa}, {Clotet}, {Cocozza}, {Collins}, {Collins}, {Costigan}, {Crifo},
  {Cross}, {Crosta}, {Crowley}, {Dafonte}, {Damerdji}, {Dapergolas}, {David},
  {David}, {De Cat}, {de Felice}, {de Laverny}, {De Luise}, {De March}, {de
  Martino}, {de Souza}, {Debosscher}, {del Pozo}, {Delbo}, {Delgado},
  {Delgado}, {di Marco}, {Di Matteo}, {Diakite}, {Distefano}, {Dolding}, {Dos
  Anjos}, {Drazinos}, {Dur{\'a}n}, {Dzigan}, {Ecale}, {Edvardsson}, {Enke},
  {Erdmann}, {Escolar}, {Espina}, {Evans}, {Eynard Bontemps}, {Fabre},
  {Fabrizio}, {Faigler}, {Falc{\~a}o}, {Farr{\`a}s Casas}, {Faye}, {Federici},
  {Fedorets}, {Fern{\'a}ndez-Hern{\'a}ndez}, {Fernique}, {Fienga}, {Figueras},
  {Filippi}, {Findeisen}, {Fonti}, {Fouesneau}, {Fraile}, {Fraser}, {Fuchs},
  {Furnell}, {Gai}, {Galleti}, {Galluccio}, {Garabato}, {Garc{\'\i}a-Sedano},
  {Gar{\'e}}, {Garofalo}, {Garralda}, {Gavras}, {Gerssen}, {Geyer}, {Gilmore},
  {Girona}, {Giuffrida}, {Gomes}, {Gonz{\'a}lez-Marcos},
  {Gonz{\'a}lez-N{\'u}{\~n}ez}, {Gonz{\'a}lez-Vidal}, {Granvik}, {Guerrier},
  {Guillout}, {Guiraud}, {G{\'u}rpide}, {Guti{\'e}rrez-S{\'a}nchez}, {Guy},
  {Haigron}, {Hatzidimitriou}, {Haywood}, {Heiter}, {Helmi}, {Hobbs},
  {Hofmann}, {Holl}, {Holland }, {Hunt}, {Hypki}, {Icardi}, {Irwin}, {Jevardat
  de Fombelle}, {Jofr{\'e}}, {Jonker}, {Jorissen}, {Julbe}, {Karampelas},
  {Kochoska}, {Kohley}, {Kolenberg}, {Kontizas}, {Koposov}, {Kordopatis},
  {Koubsky}, {Kowalczyk}, {Krone-Martins}, {Kudryashova}, {Kull}, {Bachchan},
  {Lacoste-Seris}, {Lanza}, {Lavigne}, {Le Poncin-Lafitte}, {Lebreton},
  {Lebzelter}, {Leccia}, {Leclerc}, {Lecoeur-Taibi}, {Lemaitre}, {Lenhardt},
  {Leroux}, {Liao}, {Licata}, {Lindstr{\o}m}, {Lister}, {Livanou}, {Lobel},
  {L{\"o}ffler}, {L{\'o}pez}, {Lopez-Lozano}, {Lorenz}, {Loureiro},
  {MacDonald}, {Magalh{\~a}es Fernandes}, {Managau}, {Mann}, {Mantelet},
  {Marchal}, {Marchant}, {Marconi}, {Marie}, {Marinoni}, {Marrese},
  {Marschalk{\'o}}, {Marshall}, {Mart{\'\i}n-Fleitas}, {Martino}, {Mary},
  {Matijevi{\v{c}}}, {Mazeh}, {McMillan}, {Messina}, {Mestre}, {Michalik},
  {Millar}, {Miranda}, {Molina}, {Molinaro}, {Molinaro}, {Moln{\'a}r},
  {Moniez}, {Montegriffo}, {Monteiro}, {Mor}, {Mora}, {Morbidelli}, {Morel},
  {Morgenthaler}, {Morley}, {Morris}, {Mulone}, {Muraveva}, {Musella},
  {Narbonne}, {Nelemans}, {Nicastro}, {Noval}, {Ord{\'e}novic},
  {Ordieres-Mer{\'e}}, {Osborne}, {Pagani}, {Pagano}, {Pailler}, {Palacin},
  {Palaversa}, {Parsons}, {Paulsen}, {Pecoraro}, {Pedrosa}, {Pentik{\"a}inen},
  {Pereira}, {Pichon}, {Piersimoni}, {Pineau}, {Plachy}, {Plum}, {Poujoulet},
  {Pr{\v{s}}a}, {Pulone}, {Ragaini}, {Rago}, {Rambaux}, {Ramos-Lerate},
  {Ranalli}, {Rauw}, {Read}, {Regibo}, {Renk}, {Reyl{\'e}}, {Ribeiro},
  {Rimoldini}, {Ripepi}, {Riva}, {Rixon}, {Roelens}, {Romero-G{\'o}mez},
  {Rowell}, {Royer}, {Rudolph}, {Ruiz-Dern}, {Sadowski}, {Sagrist{\`a}
  Sell{\'e}s}, {Sahlmann}, {Salgado}, {Salguero}, {Sarasso}, {Savietto},
  {Schnorhk}, {Schultheis}, {Sciacca}, {Segol}, {Segovia}, {Segransan},
  {Serpell}, {Shih}, {Smareglia}, {Smart}, {Smith}, {Solano}, {Solitro},
  {Sordo}, {Soria Nieto}, {Souchay}, {Spagna}, {Spoto}, {Stampa}, {Steele},
  {Steidelm{\"u}ller}, {Stephenson}, {Stoev}, {Suess}, {S{\"u}veges}, {Surdej},
  {Szabados}, {Szegedi-Elek}, {Tapiador}, {Taris}, {Tauran}, {Taylor},
  {Teixeira}, {Terrett}, {Tingley}, {Trager}, {Turon}, {Ulla}, {Utrilla},
  {Valentini}, {van Elteren}, {Van Hemelryck}, {van Leeuwen}, {Varadi},
  {Vecchiato}, {Veljanoski}, {Via}, {Vicente}, {Vogt}, {Voss}, {Votruba},
  {Voutsinas}, {Walmsley}, {Weiler}, {Weingrill}, {Werner}, {Wevers},
  {Whitehead}, {Wyrzykowski}, {Yoldas}, {{\v{Z}}erjal}, {Zucker}, {Zurbach},
  {Zwitter}, {Alecu}, {Allen}, {Allende Prieto}, {Amorim},
  {Anglada-Escud{\'e}}, {Arsenijevic}, {Azaz}, {Balm}, {Beck}, {Bernstein},
  {Bigot}, {Bijaoui}, {Blasco}, {Bonfigli}, {Bono}, {Boudreault}, {Bressan},
  {Brown}, {Brunet}, {Bunclark}, {Buonanno}, {Butkevich}, {Carret}, {Carrion},
  {Chemin}, {Ch{\'e}reau}, {Corcione}, {Darmigny}, {de Boer}, {de Teodoro}, {de
  Zeeuw}, {Delle Luche}, {Domingues}, {Dubath}, {Fodor}, {Fr{\'e}zouls},
  {Fries}, {Fustes}, {Fyfe}, {Gallardo}, {Gallegos}, {Gardiol}, {Gebran},
  {Gomboc}, {G{\'o}mez}, {Grux}, {Gueguen}, {Heyrovsky}, {Hoar}, {Iannicola},
  {Isasi Parache}, {Janotto}, {Joliet}, {Jonckheere}, {Keil}, {Kim},
  {Klagyivik}, {Klar}, {Knude}, {Kochukhov}, {Kolka}, {Kos}, {Kutka}, {Lainey},
  {LeBouquin}, {Liu}, {Loreggia}, {Makarov}, {Marseille}, {Martayan},
  {Martinez-Rubi}, {Massart}, {Meynadier}, {Mignot}, {Munari}, {Nguyen},
  {Nordlander}, {Ocvirk}, {O'Flaherty}, {Olias Sanz}, {Ortiz}, {Osorio},
  {Oszkiewicz}, {Ouzounis}, {Palmer}, {Park}, {Pasquato}, {Peltzer}, {Peralta},
  {P{\'e}turaud}, {Pieniluoma}, {Pigozzi}, {Poels}, {Prat}, {Prod'homme},
  {Raison}, {Rebordao}, {Risquez}, {Rocca-Volmerange}, {Rosen}, {Ruiz-Fuertes},
  {Russo}, {Sembay}, {Serraller Vizcaino}, {Short}, {Siebert}, {Silva},
  {Sinachopoulos}, {Slezak}, {Soffel}, {Sosnowska}, {Strai{\v{z}}ys}, {ter
  Linden}, {Terrell}, {Theil}, {Tiede}, {Troisi}, {Tsalmantza}, {Tur},
  {Vaccari}, {Vachier}, {Valles}, {Van Hamme}, {Veltz}, {Virtanen}, {Wallut},
  {Wichmann}, {Wilkinson}, {Ziaeepour}, \& {Zschocke}}]{Gaia}
{Gaia Collaboration}, {Prusti}, T., {de Bruijne}, J.~H.~J., {et~al.} 2016,
  \aap, 595, A1

\bibitem[{{Gallart} {et~al.}(2019){Gallart}, {Bernard}, {Brook}, {Ruiz-Lara},
  {Cassisi}, {Hill}, \& {Monelli}}]{Gallart2019}
{Gallart}, C., {Bernard}, E.~J., {Brook}, C.~B., {et~al.} 2019, Nature
  Astronomy, 3, 932

\bibitem[{{Gilmore} {et~al.}(2022){Gilmore}, {Randich}, {Worley}, {Hourihane},
  {Gonneau}, {Sacco}, {Lewis}, {Magrini}, {Fran{\c{c}}ois}, {Jeffries},
  {Koposov}, {Bragaglia}, {Alfaro}, {Allende Prieto}, {Blomme}, {Korn},
  {Lanzafame}, {Pancino}, {Recio-Blanco}, {Smiljanic}, {Van Eck}, {Zwitter},
  {Bensby}, {Flaccomio}, {Irwin}, {Franciosini}, {Morbidelli}, {Damiani},
  {Bonito}, {Friel}, {Vink}, {Prisinzano}, {Abbas}, {Hatzidimitriou}, {Held},
  {Jordi}, {Paunzen}, {Spagna}, {Jackson}, {Ma{\'\i}z Apell{\'a}niz},
  {Asplund}, {Bonifacio}, {Feltzing}, {Binney}, {Drew}, {Ferguson}, {Micela},
  {Negueruela}, {Prusti}, {Rix}, {Vallenari}, {Bergemann}, {Casey}, {de
  Laverny}, {Frasca}, {Hill}, {Lind}, {Sbordone}, {Sousa}, {Adibekyan},
  {Caffau}, {Daflon}, {Feuillet}, {Gebran}, {Gonzalez Hernandez}, {Guiglion},
  {Herrero}, {Lobel}, {Merle}, {Mikolaitis}, {Montes}, {Morel}, {Ruchti},
  {Soubiran}, {Tabernero}, {Tautvai{\v{s}}ien{\.{e}}}, {Traven}, {Valentini},
  {Van der Swaelmen}, {Villanova}, {Viscasillas V{\'a}zquez}, {Bayo}, {Biazzo},
  {Carraro}, {Edvardsson}, {Heiter}, {Jofr{\'e}}, {Marconi}, {Martayan},
  {Masseron}, {Monaco}, {Walton}, {Zaggia}, {Aguirre B{\o}rsen-Koch}, {Alves},
  {Balaguer-Nunez}, {Barklem}, {Barrado}, {Bellazzini}, {Berlanas}, {Binks},
  {Bressan}, {Capuzzo-Dolcetta}, {Casagrande}, {Casamiquela}, {Collins},
  {D'Orazi}, {Dantas}, {Debattista}, {Delgado-Mena}, {Di Marcantonio},
  {Drazdauskas}, {Evans}, {Famaey}, {Franchini}, {Fr{\'e}mat}, {Fu}, {Geisler},
  {Gerhard}, {Gonz{\'a}lez Solares}, {Grebel}, {Guti{\'e}rrez Albarr{\'a}n},
  {Jim{\'e}nez-Esteban}, {J{\"o}nsson}, {Khachaturyants}, {Kordopatis}, {Kos},
  {Lagarde}, {Ludwig}, {Mahy}, {Mapelli}, {Marfil}, {Martell}, {Messina},
  {Miglio}, {Minchev}, {Moitinho}, {Montalban}, {Monteiro}, {Morossi},
  {Mowlavi}, {Mucciarelli}, {Murphy}, {Nardetto}, {Ortolani}, {Paletou},
  {Palou{\v{s}}}, {Pickering}, {Quirrenbach}, {Re Fiorentin}, {Read}, {Romano},
  {Ryde}, {Sanna}, {Santos}, {Seabroke}, {Spina}, {Steinmetz}, {Stonkut{\'e}},
  {Sutorius}, {Th{\'e}venin}, {Tosi}, {Tsantaki}, {Wright}, {Wyse}, {Zoccali},
  {Zorec}, \& {Zucker}}]{Gilmore2022}
{Gilmore}, G., {Randich}, S., {Worley}, C.~C., {et~al.} 2022, \aap, 666, A120

\bibitem[{{Gilmore} {et~al.}(2002){Gilmore}, {Wyse}, \& {Norris}}]{Gilmore2002}
{Gilmore}, G., {Wyse}, R. F.~G., \& {Norris}, J.~E. 2002, \apjl, 574, L39

\bibitem[{{Giribaldi} {et~al.}(2021){Giribaldi}, {da Silva}, {Smiljanic}, \&
  {Cornejo Espinoza}}]{Giribaldi2021}
{Giribaldi}, R.~E., {da Silva}, A.~R., {Smiljanic}, R., \& {Cornejo Espinoza},
  D. 2021, \aap, 650, A194

\bibitem[{{Giribaldi} \& {Smiljanic}(2023)}]{GiribaldiSmiljanic23}
{Giribaldi}, R.~E. \& {Smiljanic}, R. 2023, A\&A, 673, A18

\bibitem[{{Grand} {et~al.}(2020){Grand}, {Kawata}, {Belokurov}, {Deason},
  {Fattahi}, {Fragkoudi}, {G{\'o}mez}, {Marinacci}, \& {Pakmor}}]{Grand2020}
{Grand}, R. J.~J., {Kawata}, D., {Belokurov}, V., {et~al.} 2020, \mnras, 497,
  1603

\bibitem[{{Gratton} {et~al.}(2003){Gratton}, {Carretta}, {Desidera},
  {Lucatello}, {Mazzei}, \& {Barbieri}}]{Gratton2003}
{Gratton}, R.~G., {Carretta}, E., {Desidera}, S., {et~al.} 2003, \aap, 406, 131

\bibitem[{{Hasselquist} {et~al.}(2021){Hasselquist}, {Hayes}, {Lian},
  {Weinberg}, {Zasowski}, {Horta}, {Beaton}, {Feuillet}, {Garro}, {Gallart},
  {Smith}, {Holtzman}, {Minniti}, {Lacerna}, {Shetrone}, {J{\"o}nsson},
  {Cioni}, {Fillingham}, {Cunha}, {O'Connell}, {Fern{\'a}ndez-Trincado},
  {Mu{\~n}oz}, {Schiavon}, {Almeida}, {Anguiano}, {Beers}, {Bizyaev},
  {Brownstein}, {Cohen}, {Frinchaboy}, {Garc{\'\i}a-Hern{\'a}ndez}, {Geisler},
  {Lane}, {Majewski}, {Nidever}, {Nitschelm}, {Povick}, {Price-Whelan},
  {Roman-Lopes}, {Rosado}, {Sobeck}, {Stringfellow}, {Valenzuela}, {Villanova},
  \& {Vincenzo}}]{Hasselquist2021}
{Hasselquist}, S., {Hayes}, C.~R., {Lian}, J., {et~al.} 2021, \apj, 923, 172

\bibitem[{{Hawkins} {et~al.}(2015){Hawkins}, {Jofr{\'e}}, {Masseron}, \&
  {Gilmore}}]{Hawkins2015}
{Hawkins}, K., {Jofr{\'e}}, P., {Masseron}, T., \& {Gilmore}, G. 2015, \mnras,
  453, 758

\bibitem[{{Hawkins} {et~al.}(2021){Hawkins}, {Zeimann}, {Sneden}, {Cooper},
  {Gebhardt}, {Bond}, {Carrillo}, {Casey}, {Castanheira}, {Ciardullo}, {Davis},
  {Farrow}, {Finkelstein}, {Hill}, {Kelz}, {Liu}, {Shetrone}, {Schneider},
  {Starkenburg}, {Steinmetz}, {Wheeler}, \& {Hetdex
  Collaboration}}]{Hawkins2021}
{Hawkins}, K., {Zeimann}, G., {Sneden}, C., {et~al.} 2021, \apj, 911, 108

\bibitem[{{Hayes} {et~al.}(2018){Hayes}, {Majewski}, {Shetrone},
  {Fern{\'a}ndez-Alvar}, {Allende Prieto}, {Schuster}, {Carigi}, {Cunha},
  {Smith}, {Sobeck}, {Almeida}, {Beers}, {Carrera}, {Fern{\'a}ndez-Trincado},
  {Garc{\'\i}a-Hern{\'a}ndez}, {Geisler}, {Lane}, {Lucatello}, {Matthews},
  {Minniti}, {Nitschelm}, {Tang}, {Tissera}, \& {Zamora}}]{Hayes2018}
{Hayes}, C.~R., {Majewski}, S.~R., {Shetrone}, M., {et~al.} 2018, \apj, 852, 49

\bibitem[{{Haywood} {et~al.}(2018){Haywood}, {Di Matteo}, {Lehnert}, {Snaith},
  {Khoperskov}, \& {G{\'o}mez}}]{Haywood2018}
{Haywood}, M., {Di Matteo}, P., {Lehnert}, M.~D., {et~al.} 2018, \apj, 863, 113

\bibitem[{{Helmi}(2008)}]{Helmi2008}
{Helmi}, A. 2008, \aapr, 15, 145

\bibitem[{{Helmi}(2020)}]{Helmi2020}
{Helmi}, A. 2020, \araa, 58, 205

\bibitem[{{Helmi} {et~al.}(2018){Helmi}, {Babusiaux}, {Koppelman}, {Massari},
  {Veljanoski}, \& {Brown}}]{Helmi2018}
{Helmi}, A., {Babusiaux}, C., {Koppelman}, H.~H., {et~al.} 2018, \nat, 563, 85

\bibitem[{{Helmi} {et~al.}(1999){Helmi}, {White}, {de Zeeuw}, \&
  {Zhao}}]{Helmi1999b}
{Helmi}, A., {White}, S. D.~M., {de Zeeuw}, P.~T., \& {Zhao}, H. 1999, \nat,
  402, 53

\bibitem[{{Horta} {et~al.}(2021){Horta}, {Schiavon}, {Mackereth}, {Pfeffer},
  {Mason}, {Kisku}, {Fragkoudi}, {Allende Prieto}, {Cunha}, {Hasselquist},
  {Holtzman}, {Majewski}, {Nataf}, {O'Connell}, {Schultheis}, \&
  {Smith}}]{Horta2021}
{Horta}, D., {Schiavon}, R.~P., {Mackereth}, J.~T., {et~al.} 2021, \mnras, 500,
  1385

\bibitem[{{Horta} {et~al.}(2023){Horta}, {Schiavon}, {Mackereth}, {Weinberg},
  {Hasselquist}, {Feuillet}, {O'Connell}, {Anguiano}, {Allende-Prieto},
  {Beaton}, {Bizyaev}, {Cunha}, {Geisler}, {Garc{\'\i}a-Hern{\'a}ndez},
  {Holtzman}, {J{\"o}nsson}, {Lane}, {Majewski}, {M{\'e}sz{\'a}ros}, {Minniti},
  {Nitschelm}, {Shetrone}, {Smith}, \& {Zasowski}}]{Horta2023}
{Horta}, D., {Schiavon}, R.~P., {Mackereth}, J.~T., {et~al.} 2023, \mnras, 520,
  5671

\bibitem[{{Hughes} {et~al.}(2022){Hughes}, {Spitler}, {Zucker}, {Nordlander},
  {Simpson}, {da Costa}, {Ting}, {Li}, {Bland-Hawthorn}, {Buder}, {Casey}, {de
  Silva}, {D'Orazi}, {Freeman}, {Hayden}, {Kos}, {Lewis}, {Lin}, {Lind},
  {Martell}, {Schlesinger}, {Sharma}, {Zwitter}, \& {GALAH
  Collaboration}}]{Hughes2022}
{Hughes}, A. C.~N., {Spitler}, L.~R., {Zucker}, D.~B., {et~al.} 2022, \apj,
  930, 47

\bibitem[{{Ibata} {et~al.}(2021){Ibata}, {Malhan}, {Martin}, {Aubert},
  {Famaey}, {Bianchini}, {Monari}, {Siebert}, {Thomas}, {Bellazzini},
  {Bonifacio}, {Caffau}, \& {Renaud}}]{Ibata2021}
{Ibata}, R., {Malhan}, K., {Martin}, N., {et~al.} 2021, \apj, 914, 123

\bibitem[{{Ibata} {et~al.}(1994){Ibata}, {Gilmore}, \& {Irwin}}]{Ibata1994}
{Ibata}, R.~A., {Gilmore}, G., \& {Irwin}, M.~J. 1994, \nat, 370, 194

\bibitem[{{Jean-Baptiste} {et~al.}(2017){Jean-Baptiste}, {Di Matteo},
  {Haywood}, {G{\'o}mez}, {Montuori}, {Combes}, \&
  {Semelin}}]{JeanBaptiste2017}
{Jean-Baptiste}, I., {Di Matteo}, P., {Haywood}, M., {et~al.} 2017, \aap, 604,
  A106

\bibitem[{{Juri{\'c}} {et~al.}(2008){Juri{\'c}}, {Ivezi{\'c}}, {Brooks},
  {Lupton}, {Schlegel}, {Finkbeiner}, {Padmanabhan}, {Bond}, {Sesar},
  {Rockosi}, {Knapp}, {Gunn}, {Sumi}, {Schneider}, {Barentine}, {Brewington},
  {Brinkmann}, {Fukugita}, {Harvanek}, {Kleinman}, {Krzesinski}, {Long},
  {Neilsen}, {Nitta}, {Snedden}, \& {York}}]{Juric2008}
{Juri{\'c}}, M., {Ivezi{\'c}}, {\v{Z}}., {Brooks}, A., {et~al.} 2008, \apj,
  673, 864

\bibitem[{{Kielty} {et~al.}(2021){Kielty}, {Venn}, {Sestito}, {Starkenburg},
  {Martin}, {Aguado}, {Arentsen}, {Fabbro}, {Gonz{\'a}lez Hern{\'a}ndez},
  {Hill}, {Jablonka}, {Lardo}, {Mashonkina}, {Navarro}, {Sneden}, {Thomas},
  {Youakim}, {Bialek}, \& {S{\'a}nchez-Janssen}}]{Kielty2021}
{Kielty}, C.~L., {Venn}, K.~A., {Sestito}, F., {et~al.} 2021, \mnras, 506, 1438

\bibitem[{{Kirby} {et~al.}(2011){Kirby}, {Cohen}, {Smith}, {Majewski}, {Sohn},
  \& {Guhathakurta}}]{Kirby2011}
{Kirby}, E.~N., {Cohen}, J.~G., {Smith}, G.~H., {et~al.} 2011, \apj, 727, 79

\bibitem[{{Kirby} {et~al.}(2010){Kirby}, {Guhathakurta}, {Simon}, {Geha},
  {Rockosi}, {Sneden}, {Cohen}, {Sohn}, {Majewski}, \& {Siegel}}]{Kirby2010}
{Kirby}, E.~N., {Guhathakurta}, P., {Simon}, J.~D., {et~al.} 2010, \apjs, 191,
  352

\bibitem[{{Kobayashi} {et~al.}(2020){Kobayashi}, {Karakas}, \&
  {Lugaro}}]{Kobayashi2020}
{Kobayashi}, C., {Karakas}, A.~I., \& {Lugaro}, M. 2020, \apj, 900, 179

\bibitem[{{Koch-Hansen} {et~al.}(2021){Koch-Hansen}, {Hansen}, \&
  {McWilliam}}]{KochHansen2021}
{Koch-Hansen}, A.~J., {Hansen}, C.~J., \& {McWilliam}, A. 2021, \aap, 653, A2

\bibitem[{{Koppelman} {et~al.}(2020){Koppelman}, {Bos}, \&
  {Helmi}}]{Koppelman2020}
{Koppelman}, H.~H., {Bos}, R. O.~Y., \& {Helmi}, A. 2020, \aap, 642, L18

\bibitem[{{Koppelman} {et~al.}(2019){Koppelman}, {Helmi}, {Massari},
  {Price-Whelan}, \& {Starkenburg}}]{Koppelman2019}
{Koppelman}, H.~H., {Helmi}, A., {Massari}, D., {Price-Whelan}, A.~M., \&
  {Starkenburg}, T.~K. 2019, \aap, 631, L9

\bibitem[{{Kordopatis} {et~al.}(2020){Kordopatis}, {Recio-Blanco},
  {Schultheis}, \& {Hill}}]{Kordopatis2020}
{Kordopatis}, G., {Recio-Blanco}, A., {Schultheis}, M., \& {Hill}, V. 2020,
  \aap, 643, A69

\bibitem[{{Kos} {et~al.}(2018){Kos}, {Bland-Hawthorn}, {Freeman}, {Buder},
  {Traven}, {De Silva}, {Sharma}, {Asplund}, {Duong}, {Lin}, {Lind}, {Martell},
  {Simpson}, {Stello}, {Zucker}, {Zwitter}, {Anguiano}, {Da Costa}, {D'Orazi},
  {Horner}, {Kafle}, {Lewis}, {Munari}, {Nataf}, {Ness}, {Reid}, {Schlesinger},
  {Ting}, \& {Wyse}}]{Kos2018}
{Kos}, J., {Bland-Hawthorn}, J., {Freeman}, K., {et~al.} 2018, \mnras, 473,
  4612

\bibitem[{{Kruijssen} {et~al.}(2020){Kruijssen}, {Pfeffer}, {Chevance},
  {Bonaca}, {Trujillo-Gomez}, {Bastian}, {Reina-Campos}, {Crain}, \&
  {Hughes}}]{Kruijssen2020}
{Kruijssen}, J.~M.~D., {Pfeffer}, J.~L., {Chevance}, M., {et~al.} 2020, \mnras,
  498, 2472

\bibitem[{{Kruijssen} {et~al.}(2019{\natexlab{a}}){Kruijssen}, {Pfeffer},
  {Reina-Campos}, {Crain}, \& {Bastian}}]{Kruijssen2019a}
{Kruijssen}, J.~M.~D., {Pfeffer}, J.~L., {Reina-Campos}, M., {Crain}, R.~A., \&
  {Bastian}, N. 2019{\natexlab{a}}, \mnras, 486, 3180

\bibitem[{{Kruijssen} {et~al.}(2019{\natexlab{b}}){Kruijssen}, {Pfeffer},
  {Reina-Campos}, {Crain}, \& {Bastian}}]{Kruijssen2019b}
{Kruijssen}, J.~M.~D., {Pfeffer}, J.~L., {Reina-Campos}, M., {Crain}, R.~A., \&
  {Bastian}, N. 2019{\natexlab{b}}, \mnras, 486, 3180

\bibitem[{Kullback \& Leibler(1951)}]{kullback1951}
Kullback, S. \& Leibler, R.~A. 1951, The annals of mathematical statistics, 22,
  79

\bibitem[{{Lane} {et~al.}(2023){Lane}, {Bovy}, \& {Mackereth}}]{Lane2023}
{Lane}, J., {Bovy}, J., \& {Mackereth}, T. 2023, arXiv e-prints,
  arXiv:2306.03084

\bibitem[{{Lane} {et~al.}(2022){Lane}, {Bovy}, \& {Mackereth}}]{Lane2022}
{Lane}, J. M.~M., {Bovy}, J., \& {Mackereth}, J.~T. 2022, \mnras, 510, 5119

\bibitem[{{Li} {et~al.}(2018){Li}, {Zhao}, {Zhai}, \& {Jia}}]{Li2018}
{Li}, C., {Zhao}, G., {Zhai}, M., \& {Jia}, Y. 2018, \apj, 860, 53

\bibitem[{{Limberg} {et~al.}(2021){Limberg}, {Santucci}, {Rossi}, {Queiroz},
  {Chiappini}, {Souza}, {Perottoni}, {P{\'e}rez-Villegas}, \&
  {Barbosa}}]{Limberg2021c}
{Limberg}, G., {Santucci}, R.~M., {Rossi}, S., {et~al.} 2021, \apjl, 913, L28

\bibitem[{{Limberg} {et~al.}(2022){Limberg}, {Souza}, {P{\'e}rez-Villegas},
  {Rossi}, {Perottoni}, \& {Santucci}}]{Limberg2022}
{Limberg}, G., {Souza}, S.~O., {P{\'e}rez-Villegas}, A., {et~al.} 2022, \apj,
  935, 109

\bibitem[{{L{\"o}vdal} {et~al.}(2022){L{\"o}vdal}, {Ruiz-Lara}, {Koppelman},
  {Matsuno}, {Dodd}, \& {Helmi}}]{Lovdal2022}
{L{\"o}vdal}, S.~S., {Ruiz-Lara}, T., {Koppelman}, H.~H., {et~al.} 2022, \aap,
  665, A57

\bibitem[{{Mackereth} \& {Bovy}(2020)}]{MackerethBovy2020}
{Mackereth}, J.~T. \& {Bovy}, J. 2020, \mnras, 492, 3631

\bibitem[{{Mackereth} {et~al.}(2019){Mackereth}, {Schiavon}, {Pfeffer},
  {Hayes}, {Bovy}, {Anguiano}, {Allende Prieto}, {Hasselquist}, {Holtzman},
  {Johnson}, {Majewski}, {O'Connell}, {Shetrone}, {Tissera}, \&
  {Fern{\'a}ndez-Trincado}}]{Mackereth2019}
{Mackereth}, J.~T., {Schiavon}, R.~P., {Pfeffer}, J., {et~al.} 2019, \mnras,
  482, 3426

\bibitem[{{Majewski} {et~al.}(1996){Majewski}, {Munn}, \&
  {Hawley}}]{Majewski1996}
{Majewski}, S.~R., {Munn}, J.~A., \& {Hawley}, S.~L. 1996, \apjl, 459, L73

\bibitem[{{Majewski} {et~al.}(2003){Majewski}, {Skrutskie}, {Weinberg}, \&
  {Ostheimer}}]{Majewski2003}
{Majewski}, S.~R., {Skrutskie}, M.~F., {Weinberg}, M.~D., \& {Ostheimer}, J.~C.
  2003, \apj, 599, 1082

\bibitem[{{Malhan} {et~al.}(2022){Malhan}, {Ibata}, {Sharma}, {Famaey},
  {Bellazzini}, {Carlberg}, {D'Souza}, {Yuan}, {Martin}, \&
  {Thomas}}]{Malhan2022}
{Malhan}, K., {Ibata}, R.~A., {Sharma}, S., {et~al.} 2022, \apj, 926, 107

\bibitem[{{Martin} {et~al.}(2022){Martin}, {Ibata}, {Starkenburg}, {Yuan},
  {Malhan}, {Bellazzini}, {Viswanathan}, {Aguado}, {Arentsen}, {Bonifacio},
  {Carlberg}, {Gonz{\'a}lez Hern{\'a}ndez}, {Hill}, {Jablonka}, {Kordopatis},
  {Lardo}, {McConnachie}, {Navarro}, {S{\'a}nchez-Janssen}, {Sestito},
  {Thomas}, {Venn}, {Vitali}, \& {Voggel}}]{Martin2022}
{Martin}, N.~F., {Ibata}, R.~A., {Starkenburg}, E., {et~al.} 2022, \mnras, 516,
  5331

\bibitem[{{Massari} {et~al.}(2019){Massari}, {Koppelman}, \&
  {Helmi}}]{Massari2019}
{Massari}, D., {Koppelman}, H.~H., \& {Helmi}, A. 2019, \aap, 630, L4

\bibitem[{{Matijevi{\v{c}}} {et~al.}(2017){Matijevi{\v{c}}}, {Chiappini},
  {Grebel}, {Wyse}, {Zwitter}, {Bienaym{\'e}}, {Bland-Hawthorn}, {Freeman},
  {Gibson}, {Gilmore}, {Helmi}, {Kordopatis}, {Kunder}, {Munari}, {Navarro},
  {Parker}, {Reid}, {Seabroke}, {Siviero}, {Steinmetz}, \&
  {Watson}}]{Matijevic2017}
{Matijevi{\v{c}}}, G., {Chiappini}, C., {Grebel}, E.~K., {et~al.} 2017, \aap,
  603, A19

\bibitem[{{Matsuno} {et~al.}(2019){Matsuno}, {Aoki}, \& {Suda}}]{Matsuno2019}
{Matsuno}, T., {Aoki}, W., \& {Suda}, T. 2019, \apjl, 874, L35

\bibitem[{{Matsuno} {et~al.}(2021){Matsuno}, {Hirai}, {Tarumi}, {Hotokezaka},
  {Tanaka}, \& {Helmi}}]{Matsuno2021}
{Matsuno}, T., {Hirai}, Y., {Tarumi}, Y., {et~al.} 2021, \aap, 650, A110

\bibitem[{{Matteucci}(2021)}]{Matteucci2021}
{Matteucci}, F. 2021, \aapr, 29, 5

\bibitem[{{McMillan}(2017)}]{McMillan2017}
{McMillan}, P.~J. 2017, \mnras, 465, 76

\bibitem[{{Montalb{\'a}n} {et~al.}(2021){Montalb{\'a}n}, {Mackereth}, {Miglio},
  {Vincenzo}, {Chiappini}, {Buldgen}, {Mosser}, {Noels}, {Scuflaire}, {Vrard},
  {Willett}, {Davies}, {Hall}, {Nielsen}, {Khan}, {Rendle}, {van Rossem},
  {Ferguson}, \& {Chaplin}}]{Montalban2021}
{Montalb{\'a}n}, J., {Mackereth}, J.~T., {Miglio}, A., {et~al.} 2021, Nature
  Astronomy, 5, 640

\bibitem[{{Monty} {et~al.}(2020){Monty}, {Venn}, {Lane}, {Lokhorst}, \&
  {Yong}}]{Monty2020}
{Monty}, S., {Venn}, K.~A., {Lane}, J. M.~M., {Lokhorst}, D., \& {Yong}, D.
  2020, \mnras, 497, 1236

\bibitem[{{Morrison} {et~al.}(1990){Morrison}, {Flynn}, \&
  {Freeman}}]{Morrison1990}
{Morrison}, H.~L., {Flynn}, C., \& {Freeman}, K.~C. 1990, \aj, 100, 1191

\bibitem[{{Myeong} {et~al.}(2022){Myeong}, {Belokurov}, {Aguado}, {Evans},
  {Caldwell}, \& {Bradley}}]{Myeong2022}
{Myeong}, G.~C., {Belokurov}, V., {Aguado}, D.~S., {et~al.} 2022, \apj, 938, 21

\bibitem[{{Myeong} {et~al.}(2018){Myeong}, {Evans}, {Belokurov}, {Sand ers}, \&
  {Koposov}}]{Myeong2018}
{Myeong}, G.~C., {Evans}, N.~W., {Belokurov}, V., {Sand ers}, J.~L., \&
  {Koposov}, S.~E. 2018, \apjl, 863, L28

\bibitem[{{Myeong} {et~al.}(2019){Myeong}, {Vasiliev}, {Iorio}, {Evans}, \&
  {Belokurov}}]{Myeong2019}
{Myeong}, G.~C., {Vasiliev}, E., {Iorio}, G., {Evans}, N.~W., \& {Belokurov},
  V. 2019, \mnras, 488, 1235

\bibitem[{{Naidu} {et~al.}(2020){Naidu}, {Conroy}, {Bonaca}, {Johnson}, {Ting},
  {Caldwell}, {Zaritsky}, \& {Cargile}}]{Naidu2020}
{Naidu}, R.~P., {Conroy}, C., {Bonaca}, A., {et~al.} 2020, \apj, 901, 48

\bibitem[{{Naidu} {et~al.}(2022){Naidu}, {Ji}, {Conroy}, {Bonaca}, {Ting},
  {Zaritsky}, {van Son}, {Broekgaarden}, {Tacchella}, {Chandra}, {Caldwell},
  {Cargile}, \& {Speagle}}]{Naidu2022}
{Naidu}, R.~P., {Ji}, A.~P., {Conroy}, C., {et~al.} 2022, \apjl, 926, L36

\bibitem[{{Necib} {et~al.}(2020){Necib}, {Ostdiek}, {Lisanti}, {Cohen},
  {Freytsis}, {Garrison-Kimmel}, {Hopkins}, {Wetzel}, \&
  {Sanderson}}]{Necib2020}
{Necib}, L., {Ostdiek}, B., {Lisanti}, M., {et~al.} 2020, Nature Astronomy, 4,
  1078

\bibitem[{{Newberg} {et~al.}(2002){Newberg}, {Yanny}, {Rockosi}, {Grebel},
  {Rix}, {Brinkmann}, {Csabai}, {Hennessy}, {Hindsley}, {Ibata}, {Ivezi{\'c}},
  {Lamb}, {Nash}, {Odenkirchen}, {Rave}, {Schneider}, {Smith}, {Stolte}, \&
  {York}}]{Newberg2002}
{Newberg}, H.~J., {Yanny}, B., {Rockosi}, C., {et~al.} 2002, \apj, 569, 245

\bibitem[{{Newberg} {et~al.}(2009){Newberg}, {Yanny}, \&
  {Willett}}]{Newberg2009}
{Newberg}, H.~J., {Yanny}, B., \& {Willett}, B.~A. 2009, \apjl, 700, L61

\bibitem[{{Norris} {et~al.}(1985){Norris}, {Bessell}, \&
  {Pickles}}]{Norris1985}
{Norris}, J., {Bessell}, M.~S., \& {Pickles}, A.~J. 1985, \apjs, 58, 463

\bibitem[{{Ou} {et~al.}(2023){Ou}, {Necib}, \& {Frebel}}]{Ou2023}
{Ou}, X., {Necib}, L., \& {Frebel}, A. 2023, \mnras, 521, 2623

\bibitem[{{Pagnini} {et~al.}(2022){Pagnini}, {Di Matteo}, {Khoperskov},
  {Mastrobuono-Battisti}, {Haywood}, {Renaud}, \& {Combes}}]{Pagnini22}
{Pagnini}, G., {Di Matteo}, P., {Khoperskov}, S., {et~al.} 2022, arXiv
  e-prints, arXiv:2210.04245

\bibitem[{Pedregosa {et~al.}(2011)Pedregosa, Varoquaux, Gramfort, Michel,
  Thirion, Grisel, Blondel, Prettenhofer, Weiss, Dubourg,
  {et~al.}}]{pedregosa2011}
Pedregosa, F., Varoquaux, G., Gramfort, A., {et~al.} 2011, the Journal of
  machine Learning research, 12, 2825

\bibitem[{{Perottoni} {et~al.}(2019){Perottoni}, {Martin}, {Newberg},
  {Rocha-Pinto}, {Almeida-Fernandes}, \& {Gomes-J{\'u}nior}}]{Perottoni2019}
{Perottoni}, H.~D., {Martin}, C., {Newberg}, H.~J., {et~al.} 2019, \mnras, 486,
  843

\bibitem[{{Queiroz} {et~al.}(2023){Queiroz}, {Anders}, {Chiappini},
  {Khalatyan}, {Santiago}, {Nepal}, {Steinmetz}, {Gallart}, {Valentini}, {Dal
  Ponte}, {Barbuy}, {P{\'e}rez-Villegas}, {Masseron}, {Fern{\'a}ndez-Trincado},
  {Khoperskov}, {Minchev}, {Fern{\'a}ndez-Alvar}, {Lane}, \&
  {Nitschelm}}]{Queiroz2023}
{Queiroz}, A. B.~A., {Anders}, F., {Chiappini}, C., {et~al.} 2023, arXiv
  e-prints, arXiv:2303.09926

\bibitem[{{Randich} {et~al.}(2022){Randich}, {Gilmore}, {Magrini}, {Sacco},
  {Jackson}, {Jeffries}, {Worley}, {Hourihane}, {Gonneau}, {Viscasillas
  Vazquez}, {Franciosini}, {Lewis}, {Alfaro}, {Allende Prieto}, {Bensby},
  {Blomme}, {Bragaglia}, {Flaccomio}, {Fran{\c{c}}ois}, {Irwin}, {Koposov},
  {Korn}, {Lanzafame}, {Pancino}, {Recio-Blanco}, {Smiljanic}, {Van Eck},
  {Zwitter}, {Asplund}, {Bonifacio}, {Feltzing}, {Binney}, {Drew}, {Ferguson},
  {Micela}, {Negueruela}, {Prusti}, {Rix}, {Vallenari}, {Bayo}, {Bergemann},
  {Biazzo}, {Carraro}, {Casey}, {Damiani}, {Frasca}, {Heiter}, {Hill},
  {Jofr{\'e}}, {de Laverny}, {Lind}, {Marconi}, {Martayan}, {Masseron},
  {Monaco}, {Morbidelli}, {Prisinzano}, {Sbordone}, {Sousa}, {Zaggia},
  {Adibekyan}, {Bonito}, {Caffau}, {Daflon}, {Feuillet}, {Gebran}, {Gonzalez
  Hernandez}, {Guiglion}, {Herrero}, {Lobel}, {Maiz Apellaniz}, {Merle},
  {Mikolaitis}, {Montes}, {Morel}, {Soubiran}, {Spina}, {Tabernero},
  {Tautvai{\v{s}}iene}, {Traven}, {Valentini}, {Van der Swaelmen}, {Villanova},
  {Wright}, {Abbas}, {Aguirre B{\o}rsen-Koch}, {Alves}, {Balaguer-Nunez},
  {Barklem}, {Barrado}, {Berlanas}, {Binks}, {Bressan}, {Capuzzo-Dolcetta},
  {Casagrande}, {Casamiquela}, {Collins}, {D'Orazi}, {Dantas}, {Debattista},
  {Delgado-Mena}, {Di Marcantonio}, {Drazdauskas}, {Evans}, {Famaey},
  {Franchini}, {Fr{\'e}mat}, {Friel}, {Fu}, {Geisler}, {Gerhard}, {Gonzalez
  Solares}, {Grebel}, {Gutierrez Albarran}, {Hatzidimitriou}, {Held},
  {Jim{\'e}nez-Esteban}, {J{\"o}nsson}, {Jordi}, {Khachaturyants},
  {Kordopatis}, {Kos}, {Lagarde}, {Mahy}, {Mapelli}, {Marfil}, {Martell},
  {Messina}, {Miglio}, {Minchev}, {Moitinho}, {Montalban}, {Monteiro},
  {Morossi}, {Mowlavi}, {Mucciarelli}, {Murphy}, {Nardetto}, {Ortolani},
  {Paletou}, {Palou{\v{s}}}, {Paunzen}, {Pickering}, {Quirrenbach}, {Re
  Fiorentin}, {Read}, {Romano}, {Ryde}, {Sanna}, {Santos}, {Seabroke},
  {Spagna}, {Steinmetz}, {Stonkut{\'e}}, {Sutorius}, {Th{\'e}venin}, {Tosi},
  {Tsantaki}, {Vink}, {Wright}, {Wyse}, {Zoccali}, {Zorec}, {Zucker}, \&
  {Walton}}]{Randich2022}
{Randich}, S., {Gilmore}, G., {Magrini}, L., {et~al.} 2022, \aap, 666, A121

\bibitem[{{Recio-Blanco} {et~al.}(2014){Recio-Blanco}, {de Laverny},
  {Kordopatis}, {Helmi}, {Hill}, {Gilmore}, {Wyse}, {Adibekyan}, {Randich},
  {Asplund}, {Feltzing}, {Jeffries}, {Micela}, {Vallenari}, {Alfaro}, {Allende
  Prieto}, {Bensby}, {Bragaglia}, {Flaccomio}, {Koposov}, {Korn}, {Lanzafame},
  {Pancino}, {Smiljanic}, {Jackson}, {Lewis}, {Magrini}, {Morbidelli},
  {Prisinzano}, {Sacco}, {Worley}, {Hourihane}, {Bergemann}, {Costado},
  {Heiter}, {Joffre}, {Lardo}, {Lind}, \& {Maiorca}}]{Recio-Blanco2014}
{Recio-Blanco}, A., {de Laverny}, P., {Kordopatis}, G., {et~al.} 2014, \aap,
  567, A5

\bibitem[{{Rey} {et~al.}(2023){Rey}, {Agertz}, {Starkenburg}, {Renaud},
  {Joshi}, {Pontzen}, {Martin}, {Feuillet}, \& {Read}}]{Rey2023}
{Rey}, M.~P., {Agertz}, O., {Starkenburg}, T.~K., {et~al.} 2023, \mnras, 521,
  995

\bibitem[{{Ruiz-Lara} {et~al.}(2022){Ruiz-Lara}, {Helmi}, {Gallart}, {Surot},
  \& {Cassisi}}]{RuizLara2022}
{Ruiz-Lara}, T., {Helmi}, A., {Gallart}, C., {Surot}, F., \& {Cassisi}, S.
  2022, \aap, 668, L10

\bibitem[{{Searle} \& {Zinn}(1978)}]{Searle1978}
{Searle}, L. \& {Zinn}, R. 1978, \apj, 225, 357

\bibitem[{{Shank} {et~al.}(2023){Shank}, {Beers}, {Placco}, {Gudin},
  {Catapano}, {Holmbeck}, {Ezzeddine}, {Roederer}, {Sakari}, {Frebel}, \&
  {Hansen}}]{Shank2023}
{Shank}, D., {Beers}, T.~C., {Placco}, V.~M., {et~al.} 2023, \apj, 943, 23

\bibitem[{{Shank} {et~al.}(2022{\natexlab{a}}){Shank}, {Beers}, {Placco},
  {Limberg}, {Jaques}, {Yuan}, {Schlaufman}, {Casey}, {Huang}, {Lee},
  {Hattori}, \& {Santucci}}]{Shank2022a}
{Shank}, D., {Beers}, T.~C., {Placco}, V.~M., {et~al.} 2022{\natexlab{a}},
  \apj, 926, 26

\bibitem[{{Shank} {et~al.}(2022{\natexlab{b}}){Shank}, {Komater}, {Beers},
  {Placco}, \& {Huang}}]{Shank2022b}
{Shank}, D., {Komater}, D., {Beers}, T.~C., {Placco}, V.~M., \& {Huang}, Y.
  2022{\natexlab{b}}, \apjs, 261, 19

\bibitem[{{Sk{\'u}lad{\'o}ttir} \& {Salvadori}(2020)}]{Asa2020}
{Sk{\'u}lad{\'o}ttir}, {\'A}. \& {Salvadori}, S. 2020, \aap, 634, L2

\bibitem[{{Spergel} {et~al.}(2007){Spergel}, {Bean}, {Dor{\'e}}, {Nolta},
  {Bennett}, {Dunkley}, {Hinshaw}, {Jarosik}, {Komatsu}, {Page}, {Peiris},
  {Verde}, {Halpern}, {Hill}, {Kogut}, {Limon}, {Meyer}, {Odegard}, {Tucker},
  {Weiland}, {Wollack}, \& {Wright}}]{Spergel2007}
{Spergel}, D.~N., {Bean}, R., {Dor{\'e}}, O., {et~al.} 2007, \apjs, 170, 377

\bibitem[{{Springel} {et~al.}(2006){Springel}, {Frenk}, \&
  {White}}]{Springel2006}
{Springel}, V., {Frenk}, C.~S., \& {White}, S. D.~M. 2006, \nat, 440, 1137

\bibitem[{{Tolstoy} {et~al.}(2003){Tolstoy}, {Venn}, {Shetrone}, {Primas},
  {Hill}, {Kaufer}, \& {Szeifert}}]{Tolstoy2003}
{Tolstoy}, E., {Venn}, K.~A., {Shetrone}, M., {et~al.} 2003, \aj, 125, 707

\bibitem[{{Traven} {et~al.}(2020){Traven}, {Feltzing}, {Merle}, {Van der
  Swaelmen}, {{\v{C}}otar}, {Church}, {Zwitter}, {Ting}, {Sahlholdt},
  {Asplund}, {Bland-Hawthorn}, {De Silva}, {Freeman}, {Martell}, {Sharma},
  {Zucker}, {Buder}, {Casey}, {D'Orazi}, {Kos}, {Lewis}, {Lin}, {Lind},
  {Simpson}, {Stello}, {Munari}, \& {Wittenmyer}}]{Traven2020}
{Traven}, G., {Feltzing}, S., {Merle}, T., {et~al.} 2020, \aap, 638, A145

\bibitem[{{Traven} {et~al.}(2017){Traven}, {Matijevi{\v{c}}}, {Zwitter},
  {{\v{Z}}erjal}, {Kos}, {Asplund}, {Bland-Hawthorn}, {Casey}, {De Silva},
  {Freeman}, {Lin}, {Martell}, {Schlesinger}, {Sharma}, {Simpson}, {Zucker},
  {Anguiano}, {Da Costa}, {Duong}, {Horner}, {Hyde}, {Kafle}, {Munari},
  {Nataf}, {Navin}, {Reid}, \& {Ting}}]{Traven2017}
{Traven}, G., {Matijevi{\v{c}}}, G., {Zwitter}, T., {et~al.} 2017, \apjs, 228,
  24

\bibitem[{{Vargas} {et~al.}(2013){Vargas}, {Geha}, {Kirby}, \&
  {Simon}}]{Vargas2013}
{Vargas}, L.~C., {Geha}, M., {Kirby}, E.~N., \& {Simon}, J.~D. 2013, \apj, 767,
  134

\bibitem[{{Vasiliev}(2019)}]{Vasiliev2019}
{Vasiliev}, E. 2019, \mnras, 484, 2832

\bibitem[{{Venn} {et~al.}(2004){Venn}, {Irwin}, {Shetrone}, {Tout}, {Hill}, \&
  {Tolstoy}}]{Venn2004}
{Venn}, K.~A., {Irwin}, M., {Shetrone}, M.~D., {et~al.} 2004, \aj, 128, 1177

\bibitem[{{Vincenzo} {et~al.}(2019){Vincenzo}, {Spitoni}, {Calura},
  {Matteucci}, {Silva Aguirre}, {Miglio}, \& {Cescutti}}]{Vincenzo2019}
{Vincenzo}, F., {Spitoni}, E., {Calura}, F., {et~al.} 2019, \mnras, 487, L47

\bibitem[{{Viswanathan} {et~al.}(2023){Viswanathan}, {Starkenburg},
  {Koppelman}, {Helmi}, {Balbinot}, \& {Esselink}}]{Viswanathan2023}
{Viswanathan}, A., {Starkenburg}, E., {Koppelman}, H.~H., {et~al.} 2023,
  \mnras, 521, 2087

\bibitem[{{Ward Jr.}(1963)}]{Ward1963}
{Ward Jr.}, J.~H. 1963, Journal of the American Statistical Association, 58,
  236

\bibitem[{Wattenberg {et~al.}(2016)Wattenberg, Viégas, \&
  Johnson}]{wattenberg2016}
Wattenberg, M., Viégas, F., \& Johnson, I. 2016, Distill

\bibitem[{{White} \& {Frenk}(1991)}]{WhiteFrenk1991}
{White}, S. D.~M. \& {Frenk}, C.~S. 1991, \apj, 379, 52

\bibitem[{{Yuan} {et~al.}(2020){Yuan}, {Chang}, {Beers}, \& {Huang}}]{Yuan2020}
{Yuan}, Z., {Chang}, J., {Beers}, T.~C., \& {Huang}, Y. 2020, \apjl, 898, L37

\bibitem[{{Zepeda} {et~al.}(2023){Zepeda}, {Beers}, {Placco}, {Shank}, {Gudin},
  {Hirai}, {Mardini}, {Pifer}, {Catapano}, \& {Calagna}}]{Zepeda2023}
{Zepeda}, J., {Beers}, T.~C., {Placco}, V.~M., {et~al.} 2023, \apj, 947, 23

\end{thebibliography}

\begin{appendix} 
\section{Additional figures}\label{sec:appendix}

This Appendix includes some additional figures that are useful for comparison with results displayed and discussed in the main body of the text.

\begin{figure*}
    \centering
    \includegraphics[width=.45\linewidth]{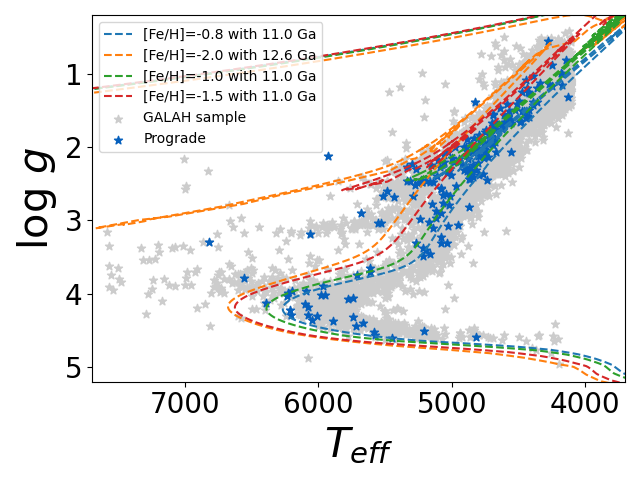}
    \includegraphics[width=.45\linewidth]{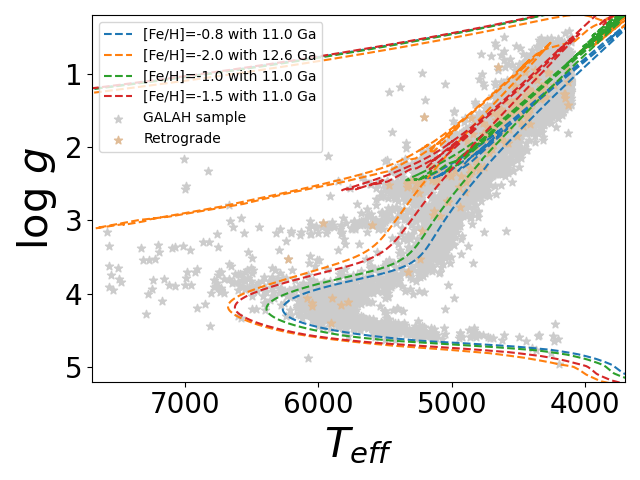}\\
    \includegraphics[width=.45\linewidth]{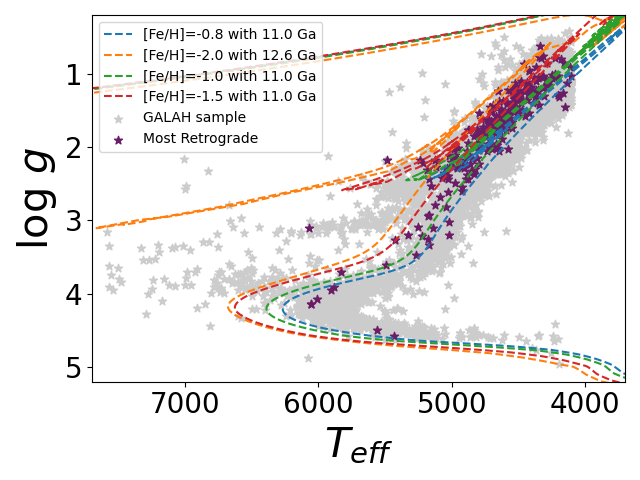}
    \caption{Kiel diagrams of the stars in the prograde, retrograde, and most retrograde groups in the top left, top right, and bottom panels, respectively. The isochrones are the same as in Fig.~\ref{fig:GE_met_isochrones}. The grey stars on the background are the sample of metal-poor stars selected from GALAH DR3.}
    \label{fig:appendix_kiel}
\end{figure*}

Figure \ref{fig:appendix_kiel} shows the Kiel diagram for the stars included in the prograde, retrograde, and most retrograde groups (top left, top right, and bottom panels, respectively). It should be compared to the Kiel diagram of the GE-dominated group, left panel of Fig.~\ref{fig:GE_met_isochrones}. All four groups include stars from the main sequence to the red-giant branch, with the largest fraction being giants. The most retrograde group has the smallest number of dwarfs. Considering as dwarfs those stars with $\log~g$ $\geq$ 3.5, the numbers for the prograde, GE-dominated, retrograde, and most retrograde groups are: 30 dwarfs and 168 giants; 43 dwarfs and 274 giants; 10 dwarfs and 166 giants; 9 dwarfs and 244 giants, respectively.

\begin{figure*}
    \centering
    \includegraphics[width=\linewidth]{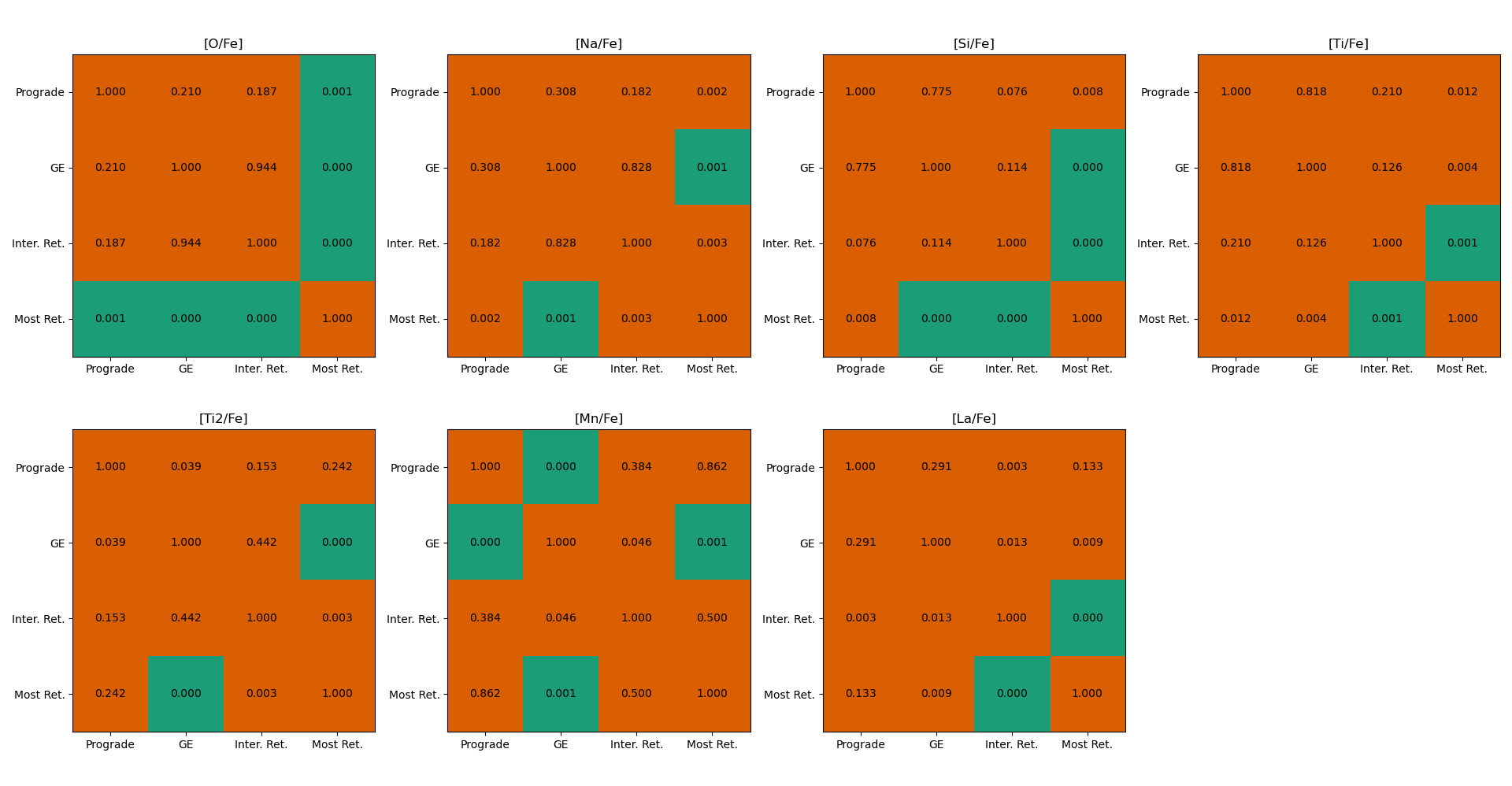}
    \caption{Results of the KS test applied to the chemical abundances of only the giant stars ($1.5\leq$\logg$\leq 2.5$) in all four halo groups defined in this work. Only elements for which abundances were available for at least 10 stars were tested. The figure displays only the elements for which differences were found.}
    \label{fig:appendix_ks_giants}
\end{figure*}

Figure \ref{fig:appendix_ks_giants} presents results of KS tests applied to the chemical abundances of only giants. For this exercise, we defined as giants those stars with 1.5 $\leq$ $\log~g$ $\leq$ 2.5, the interval with the largest number of stars. The restriction was also motivated as an effort to decrease the influence of possible systematic errors that can depend on atmospheric parameters. The results should be compared with those in Fig. \ref{fig:KStest}, discussed in Section \ref{sec:ks_abun}.

For the giant sample, some of the chemical differences are the same detected in the analysis of the larger sample: O, Na, Si, and Mn. However, differences are now also seen in Ti and La. Again, the differences are seen strongly in the most retrograde group in comparison to the others. The differences detected before in the abundances of Y and Ba disappeared. We also performed KS tests restricting the sample to dwarf stars (3.5 $\leq$ $\log~g$ $\leq$ 4.5). For dwarfs, there were no differences in any of the abundances. The different results obtained for dwarfs and giants hint at systematic effects that affect the chemical abundances in the GALAH DR3 results. Nevertheless, we note that the sample of dwarfs is considerably smaller, which can make potential differences in the distributions harder to detect. In summary, these tests support our conclusion that the most retrograde group is the one that shows clear chemical differences with respect to the others. However, the fact that the behaviour is different between dwarfs and giants, and also between the full sample and the restricted giant sample, also stresses our point that we need to strive for abundances of higher quality to firmly establish chemical differences between the different accreted and in situ halo populations.

\end{appendix}
\end{document}